\documentclass{aa}  
\AddToHook{begindocument/before}{\usepackage{nameref}}

\usepackage{xcolor}
\usepackage[colorlinks=true,allcolors=blue]{hyperref}
\usepackage{graphicx}
\usepackage{txfonts}

\def\teff{$\textit{T}_{\rm{eff}}$}
\def\logg{\mbox{log~{\it g}}}

\begin{document}

   \title{Playing CHESS with stars\thanks{Based on data obtained from the ESO Science Archive Facility}}

   \subtitle{I. Search for similar stars in large spectroscopic data sets}

 \authorrunning{Mart\'{\i}nez Fern\'andez, \"Ozdemir, et al.}

    \titlerunning{CHESS I. Search for similar stars in large spectroscopic datasets}

   \author{J.~E.~Mart\'{\i}nez Fern\'andez
          \inst{1}
          \and
          S.~\"Ozdemir\inst{1}
          \and
          R.~Smiljanic\inst{1}
          \and
          M.~L.~L.~Dantas\inst{2,3}
          \and
          A.~R.~da Silva\inst{1}
          }

   \institute{
    Nicolaus Copernicus Astronomical Center, Polish Academy of Sciences, ul. Bartycka 18, 00-716, Warsaw, Poland\\\email{johnedmf@camk.edu.pl; sergen@camk.edu.pl}
    \and
    Instituto de Astrofísica, Pontificia Universidad Católica de Chile, Av. Vicuña Mackenna 4860, Santiago, Chile 
    \and
    Centro de Astro-Ingeniería, Pontificia Universidad Católica de Chile, Av. Vicuña Mackenna 4860, Santiago, Chile
    }

   \date{Received XXX, 2025; accepted YYY, 2025}

  \abstract
   {Massive amounts of spectroscopic data obtained by stellar surveys are feeding an ongoing revolution in our knowledge of stellar and Galactic astrophysics. Analysing these data sets to extract the best possible astrophysical parameters on short time scales represents a considerable challenge.}
   {The differential analysis method is known to return the most precise results in the spectroscopic analyses of F-, G-, and K-type stars. However, it can only be applied to stars with similar parameters. Our goal is to present a procedure that significantly simplifies the identification of spectra from stars with similar atmospheric parameters within extensive spectral datasets. This approach allows for the quick application of differential analyses in these samples, thus enhancing the precision of the results.}
   {We used projection maps created by the t-SNE dimensionality reduction algorithm applied directly to the spectra using pixels as dimensions. For testing the method, we used more than 7300 high-resolution UVES spectra of about 3000 stars in the field-of-view towards open and globular clusters. As reference, we used 1244 spectra of 274 stars with well-determined and high-quality atmospheric parameters.}
   {We calibrated a spectral similarity metric that can identify stars in a t-SNE projection map with parameters that differ by $\pm$ 200 K, $\pm$ 0.3 dex, and $\pm$ 0.2 dex in effective temperatures, surface gravities, and metallicities, respectively. We achieved completeness between 74-98 \% and typical purity between 39-54\% in this selection. With this, we will drastically facilitate the detection of stars with similar spectra for a successful differential analysis. We used this method to evaluate the accuracy and precision of four atmospheric parameter catalogues, identifying the regions of the parameter space where spectral analysis methods needs improvement.}
   {}


   \maketitle
%
\nolinenumbers
\section{Introduction}
\label{sec:introduction}

Stellar spectra encode information on properties such as detailed surface chemical abundances, rotation, magnetic activity, and mass accretion or loss. Having such data for large samples of stars of different ages, formed in a wide variety of environments, is essential to clarify the origins of the elements, the formation and evolution of planets and stars, and to better understand the complexities of the formation and evolution of the Milky Way. 

Recognising this need for extensive spectroscopic data, several large stellar spectroscopic surveys have been designed and are now feeding an on-going revolution in stellar and Galactic astrophysics. Examples include the \textit{Gaia}-ESO Survey \cite[GES,][]{2022A&A...666A.120G, 2022A&A...666A.121R}, the Galactic Archaeology with HERMES Survey \citep[GALAH,][]{DeSilva2015}, the Large Sky Area Multi-Object Fibre Spectroscopic Telescope \cite[LAMOST,][]{lamost, lamost-mrs}, and the Apache Point Observatory Galactic Evolution Experiment \cite[APOGEE,][]{2017AJ....154...94M}. In the near future, projects such as the 4-metre Multi-Object Spectroscopic Telescope \cite[4MOST,][]{2019Msngr.175....3D} and the WHT Enhanced Area Velocity Explorer \cite[WEAVE,][]{2024MNRAS.530.2688J} will further expand previous efforts in unprecedented ways.

With spectra available for tens of millions of stars, the challenge shifts to efficiently analysing the data on short timescales while maintaining the high quality of the results. This can only be attempted by automatic routines and pipelines \citep[e.g.][to mention only a few]{RecioBlanco2006, Magrini2013, ASPCAP, Hanke2018, Tabernero2022, Li2024}. The most common analysis approach involves radiative transfer calculations based on model atmospheres to fit either the equivalent widths or the profiles of spectral lines. An alternative but related approach involves the matching or interpolation among a grid of template spectra (which can be of observational or synthetic origin).

More recently, data-driven approaches have been used to determine stellar parameters and abundances from large data sets. In a first phase, these methods use an input training grid of spectra to learn the complex relations between the so-called `labels' (stellar parameters and abundances in this case) and the input data. After training, the algorithms can apply what has been learnt to infer the labels of the new data \citep[see e.g.][]{Ness2015, Fabbro18, LeungBovy2019, Ting2019, Xiang2019, Guiglion2020, OBriain2021, Li2022, Nepal2023, Candebat2024, Sizemore24}. These data-driven methods are fast and can reach very high precision. However, creating a high-quality training sample still requires the use of more traditional methods. 

Among the different implementations of the traditional approach to spectroscopic analysis, the differential method produces the most precise results \citep[see e.g.][]{Bedell2014, Nissen2018}. Radiative transfer analyses may suffer systematic errors coming from typical assumptions, such as reducing physical processes to one-dimensional (1D) problems, using the local thermodynamic equilibrium (LTE) approximation, and adopting a simplified treatment of convection \citep[see e.g.][for reviews on these topics]{KupkaMuthsam2017, JoyceTayar2023, LindAmarsi2024}. In a differential analysis, stars of similar atmospheric parameters are analysed with respect to each other. In such a comparison, the systematic errors are assumed to be the same and to cancel out. In this case, the differences in parameters and abundances are determined with very high precision \citep[below 0.01 dex for abundances, see e.g.][]{Bedell2014}, in particular, if the data are of high resolution and have a high signal-to-noise ratio (S/N). 

However, the accuracy of the results is harder to establish. The fundamental determination of effective temperatures (\teff) and surface gravities (\logg) is possible with minimum assumptions, if independent measurements of the stellar bolometric flux, angular diameter, distance, and mass are available \citep[e.g.][]{Boyajian2013, Creevey2015, Kiman2024, 2024A&A...682A.145S, Pinsonneault2024}. Metallicities and abundances, on the other hand, need to be inferred from modelling, with the exception of the Solar System, where meteorites can be used for independent (but not assumption-free) measurements \citep{Alexander2019, Lodders2021}. Accurate determination of stellar parameters is also possible with state-of-the-art three-dimensional (3D) model atmospheres, as demonstrated, for example, in \citet{2021A&A...650A.194G, 2023A&A...679A.110G}.

Finally, it is worth mentioning that the variety of approaches, method implementations, codes, physical assumptions, and even somewhat subjective decisions (such as continuum placement) create important discrepancies among the results obtained for the same stars in different works \citep{Lebzelter2012, Smiljanic2014, Hinkel2016, Jofre2017, Jonsson2018, Brucalassi2022b}. Systematic biases between surveys remain a concern \citep[see e.g.][]{Soubiran2022, Hegedus2023, VanderSwaelmen2024}. Moreover, since the vast majority of the spectra obtained in surveys is of low resolution, the validation of their analysis in a comparison with observations of higher resolution is important \citep{Karinkuzhi2021, Sandford2023}.

Amid the variety of surveys, spectral databases, and analysis methods, it is a concern that the most and the best possible information are extracted from each spectrum. With this motivation, we are developing a new pipeline called {\sf CHESS} (CHEmical Survey analysis System) which aims to derive precise, accurate, and complete chemical information from large samples of spectra of F-, G-, and K-type stars. It uses the differential analysis method for high precision and a set of reference stars to anchor the parameter scale with high accuracy. 

To identify similar stars for differential analysis, {\sf CHESS} uses unsupervised machine learning for a similarity analysis using the spectra. In this paper, we describe and validate the steps and results of this similarity analysis. We also show that this step can be useful for a blind test of large survey results. Future articles in this series will discuss the results of the spectral analysis using {\sf CHESS} in a variety of science cases. This paper is divided as follows. In Sect. \ref{sec:data} we present the spectroscopic data and the preliminary processing steps. The methodology applied for the similarity analysis is discussed in Sect. \ref{sec:method}. We present and discuss the results in Sect. \ref{sec:results} and summarise our conclusions in Sect. \ref{sec:conclusion}.

\section{Data and preliminary data processing}\label{sec:data}

This work used spectroscopic data publicly available at the Science Archive Facility \citep[SAF,][]{Romaniello2023} of the European Southern Observatory (ESO). The spectra were obtained with the Ultraviolet and Visual Echelle Spectrograph \citep[UVES,][]{UVES} at the Very Large Telescope, Paranal Observatory, Chile. We focused on spectra of FGK-type stars observed in the direction of open and globular clusters. Stars in clusters are very useful for testing methods to determine astrophysical parameters \citep[e.g.][]{Holtzman2015, Pancino2017, Anders2022}. Moreover, focussing on cluster stars maximises the chances of including similar stars in the sample. 

\subsection{Sample of stars in clusters}
\label{subsec:sample_clusters}

As reference for globular cluster coordinates we used the 2010 edition of the `Catalog of parameters for Milky Way globular clusters', originally compiled by \cite{Harris96}, which contains 157 globular clusters in the Milky Way. For open clusters, the catalogue by \cite{2023A&A...673A.114H, 2024A&A...686A..42H} was used. It provides a list of 5647 groups of stars identified as open clusters from an analysis of sources in Data Release 3 (DR3) of \textit{Gaia} \citep{GaiaDR3}.

We performed a cone search at the SAF with a diameter of 20 arcmin centred in the coordinates of each cluster. All spectra of sources in this area were downloaded; membership information was not considered for selection. The selection was restricted to spectra with resolution R $\geq$ 20\,000 and excluded those with a signal-to-noise ratio S/N $<$ 10 (for the current selection --- for future analyses, a higher S/N threshold will be preferred). In total, we downloaded 2557 processed slit UVES spectra from stars in the field of 50 globular clusters and 8262 from stars in the field of 371 open clusters. We note here that several of the collected spectra are data from GES, which observed stars in several open clusters \citep{Bragaglia2022}. This makes the parameters from GES very useful for testing our method of finding similar stars.

\subsection{Cross-matching with catalogues and databases}
\label{subsec:crossmatch}

Properly identifying the object to which a spectrum belongs is paramount to search for complementary data (such as photometry, astrometry, or asteroseismology); to investigate the properties of cluster members at a later stage; and for general legacy purposes. Therefore, we devoted considerable effort to identifying the stars to which our spectra are linked.

Catalogue cross-matching was challenging. One complication was that the format and content of some keywords in the header of the FITS files were found to be different depending on when the observations were taken. In addition, some observations only list pointing coordinates, which indicate the direction in the sky targeted by the telescope (and are affected by the telescope system uncertainties). Differences between such coordinates and the real object positions complicate star identification, particularly in crowded fields. 

Furthermore, object names are frequently non-informative; real examples are \texttt{Tar\_1} or \texttt{star53}. Even in cases where the name is meaningful, it is not necessarily given as an interpretable string that can be used to query a service like the Set of Identifications, Measurements, and Bibliography for Astronomical Data \citep[SIMBAD,][]{Simbad}. An example is the object name \texttt{EGGEN-45}: knowing the cluster around the target, it was possible to search SIMBAD for \texttt{Cl* IC 4651 EGG 45}, which returned an entry with very similar coordinates and confirmed the match.

All of these points limit how findable and reusable the data are. The ever-increasing volume of data from current and future instrumentation requires steps to make data mining as straightforward as possible. Therefore, we urge observatories and data archives to use standardised practices for recording object coordinates and identifiers, and thus ensure the long-term usage and legacy of their data sets.

\begin{figure*}[htbp]
  \centering
  \includegraphics[width=1.0\textwidth]{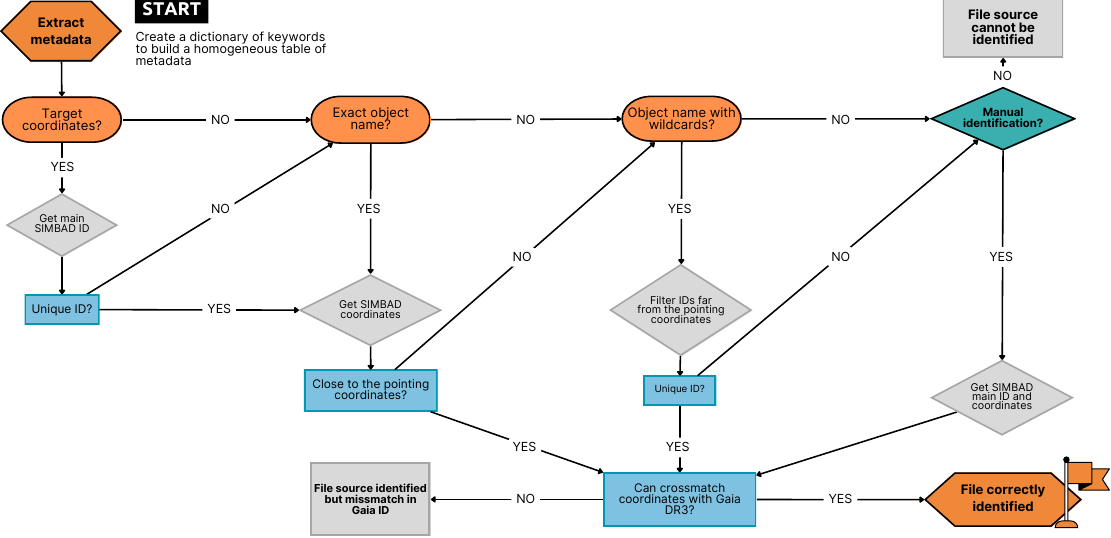}
  \caption{Object identification workflow for discovering the source ID of each spectrum. The steps for crossmatching with SIMBAD and with the \textit{Gaia} DR3 catalogue are depicted in detail.}
  \label{fig:crossmatch_workflow}
\end{figure*}

To overcome these obstacles, we implemented rules to automate stellar identification using a dedicated \textsc{Python} script. However, many cases still required manual intervention. Even then, there were cases where proper identification of a spectrum was not feasible. In the following subsections, we detail the steps in the object identification workflow (summarised in Fig.~\ref{fig:crossmatch_workflow}).

\subsubsection{Cross-match with SIMBAD}

The script queries SIMBAD to retrieve each star's correct identifier and coordinates (RA, DEC). The steps are listed below.

\begin{enumerate}[I.]
    \item If present, the keywords \texttt{ESO TEL TARG ALPHA} and \texttt{ESO TEL TARG DELTA} are used for a cone search in SIMBAD with a radius of 2 arcsec. These coordinates tend to be very accurate. We proceed differently depending on the results:
    
    \begin{itemize}
        \item If more than one source is detected, the correct identification cannot be established unless it refers to a GES object and a \textit{Gaia} DR3 ID.
        
        \item In the case of a single source identification, the pointing coordinates (RA, DEC) are checked to confirm whether the source coordinates are in agreement with those in SIMBAD (within 20 arcsec).
    \end{itemize}
        
    \item If the source identification is uncertain in the previous step, the keyword \texttt{OBJECT} is checked and passed directly to SIMBAD, unless it is a GES spectrum, in which case the name is set to start with `GES J'. If an exact match is found, we verify the pointing coordinates against SIMBAD to confirm the match and detect typos in the object name. 
    \item For unidentified objects in the previous steps, we performed a final search using the header's object name. We excluded names with fewer than five characters or purely numeric values. For the remaining cases, we added asterisk wildcards (*) between letters and numbers and at both ends of the name. This search used a 30 arcsec radius around the pointing coordinates and required complete character matches with the header name. If a single source match is found, it was accepted as the correct identification.
\end{enumerate}

If the spectrum being analysed was still unidentified following the aforementioned steps, we attempted a manual identification. For example, by performing a systematic literature search to find publications that used the original data and provided a list of targets. This process was occasionally unsuccessful when the relevant source catalogues were unpublished or inaccessible.

For stars matched to an entry in SIMBAD, we included a check of the fields \texttt{Object type} and \texttt{Spectral type}. The goal was to exclude sources that are not of our interest. The following categories were excluded from the sample:

\begin{itemize}
    \item \textbf{Object types}\footnote{See \url{http://simbad.cds.unistra.fr/guide/otypes.htx}}: Ae*, Be*, Cl*, HXB, HS*, HS?, Mi*, No*, Or*, PN, PN?, QSO, Sy1, TT*, WD*, WR*, Y*O, Y*?, a2*, pA*
    \item \textbf{Spectral types}: B, O, WC, WN, WR
\end{itemize}

Of the 10\,819 spectra we initially downloaded, we excluded 76 based on object names that clearly indicated that they are a completely different class of objects (e.g. white dwarfs or pulsars). From the remaining spectra, we were able to identify with high confidence 10\,343. From these we excluded all spectra with the spectral types mentioned above. On the other hand, we kept for analysis the spectra that, although not successfully identified, meet our S/N and resolution criteria. After applying these criteria, the final sample used in our analysis consists of 7\,182 spectra.

\subsubsection{Cross-match with \textit{Gaia} DR3}

Having completed the SIMBAD cross-match, we proceeded to make a cross-identification with the \textit{Gaia} DR3 catalogue, through a cone search of 2 arcsec. This is useful as another check of the SIMBAD coordinates, since we can confirm whether the \textit{Gaia} ID found in SIMBAD is the same as the one directly found in Gaia. Of the 7\,182 spectra kept for analysis, we had coordinates for 6\,743. Among those, 5\,997 spectra were successfully matched with \textit{Gaia}. We found a total of 24 spectra (from 5 different objects) with mismatched \textit{Gaia} IDs. The remaining stars do not have a \textit{Gaia} DR3 ID that matches their entry in SIMBAD.

\subsubsection{Cross-match with external catalogues}
\label{sec:externalcats}

Following the identification of the stars within our sample, we subsequently obtained the available stellar parameters from various sources in the literature, as follows: \noindent i) the catalogue of stellar parameters obtained by \citet{Andrae23xgboost} using the \textsc{XGBoost} algorithm for 175 million stars from \textit{Gaia} DR3 (hereafter GaiaXGBoost); ii) the \texttt{StarHorse 2} catalogue by \citet{Anders2022}, based on the isochrone fitting code created by \citet{2018MNRAS.476.2556Q} (hereafter StarHorse2); iii) the \textit{Gaia}-ESO spectroscopic survey with parameters for almost 7\,000 stars observed with the UVES instrument \citep{Hourihane2023, Worley2024} (hereafter GES); iv) the 220 million \textit{Gaia} XP sources analysed by \citet{2023MNRAS.524.1855Z} with a data-driven model trained on the LAMOST stellar parameters \citep{LAMOSTstellarparams} (hereafter XP-LAMOST). These catalogues cover most of our sample and we can use them to check the consistency of the results of our analysis.

\subsection{Reference stars}\label{sec:referencestars}

\begin{figure}[t]
  \centering
  \includegraphics[width=0.9\linewidth]{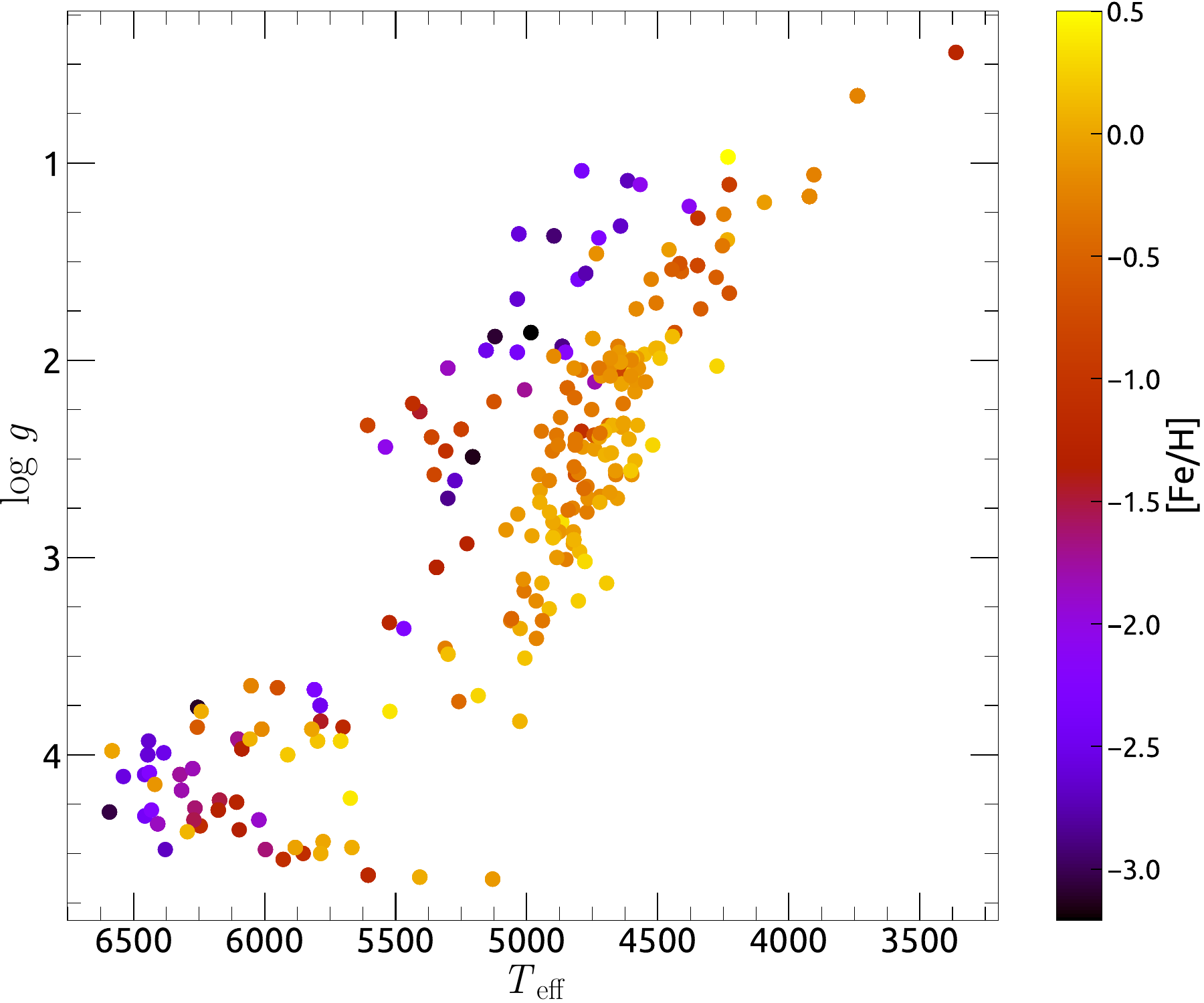}
  \caption{Kiel diagram, colour-coded by metallicity, showing the stellar parameters of the reference stars used in this work. The O- and B-type stars are not shown to facilitate the visualisation.}
  \label{fig:kieldiag_bench}
\end{figure}

As briefly mentioned in Sect. \ref{sec:introduction}, the {\sf CHESS} methodology relies on a set of reference stars that we assume to have accurate stellar parameters. In total, we collected 1244 spectra from 274 reference stars with different values of S/N and wavelength coverage. The distribution of these stars in the Kiel diagram can be seen in Fig.~\ref{fig:kieldiag_bench}. The selected reference stars include:

\begin{itemize}

    \item The Sun: We included two twilight spectra observed with FLAMES-UVES by GES, the FLAMES-UVES solar spectrum made available by ESO\footnote{\url{https://www.eso.org/observing/dfo/quality/GIRAFFE/pipeline/solar.html}}, also from observations of the twilight sky, and a spectrum of reflected solar light on Ganymede observed with the High Accuracy Radial velocity Planet Searcher \citep[HARPS;][]{Mayor2003} instrument\footnote{\url{https://www.eso.org/sci/facilities/lasilla/instruments/harps/inst/monitoring/sun.html}}. The latter is the only non-UVES spectrum included in the current analysis and is used only for testing purposes. \\

    \item TITANS I \citep{2021A&A...650A.194G}: comprises 165 UVES spectra from 34 stars. Part I of the TITANS is a sample of 41 metal-poor dwarf and subgiants ([Fe/H] $\leq$ $-$1.0 dex) with atmospheric parameters derived with high accuracy (maximum errors in \teff\ of $\pm$ 50 K, for \logg\ the typical error is $\leq$ 0.04 dex, and for [Fe/H] the dispersion is $\pm$ 0.05 dex). 
    
    \item TITANS II \citep{2023A&A...679A.110G}: comprises 164 UVES spectra from 32 stars. Part II of the TITANS presents a sample of 47 metal-poor red giants, some of which are carbon-enhanced metal-poor (CEMP) stars ([Fe/H] $\leq$ $-$0.8 dex). The estimated errors are: $\sim$ 50 K for \teff; $\pm$ 0.15 dex for \logg; and $\pm$ 0.09 dex for [Fe/H].
        
    \item GES K2 stars \citep{2020A&A...643A..83W}: 81 UVES spectra of 44 stars. This is a sample of 90 red giants whose analysis combined spectroscopy from the \textit{Gaia}-ESO Survey with asteroseismic constraints from the K2 mission \citep{2014PASP..126..398H}. The estimated errors are: $\sim$ 85 K for \teff; $\sim$ 0.02 dex for \logg; and $\sim$ 0.12 dex for [Fe/H].
        
    \item GES set of hot and cool reference stars \citep[Tables 5 and 6 in][]{Pancino2017}: 274 UVES spectra from 16 stars. \citet{Pancino2017} extended the GES list of reference stars with 17 well-studied OBAM-type stars. Although these are not the types of stars analysed here, they were added to help in distinguishing any such kind of star that may have been included in our target sample unknowingly. 
    
    \item \textit{Gaia} FGK benchmark stars \citep{2024A&A...682A.145S}: 252 UVES spectra of 37 stars. The \textit{Gaia} benchmark stars are a sample of 192 FGK-type stars with \teff\ and \logg\ derived from fundamental methods, independent of spectroscopy, using angular diameter, bolometric fluxes, and parallaxes. The uncertainties in these parameters are usually below 2$\%$. The metallicity values come from the literature. 
    
    \item \textit{Gaia} `golden sample' \citep{2023A&A...674A..39G}: 304 UVES spectra from 110 stars.
    As part of \textit{Gaia} DR3, six samples of stars with high-quality stellar parameters were provided. The parameters of these samples were validated in detail, using both \textit{Gaia} data and literature results. We focus here on the subsample of FGKM-type stars.
    
\end{itemize}

\subsection{Data homogenisation}
\label{subsec:homogenization}

Before comparing the spectra, we performed the homogenisation of the data, which entails correcting for the Doppler shift induced by the stellar radial velocities, normalising the flux scale, and setting all data to the same resolution and binning. If this is not done, the differences in such characteristics dominate the comparison between the spectra, as the pixels would not correspond to the same wavelength interval and the relative fluxes can vary a lot depending on exposure time and target brightness. 

Originally, the spectral resolution in our sample ranged from 20\,000 to $\sim$ 110\,000. Since there were very few spectra with R$<$ 30\,000 (63), we excluded them from further analysis. Therefore, we degraded the data to R = 30\,000. For rebinning, we chose the maximum value of the sample, a pixel size of 0.0043 nm. Both steps were performed using \textsc{iSpec} \citep{2014A&A...569A.111B, 2019MNRAS.486.2075B}.

For continuum normalisation, we used \textsc{SUPPNet}, a neural network software for stellar spectrum normalisation \footnote{See https://rozanskit.com/suppnet/}, developed by \cite{SUPPNet}. At this step, we decided to focus on the wavelength range between 400 - 700 nm. This choice was made to avoid the crowded regions in the blue region and the contamination by telluric lines in the red wavelengths. 

We used cross-correlation against a set of 12 synthetic spectra templates (see Table \ref{tab:templates4rv}) to compute the relative radial velocity and correct the observed data to a rest-frame wavelength solution. Synthetic spectra were calculated using the code \textsc{Turbospectrum} \citep{Turbospectrum} in the 300-900 nm range, with steps of 0.0015 nm, and a resolution of R = 30\,000 at 450 nm. We used the atomic and molecular line list compiled and tested by \citet{GiribaldiSmiljanic2023}. 

The cross-correlation was run against each template in the grid. We discarded extreme values that deviate from the median by at least ten times the median absolute deviation (MAD). We adopted the radial velocity estimate with the smallest error, which was computed by \textsc{iSpec} following \citet{2003MNRAS.342.1291Z}. This procedure ensured the selection of the template most compatible with the observed spectrum, the one exhibiting the highest peak in the cross-correlation functions.

\section{Similarity analysis}
\label{sec:method}

The similarity analysis is the step that identifies stars with similar atmospheric parameters before the detailed spectroscopic analysis. Our aim is to create a method that directly uses the comparison of stellar spectra to select targets for a successful differential analysis. Specifically, our objective is to identify stars with parameters that differ by $\pm$200 K in \teff, $\pm$0.3 dex in \logg, and $\pm$0.2 dex in [Fe/H], from those of a selected reference star. We recall that the use case for this method is in the context of large spectroscopic surveys dealing with several thousands to millions of spectra of stars with yet unknown atmospheric parameters. Below, we detail the steps needed to identify similar stars. 

\begin{table}
\caption{Stellar parameter of the templates used in the radial velocity determination.}            
\label{tab:templates4rv}      
\centering                          
\begin{tabular}{c c c c}        
\hline\hline                 
Type & \teff & \logg & [Fe/H] \\    
\hline                        
Bright giant      & 4200          & 1.00          & $-$0.50        \\
Bright giant      & 4200          & 1.00          & $-$1.00        \\
Giant             & 4500          & 2.50          & 0.00         \\
Giant             & 4500          & 2.50          & $-$0.50        \\
Giant             & 4500          & 2.50          & $-$1.00        \\
Giant             & 4500          & 2.50          & $-$2.00        \\
Subgiant          & 5200          & 3.80          & 0.00         \\
Subgiant          & 5400          & 3.80          & $-$0.50        \\
Dwarf             & 6000          & 4.40          & 0.00         \\
Dwarf             & 6000          & 4.40          & $-$0.50        \\
Dwarf             & 6000          & 4.50          & $-$1.00        \\
Dwarf             & 6000          & 4.50          & $-$2.00        \\
\hline                                   
\end{tabular}
\end{table}

\subsection{Dimensionality reduction}

We used the unsupervised machine learning algorithm known as t-distributed Stochastic Neighbour Embedding \cite[t-SNE,][]{JMLR:v9:vandermaaten08a}. This algorithm has been used before in support of spectroscopic analyses. \citet{Matijevic2017} were the first to use t-SNE for that purpose as part of a method to identify candidate metal-poor stars observed by RAVE \cite[The Radial Velocity Experiment,][]{Steinmetz2020}. Similarly, \citet{Traven2017} applied t-SNE to GALAH spectra, producing a two-dimensional map that facilitated semi-automated spectral classification and the identification of peculiar or problematic spectra. 

The algorithm projects high-dimensional data into lower dimensions while trying to preserve the global structure of those data. In this process, t-SNE tries to conserve the `neighbourhood' of the data points. Thus, data points that are nearby in high-dimensional space should maintain their proximity in lower-dimensional projections, with neighbours being more alike than distant points. A key hyperparameter of t-SNE is the perplexity, which controls the number of neighbours around each data point when the lower-dimensional space is built. It controls the balance between preserving the local and global structure of the data distribution. A low perplexity value creates tightly grouped clusters in the projected space, allowing the algorithm to separate data points with subtle differences. In contrast, a higher perplexity helps to uncover broader structures in the data.

\begin{figure}[t]
  \centering
  \includegraphics[width=\linewidth]{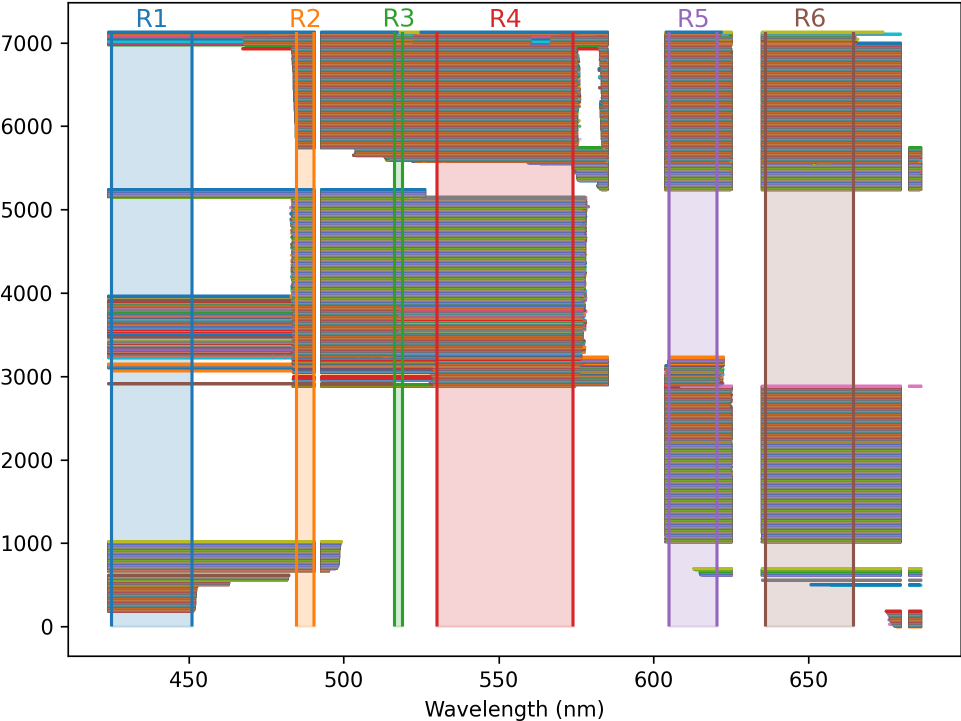}
  \caption{Wavelength range of each of the $>$ 7000 spectra in our sample, limited to the interval between 400--700 nm. Each spectra is represented as a horizontal line. The shaded regions (R1 to R6) indicate the selected regions to be analysed. Some of the empty spaces correspond to artifacts in the spectra or regions of telluric lines that were masked.}
  \label{fig:region_select}
\end{figure}

In our case, the high-dimensional data are the homogenised spectra of different types of stars. Each spectrum is a point in a multidimensional space with pixels as dimensions and their flux as coordinates. Within this space, data points representing similar spectra cluster together. With t-SNE we want to create a projection map of these data in two dimensions (2D). However, our spectra are not uniformly distributed in terms of their wavelength ranges (Fig. \ref{fig:region_select}). This hinders the use of either the entire range or a single selected range for use with t-SNE. Therefore, we had to select several short spectral regions in which to apply the similarity analysis. Our goal was to select a small number of regions, making sure that each region covers as many spectra as possible and that, together, the regions encompass all spectra in the sample. At the same time, we made sure to include some selected features sensitive to different stellar parameters (such as the H$\alpha$ line or the Mg triplet at 517 nm). The final selected spectral regions (named `R1' to `R6') are listed in Table \ref{tab:regions} and highlighted in Fig.\ \ref{fig:region_select}. Some of the sample spectra were found to contain pixels with missing values in the selected regions that cannot be used in the analysis. For small gaps of $\leq 5$ consecutive pixels, we employed a cubic-spline interpolation. For larger gaps, the corresponding pixels were excluded from the analysis.

We used the Python scikit-learn package \citep{scikit-learn} for t-SNE, testing both the exact and Barnes-Hut methods. The exact method took about 30 minutes per spectral region, while the Barnes-Hut method reduced this to 5 minutes (on a computer with a 12-core processor). Since our data set is relatively small (see Table \ref{tab:regions}), we opted for the exact method to maximise accuracy.

An example of a projection map, obtained using the spectral region R2, is shown in Fig. \ref{fig:projection_map_regions}. Depending on the spectral region used, the morphologies of the maps are different but their overall characteristics are the same (see the additional maps in the Appendix \ref{app:A}): the main separation of the points is between groups of giant stars (marked as groups B and C in Fig. \ref{fig:projection_map_regions}) and dwarfs (groups A and H in Fig. \ref{fig:projection_map_regions}). Within the groups of dwarfs, we see a gradient of \teff. Within the groups of giants, there is a gradient of [Fe/H]. These observations already hint that, while all three atmospheric parameters influence the distribution of stars in the projection map, the variations in \teff~and [Fe/H] values are most significant. The additional groups in the plot are better discussed in Sect. \ref{sec:results}. Given that our sample has many spectra from different types of stars, we can see some imbalance in how different groups appear in the projected space. To address this, we decided to apply t-SNE twice. First, we used the full data set and a high-perplexity value to divide the main groups (e.g. dwarfs, giants, and metal-poor stars). Then, we ran t-SNE on one of the groups identified before, with slightly lower perplexity, to further separate subgroups of similar spectra.

\begin{table}
\caption{Spectral regions (enumerated from R1 to  R6) used for the similarity analysis.}             
\label{tab:regions}      
\centering                          
\begin{tabular}{c c c c c}         
\hline\hline                 
Name & $\lambda_{\rm min}$  & $\lambda_{\rm max}$  & Flux       & N\\ 
     &    (nm)                      &      (nm)                    & dimensions & spectra\\
\hline                        
R1      & 425.10          & 451.00          & 4451           & 1189        \\
R2      & 484.81          & 490.44          & 970            & 3801        \\
R3      & 516.50          & 519.00          & 433            & 3526        \\
R4      & 530.10          & 574.00          & 7551           & 3662        \\
R5      & 605.05          & 620.40          & 2657           & 4278        \\
R6      & 636.20          & 664.50          & 4780           & 3967        \\
\hline                                   
\end{tabular}
\end{table}

\subsection{Defining the similarity among stars}
\label{subsec:method_similarstars}

\begin{figure*}[t]
  \centering
  \includegraphics[width=0.33\linewidth]{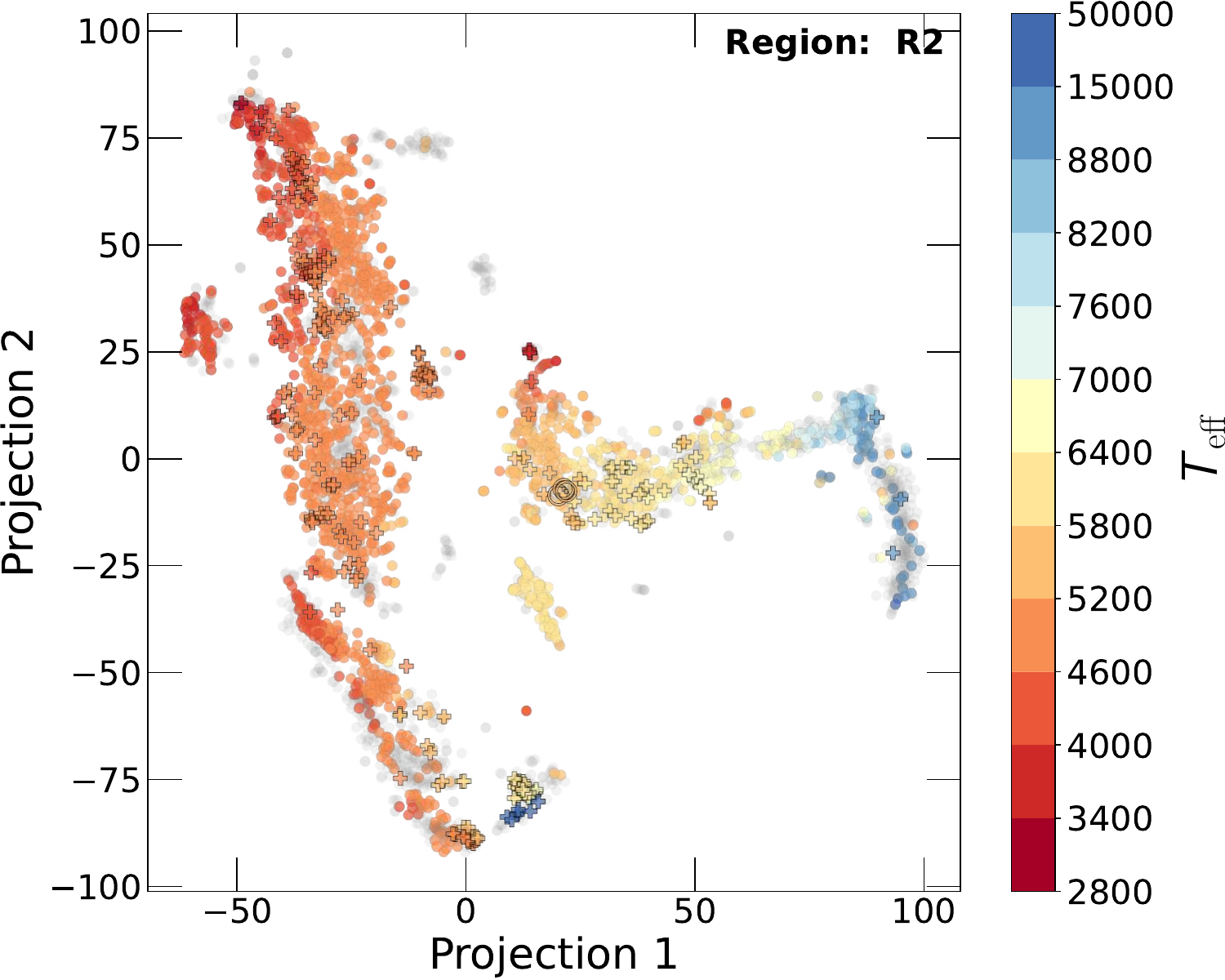}
  \includegraphics[width=0.33\linewidth]{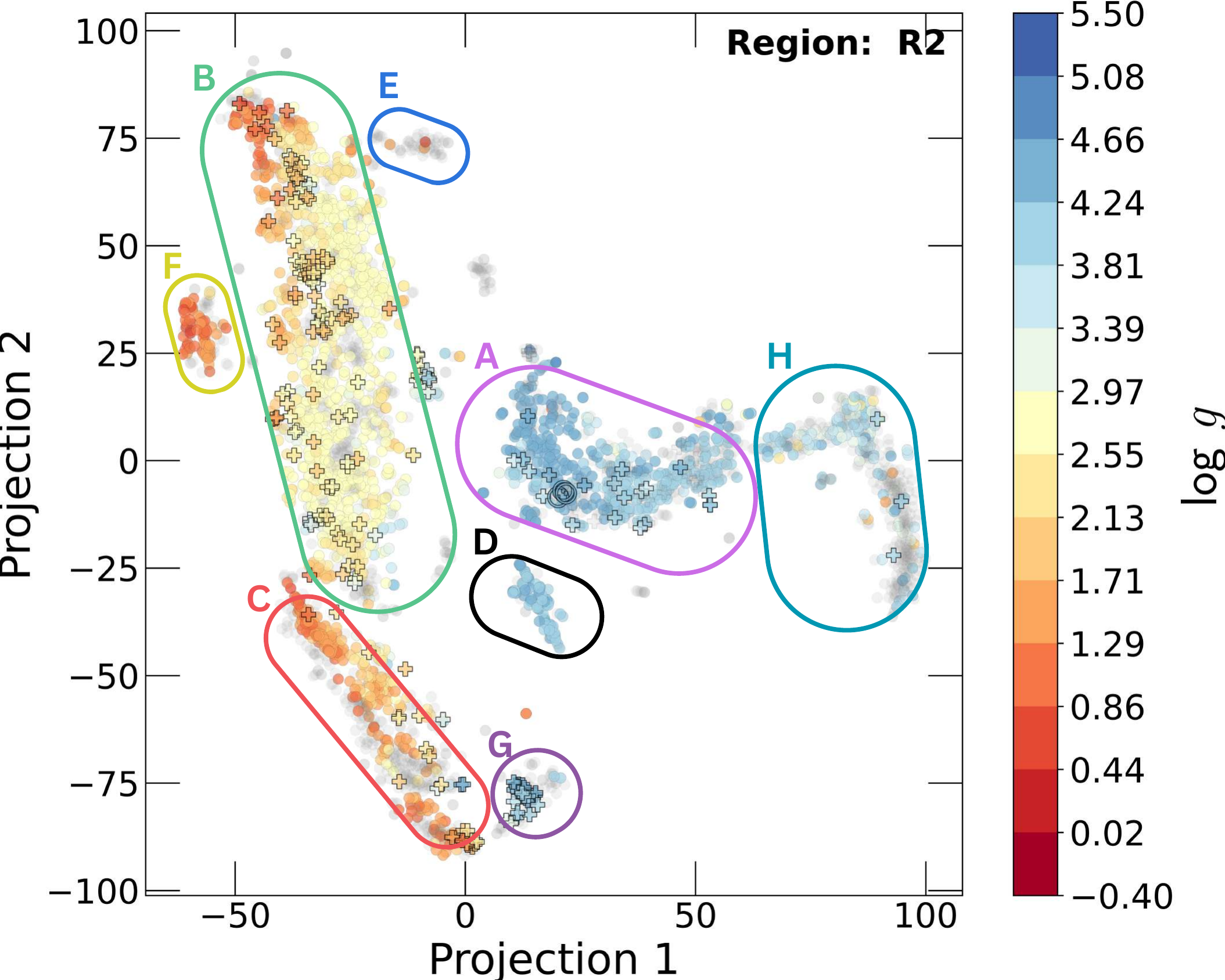}
  \includegraphics[width=0.33\linewidth]{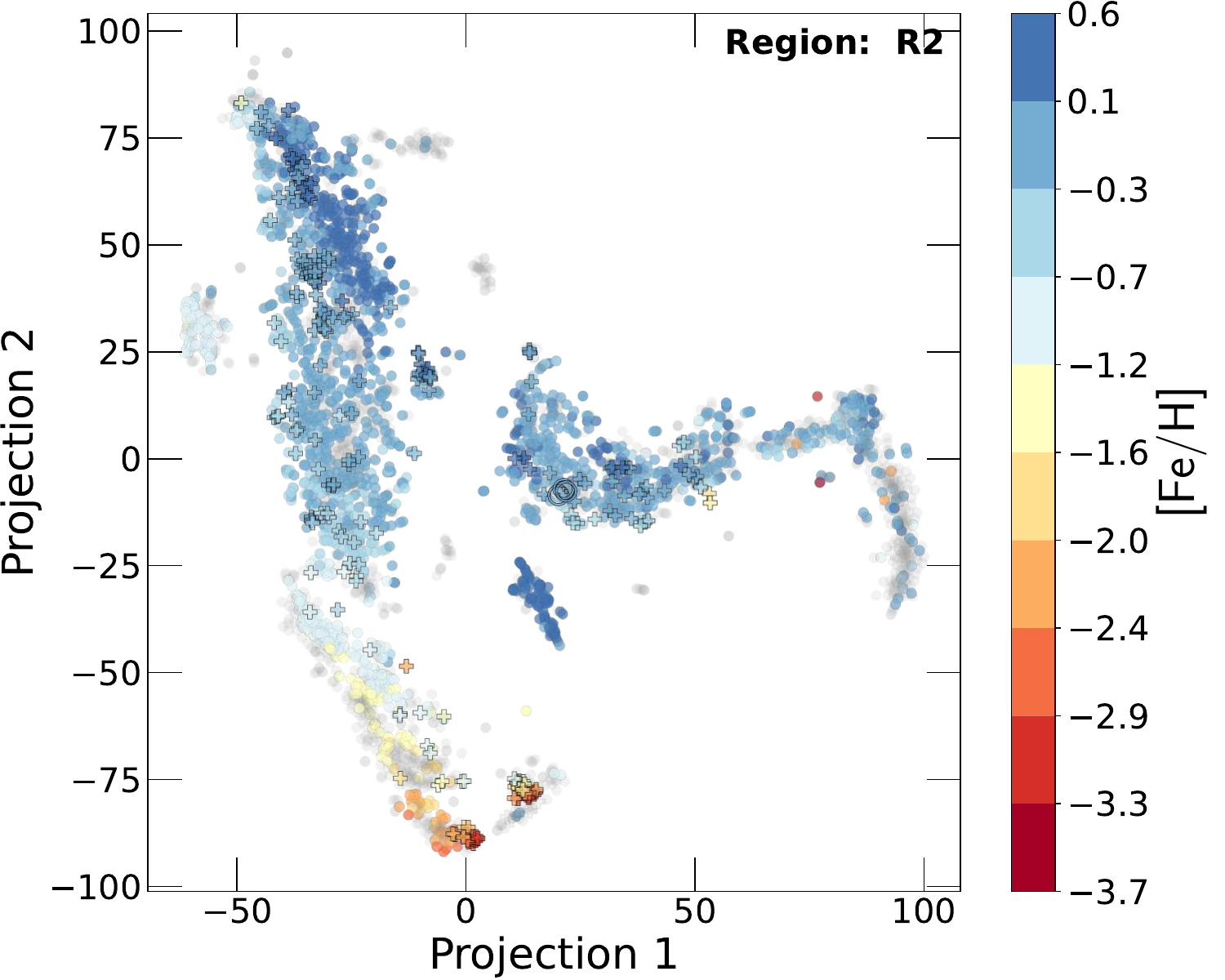}
  \caption{t-SNE projection map of the region R2 colour-coded by the atmospheric parameters from the GES catalogue, \teff\ (left), \logg\ (centre), and [Fe/H] (right). The axes do not have direct physical meaning, as they are not linear combinations of the original data; they are simply indicating relative neighbourhood of the data points in these projections.}
  \label{fig:projection_map_regions}
\end{figure*}

\begin{figure*}[t]
  \centering
  \includegraphics[width=\textwidth]{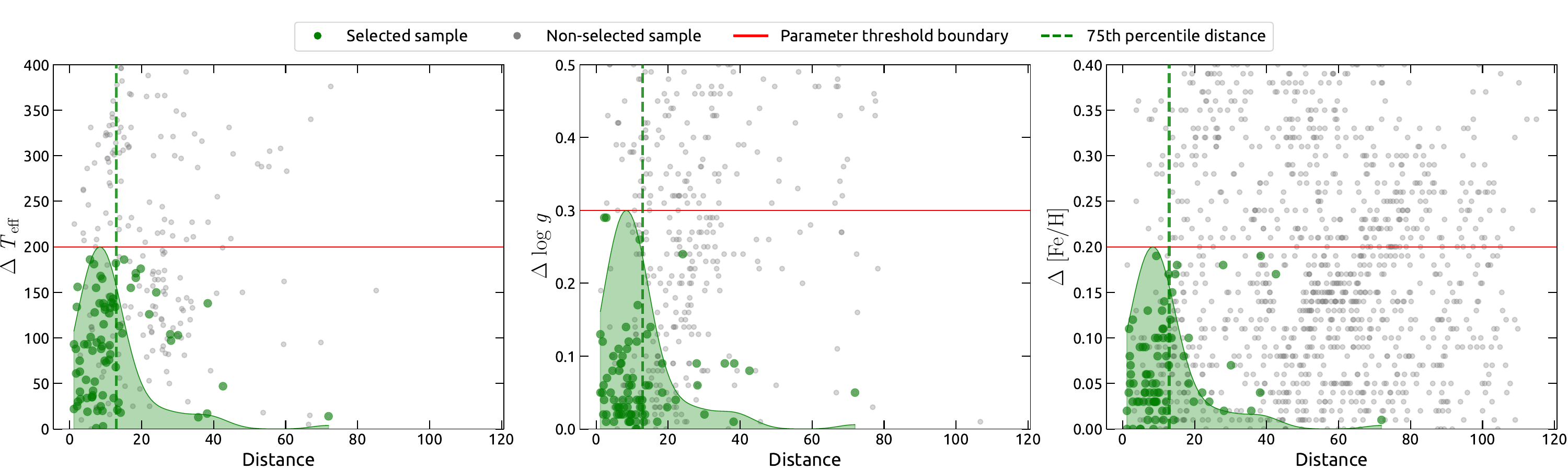}
  \caption{Distribution of distances in the t-SNE projection for stars with similar atmospheric parameters to the Sun, using the GES catalogue for the sample and spectral region R2. The absolute values of the differences in \teff, \logg, and [Fe/H] are shown in the left, centre, and right panels, respectively. In each panel, the red horizontal line indicates the adopted parameter thresholds; green points represent the similar spectra which fulfil all three of the parameter thresholds, and the green vertical dashed line marks the 75th percentile of their distances from the reference.} 
  \label{fig:dist_SUNganymharps}
\end{figure*}

In the t-SNE projection maps, similar stars are grouped together. However, the projection is obtained in a non-linear manner and the distances between points do not carry direct physical meaning. Points equally spaced in different map regions may not be equally similar. For the same reason, densities on the map are hard to interpret. However, the projections (exemplified in Fig. \ref{fig:projection_map_regions}) show clear separation and ordering of the points in terms of atmospheric parameters, with similar stars expected to be neighbours. Our goal is to define a metric for selecting a sample of neighbouring stars that satisfies our similarity criterion with high completeness. To define that metric, we start from a reference star and identify stars in the sample with parameters within $\pm$200 K in \teff, $\pm$0.3 dex in \logg, and $\pm$0.2 dex in [Fe/H]. We can then examine how the difference in parameters relates to the distance in the projection map (see Fig. \ref{fig:dist_SUNganymharps} for an example).

As seen in Fig. \ref{fig:dist_SUNganymharps}, stars with similar catalogue parameters do not necessarily cluster at the closest distances. This is caused by several factors in addition to the differences in atmospheric parameters. The main aspect for the neighbourhood is the similarity in the spectra. Variations in chemical abundances, uncertainties in the parameters, and noise in the spectra can all affect the t-SNE distance distribution. In particular, biases in the parameters can be important for Fig. \ref{fig:dist_SUNganymharps} as the methods applied to the reference stars and the neighbouring stars are not the same. These biases can cause stars with tentatively similar parameters to appear far from the reference star in the projection map because the spectra are actually different. 

For our purpose, it is important to note that a region surrounding a reference star, containing objects similar to it, can be defined. Once this is understood, one can naively conclude that the distribution of distances around all reference stars can be used to define a typical distance value to select stars with parameters within $\pm$200 K in \teff, $\pm$0.3 dex in \logg, and $\pm$0.2 dex in [Fe/H]. However, that is not the case because distances can vary significantly with position in the projection map and because the t-SNE algorithm has a stochastic component that creates different projection maps in repeated runs. The distances between points change, preventing the selection of the same sub-sample. 

As a second step, we examined how direct spectral differences correlate with t-SNE distances. For that, we defined a simple similarity score, $S_i$, which combines the median and a measurement of the dispersion of the absolute differences between the spectra of a sample star and the reference star:

\begin{equation}
    S_i = \sqrt{\left( Q_{{\rm dif}, i}^{50}\right)^2 + \left( \Delta Q_{{\rm dif}, i}^{80}\right)^2 },
    \label{eq:similarity}
\end{equation}

\begin{figure}[t]
  \centering
  \includegraphics[width=\linewidth]{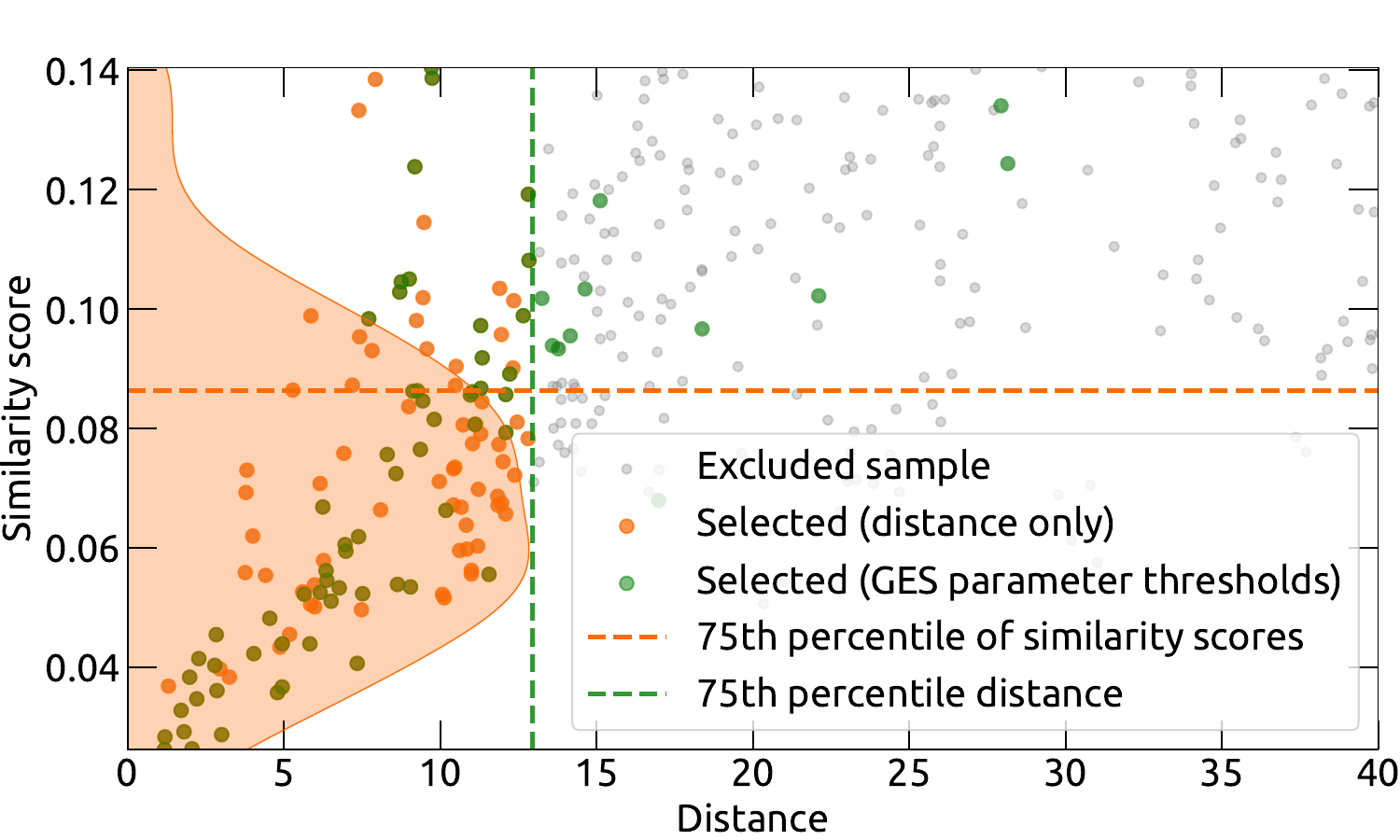}
  \caption{Distribution of similarity score values of the sample surrounding the Sun. The lower the similarity score, the more similar is the stellar spectrum to the solar spectrum. The green points have the same meaning as in Fig. \ref{fig:dist_SUNganymharps}. Orange points are all the stars within the 75th percentile of distances defined in Fig. \ref{fig:dist_SUNganymharps}. The the orange horizontal dashed line indicates the 75th percentile of the distribution of similarity scores. Note that the total range of similarity scores is much larger; the figure shows only nearby stars.}
  \label{fig:simscore}
\end{figure}

\noindent In Eq. \ref{eq:similarity}, $Q_{{\rm dif}, i}^{50}$ is the 50th quantile (median) and $\Delta Q_{{\rm dif}, i}^{80}$ represents the interquantile range between the 90th and the 10th quantiles. We adopted $\Delta Q_{{\rm dif}, i}^{80}$ as the dispersion measurement as it is more sensitive to outliers that trace small spectral differences. Such differences can be important for the determination of similarity. For this computation, we mask the wavelength regions where the normalised flux of the reference spectrum exceeds a value of 0.90. This excludes continuum regions and weak spectral lines, ensuring that $S_i$ is computed only on more informative parts of the spectra. This is important for metal-poor stars, where large continuum regions could misleadingly suggest higher spectral similarity. An example of the comparison between $S_i$ and the t-SNE distances is shown in Fig. \ref{fig:simscore}.

We tested several choices for the similarity metric: median absolute deviation (MAD), coefficient of determination (R$^2$), Euclidean distances combined with a radial basis function (RBF) kernel similarity, cosine similarity, and combinations of these. Both MAD and R$^2$ failed to provide a consistent threshold for most reference spectra. The RBF was too sensitive to outliers and required fine-tuning of the bandwidth parameter. Cosine similarity, while good at capturing overall structures, ignored subtle effects such as differences in line broadening. Combining metrics added complexity without clear benefits, so we chose the simpler option that works reliably in most cases.

Figure \ref{fig:simscore} shows that stars with atmospheric parameters within the thresholds have a narrow distribution of $S_i$ values. However, within the same range of $S_i$ values we also find stars that are not similar to the reference in terms of their parameters. This indicates that a combination of median and dispersion is too basic of a metric to capture spectral differences, emphasising the need for t-SNE in the analysis. Importantly, Fig. \ref{fig:simscore} shows that restricting the problem to the sub-sample of stars that t-SNE places nearby to the reference object, the distribution of $S_i$ values can be used to select a close to complete sample of similar stars.

\begin{figure}[t]
  \centering
  \includegraphics[width=\linewidth]{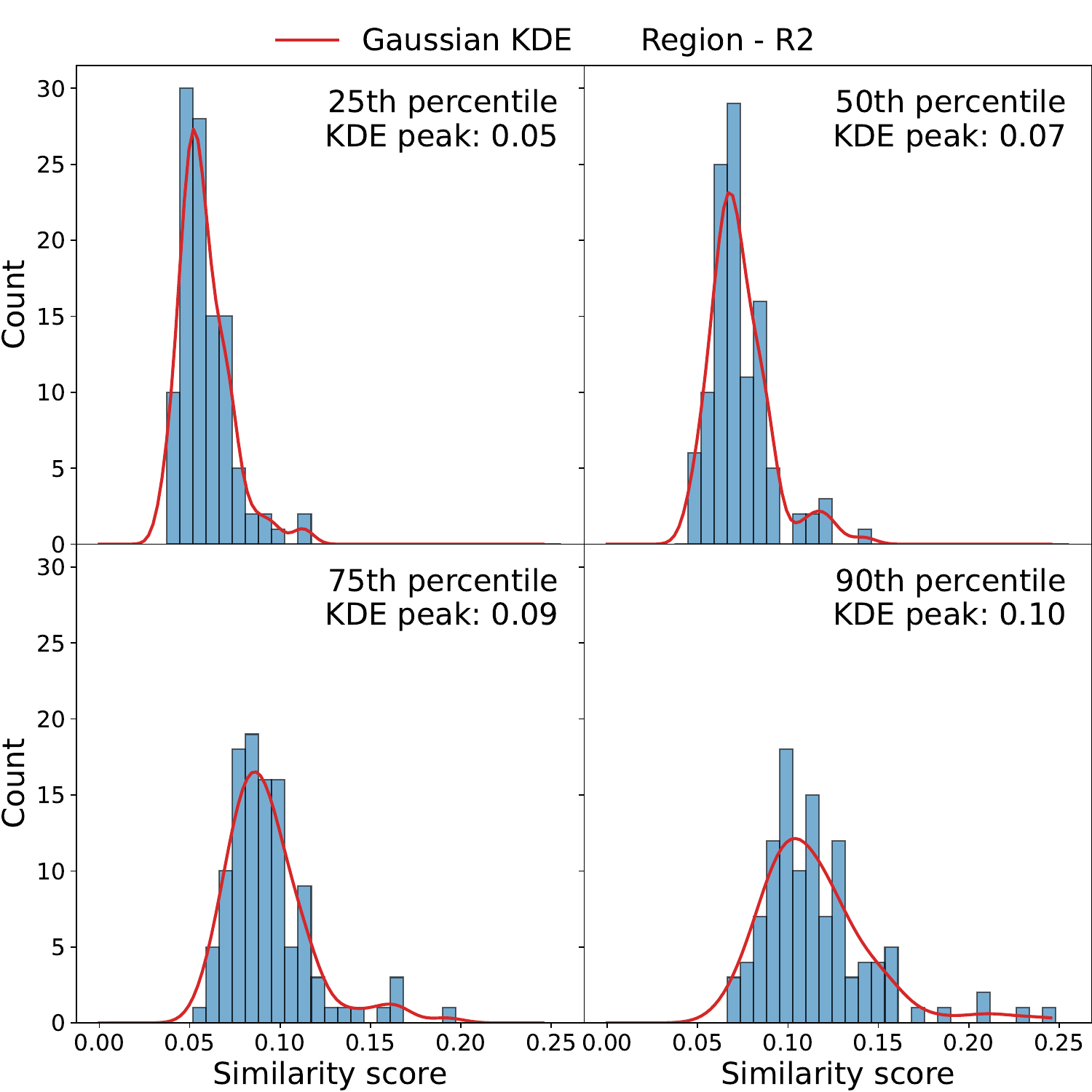}
  \caption{Each subplot shows a histogram of similarity scores thresholds obtained (as shown in Fig.~\ref{fig:simscore}) for the references in the region R2, but computed using different percentiles (25th, 50th, 75th, 90th). The red curve represents a smoothed distribution obtained using a Gaussian kernel density estimation (KDE).}
  \label{fig:R2threshold_hist}
\end{figure}

Figures \ref{fig:dist_SUNganymharps} and \ref{fig:simscore} display the comparisons between the parameters, spectra, and t-SNE distances for the case of one reference star. To generalise the comparison, we extended the investigation to the whole sample of reference stars. We first determined the distribution of distances for the stars with similar parameters (as in Fig. \ref{fig:dist_SUNganymharps}) taking the 75th percentile to represent each distribution. This value was adopted to limit the region used to calculate the distribution of the $S_i$ values (as in Fig. \ref{fig:simscore}). To compare all the distributions of $S_i$ values, we show histograms of their 25th, 50th, 75th, and 90th percentiles in Fig. \ref{fig:R2threshold_hist}. Each histogram is narrow, with a clear peak and some outliers (probably driven by biased parameters). The exception is the 90th percentile, which is broader and is the one most sensitive to stars with problems in their parameters. Figure \ref{fig:R2threshold_hist} clearly shows that all distributions are very similar. This implies that we can define a typical shape for the $S_i$ distributions that will be useful to select the most similar stars around all reference objects on the t-SNE map. 

All the discussion so far has focused on the R2 spectral region. Similar figures for other regions are included in the Appendix \ref{app:B}. The exception is region R1 (425.1-451.0 nm), which could not be analysed because the stars in this region lack parameters from the GES catalogue. Similar conclusions as above were obtained when analysing the other regions, with the exception of region R3 (516.5-519.0 nm). For R3, the histograms show large dispersions, making it difficult to select a representative shape for the distribution of $S_{i}$ values around the reference stars. This is likely due to its small wavelength coverage, which may limit its sensitiveness to variations in stellar parameters.

\subsection{Procedure for the blind search of similar stars}\label{sec:blind}

Previously, we relied on known stellar parameters to assess similar spectra around a reference object. However, what we need as part of {\sf CHESS} is a method that uses only the metric $S_i$ to blindly identify similar spectra. The core idea is that the distribution of $S_i$ values in the nearby spectra should mirror the representative distributions found in the previous discussion. To do that, we again decided to characterise the distribution of $S_i$ values around reference stars by the value of its 75th percentile. Surrounding stars are added to the distribution until that percentile reaches a threshold value, $S_{thresh}$. In the following, we describe the method that we have established for this blind search. The determination of the $S_{thresh}$ value is discussed in Section \ref{sec:testing}.

As we mentioned earlier, to address class imbalance, we apply t-SNE in two stages. First, we run t-SNE on the full dataset to identify the major groups (shown in Fig.\ \ref{fig:projection_map_regions}). Then, we perform a second iteration of t-SNE separately on each of these groups. This two-step approach allows us to better capture the local structure within each subgroup without interference of the full sample. The following method is then applied to the projection maps generated in this second iteration.

\begin{enumerate}
  
    \item Neighbourhood threshold: Similar stars are searched in the neighbourhood of a given spectra in the t-SNE map. To define what a reasonable neighbourhood threshold is, for each star in the projection map, we calculated the distance to its nearest neighbour (NN). The maximum of such a distribution is affected by some isolated points on the map. As the final threshold, we adopt twice the 95th percentile of the distribution. This ensures that the majority of points in the projection map have a neighbour closer than the threshold.

    \item Initial candidate selection: For each reference spectra, we select up to 15 sample stars within the neighbourhood threshold. This number of stars was chosen to ensure some statistical robustness when calculating the percentiles of the similarity scores even in the presence of potential outliers.

    \item Similarity evaluation: The similarity scores of the initial candidates are computed. The 75th percentile of the distribution of similarity scores is compared to the adopted metric.

    \begin{itemize}
         \item If there are no spectra within the neighbourhood threshold, the reference spectrum is no longer analysed.
        \item If the 75th percentile does not exceed the adopted metric value, $S_{thresh}$, we will continue to add candidates as described in the next step.
        \item If $S_{thresh}$ was exceeded, no similar spectrum is assigned to the reference.
    \end{itemize}

     \item Distance sorting: For the remaining sample, we sort the spectra by the sum of the Euclidean distances to the reference and to the nearest spectrum already included in the candidate list. This approach ensures that the selection of next candidates grows organically in the neighbourhood of the previous objects, minimising the effect of small spatial gaps while also taking into account the overall structure of the projection.

     \item Candidate list expansion: the next candidate is added from the sorted list if: (i) it lies within the neighbourhood threshold of at least one previously accepted spectrum, and (ii) its inclusion does not cause the 75th percentile of similarity scores to exceed the adopted metric. 

    \item Stopping criteria: the addition of new candidates is stopped if: (i) the 75th percentile of similarity scores exceeds $S_{thresh}$; or (ii) the median of the similarity scores of the last 15 stars added to the list exceeds $S_{thresh}$. This secondary criterion is needed when the sample becomes too large and the addition of new stars has little influence on the distribution. 
\end{enumerate}

\subsection{Selecting the similarity score threshold}\label{sec:testing}

Following the method explained above, we assess the completeness and purity of the selection for different values of $S_{thresh}$ using the following definitions: 

\begin{equation}
    \text{Completeness} = \frac{ \sum_{i} \text{TP}_i}{\sum \text{TP}_i + \sum \text{FN}_i}
    \label{eq:completeness}
\end{equation}

\begin{equation}
    \text{Purity} = \frac{\sum_{i} \text{TP}_i}{\sum \text{TP}_i + \sum \text{FP}_i}
    \label{eq:purity}
\end{equation}

\noindent Where for the spectrum of each reference star $i$, the true positives (TP) are the spectra that should have been selected (in terms of their atmospheric parameters) and were correctly included; the false negatives (FN) are the spectra that should have been selected but are missed by the method; and the false positives (FP) are the spectra that are incorrectly included in the selection. By summing these quantities over all reference stars, we obtain global values of completeness and purity for the sample analysed in a given spectral region.

\begin{figure}[t]
    \centering
    \includegraphics[width=\linewidth]{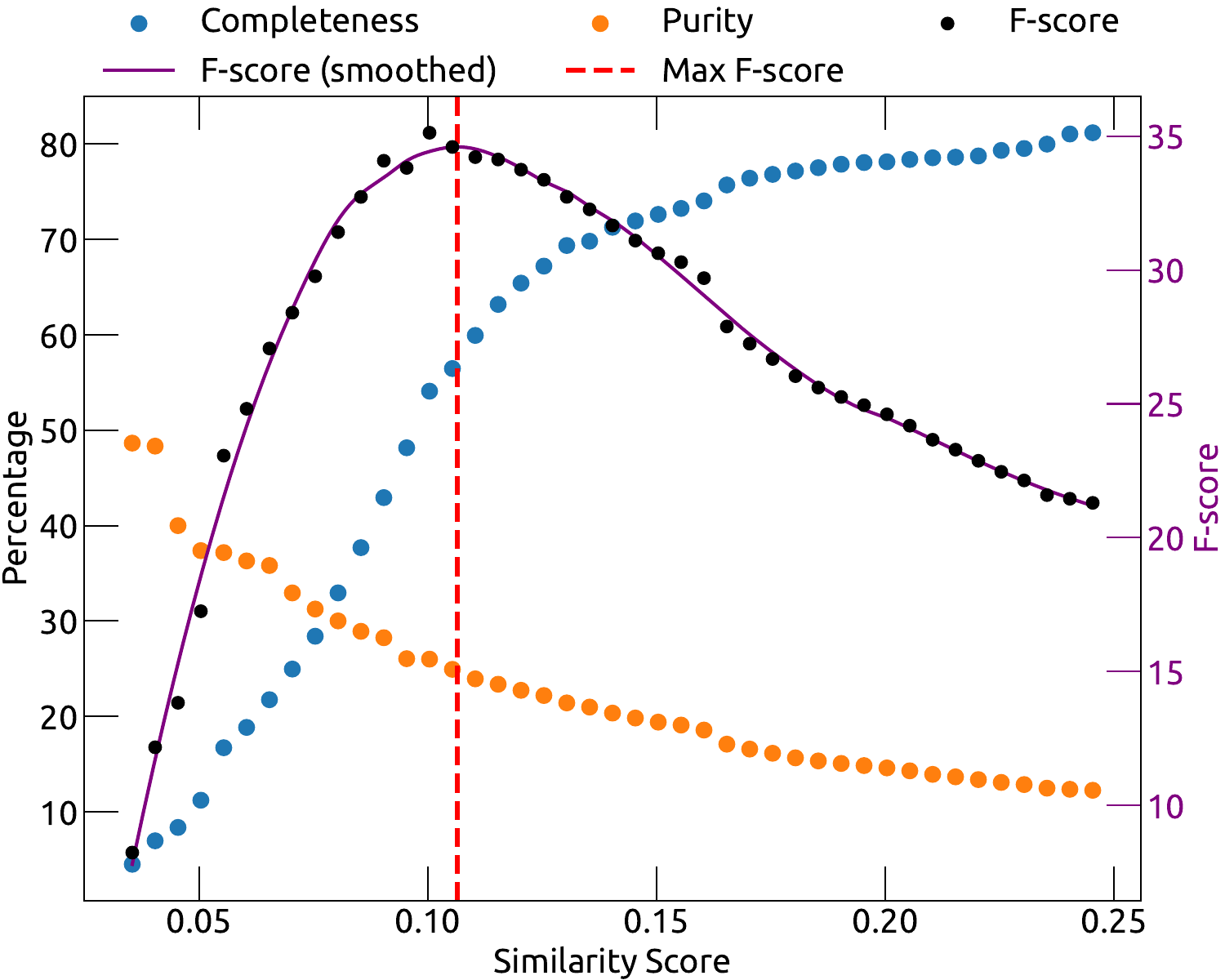}
    \caption{Completeness (blue), purity (orange), and F-score (black) as a function of the $S_{thresh}$ used to select spectra. The smoothed F-score curve (purple) is used to select its maximum value (red dashed line) representing the best trade-off between completeness and purity}
    \label{fig:score.completeness}
\end{figure}

Figure \ref{fig:score.completeness} shows how the completeness and purity of the blind selection of similar stars vary for different $S_{thresh}$ values. As expected, there is a trade-off between completeness and purity. For low values of $S_{thresh}$, the purity is higher at the cost of a very low completeness. On the other hand, for higher values of $S_{thresh}$ (meaning broader distributions of $S_i$ values around each reference star), higher completeness is achieved at the cost of low purity. We can use the F-score as a metric to find the optimal value of $S_{thresh}$ in this problem. The F-score is defined as:

\begin{equation}
    F_1 = 2 \times \frac{\text{Completeness} \times \text{Purity}}{\text{Completeness} + \text{Purity}}
    \label{eq:f1_score}
\end{equation}

The maximum value of F-score occurs at $S_{thresh}$ = 0.106. This value represents the optimal compromise between completeness and purity for a global selection over all reference stars in the t-SNE map. This is the value that we can adopt when implementing this method in the {\sf CHESS} pipeline. Moreover, we found that in the remaining regions, the maximum F-score consistently yields similar $S_{thresh}$ values centred around 0.090 (except for regon R3, where the method cannot be calibrated). However, we note that the above considerations are based on atmospheric parameters from external catalogues, GES in particular.

To double check these conclusions, we decided to analyse some of the spectra and investigate how well the selection indeed identified similar stars. Only five reference stars were used in this exercise, selected in different regions of the parameter space. For the analysis, we actually selected stars in a region beyond what would be chosen by the above method. The goal was to understand whether the selection stopped at an appropriate time or if it should have proceeded further. The line-by-line differential analysis was performed using the q2 code \citep{Ramirez2014}. The \ion{Fe}{i} and \ion{Fe}{ii} lines were selected from the GES line list \citep{Heiter2021}. The equivalent widths were measured with the REvIEW code \citep{McKenzie22}. The selected reference stars and the completeness and purity results are summarised in Table~\ref{tab:q2}. We can see that when our own parameters are used to evaluate the performance of the blind method, a higher completeness is obtained. This suggests that the method effectively selects the stars we want to identify. Purity seems to be mostly similar, although in two cases it changed significantly. We will better evaluate the reasons for this in future works. 

\begin{figure*}[t]
    \centering
    \includegraphics[width=\linewidth]{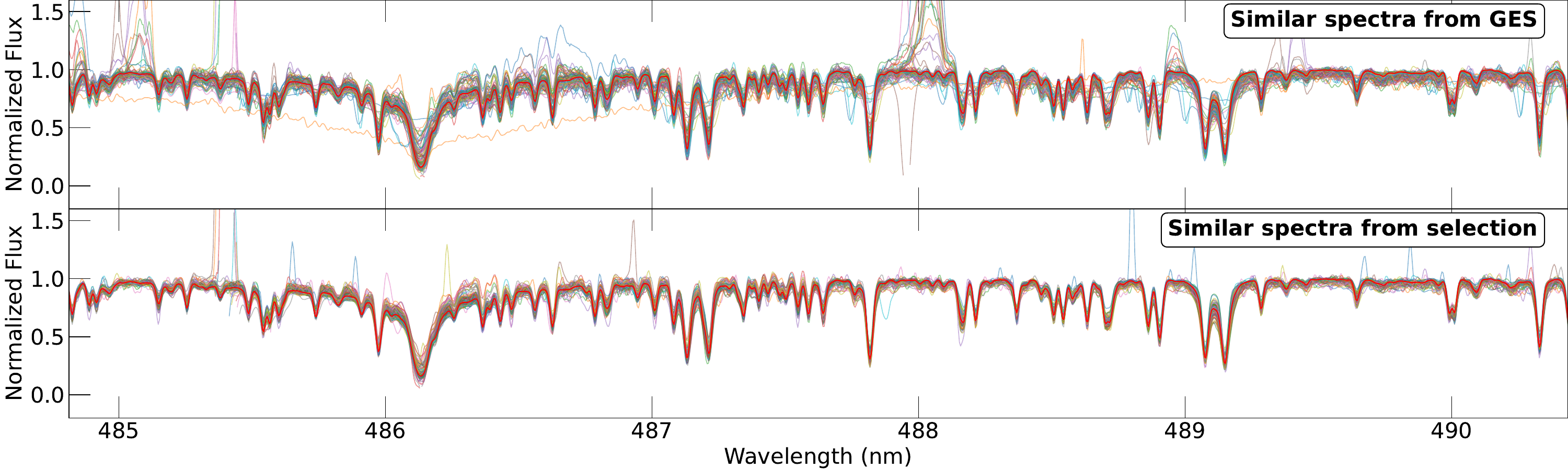}
    \caption{Comparison of similar spectra to a reference spectrum of the Sun (red line), obtained in two different ways: using the GES catalogue as reference (top panel) and the selected sample using our methodology (bottom panel). Both panels display the corresponding normalized spectra for the region R2 with different colours to show the different spectra.}
    \label{fig:SUN_similar_spectra}
\end{figure*}

These changes in purity may appear from biases between the stellar parameters in the external catalogue used for comparison and those adopted for the reference star. Thus, the purity value found in the comparison with the tabulated data is not real and is corrected after our analysis. Additionally, low purity can occur in sparsely populated regions of the parameter space. For example, in our current sample, extremely metal-poor stars are not numerous. They cluster together in a region of the t-SNE map, although their parameters are not truly similar (but their spectra are more similar among themselves than to the rest of the sample). As a result, selections around a reference star in this region may be contaminated, leading to low purity. At this point, to illustrate the differences, we show comparisons of similar spectra based on parameters of GES and our own analysis in Fig. \ref{fig:SUN_similar_spectra}. This comparison reinforces that our method is obtaining a very clean sample of similar spectra.

\begin{table*}
\caption{Comparison of completeness using GES and q2 parameters.}
\label{tab:q2}
\centering
\begin{tabular}{l r c c}
\hline\hline
Reference ID & N selected spectra & Completeness | Purity [\%] & Completeness | Purity [\%] \\
             &                    & (GES params.)              & (q2 params.)              \\
\hline
Sun                      & 138          & 74 | 43           & 91 | 44         \\
HIP 13288                & 506          & 76 | 31           & 86 | 39         \\
GES J22074730-1059405    & 610          & 79 | 23           & 98 | 8          \\
GES J22033684-1449366    & 218          & 64 | 40           & 88 | 46         \\
HIP 69746                & 54           & 50 | 11           & 74 | 54         \\
\hline
\end{tabular}
\end{table*}

\section{Discussion}
\label{sec:results}

\subsection{A deeper look at the t-SNE projection maps}

Projection maps done using the different spectral regions exhibit similar characteristics. Therefore, we focus on the map obtained using the R2 region (see Fig. \ref{fig:projection_map_regions}) for this discussion, where the details of the t-SNE projection are most visible. All maps show a clear separation between dwarf and giant stars. In addition, there are a few smaller but well distinguished groups. 

The sections marked as A and H in Fig. \ref{fig:projection_map_regions} contain dwarf stars, roughly those with $\logg$ $>$ 3.0. They are distributed according to a clear temperature gradient. The solar-like stars are found in section A while hotter stars (types O and B) occupy section H.
    
Section B includes giants with [Fe/H ] $\gtrsim$ $-$1.0. An apparent gradient in metallicity is identified, with values decreasing toward the direction of section C. There are also some secondary effects related to \teff\ and \logg. In particular, the brightest giants of low \teff\ seem to group in the same corner as the giants of high metallicity. These stars are projected together probably because both low \teff\ and high [Fe/H] values create strong spectral lines.

Section C includes low-metallicity giants with [Fe/H] $\lesssim$ $-$1.0. A metallicity gradient appears to be the main factor that causes the dispersion of the points. In the bottom tail of this section there is a concentration of the most metal-poor stars with [Fe/H] $\leq$ $-$2.0. Because there are only few of them, t-SNE interpreted that they are more similar amongst themselves than when compared to the other stars.  
Section D contains a series of metal-rich dwarfs that the algorithm detached from Section A. They all correspond to stars from NGC 6253, an open cluster observed by GES. We found these spectra to be affected by a series of artefacts in the form of spikes that were probably caused by failures in the data reduction pipeline, which was enough for the algorithm to isolate them. Section D exemplifies the ability of t-SNE to automatically detect anomalies within an otherwise homogeneous sample. 

Section E presents an unexpected group of stars that appear in maps using all the selected wavelength regions. These objects were found to be variable stars, mostly Cepheids. The combination of relatively high \teff\ values for their relatively low \logg\ values sets them apart from all other sample stars. 

Section F is dominated by stars in globular clusters (therefore usually metal poor) that are at the same time on the cool (\teff~$\leq$ 4000-4200 K) and bright (\logg~$\leq$ 1.0-1.2) sides of the sample. It seems to include a mixture of stars close to the tip of the red giant branch (RGB) and in the asymptotic giant branch (AGB). In the projection maps, they should be expected closer to the top left corner of Section C, which includes metal-poor giants. Their actual positioning --- close to the section of metal-rich giants --- is thus somewhat surprising. It may be related to the strengthening of molecular lines because of the low \teff~values. Alternatively, they could have some extreme chemical composition, being part of population II stars in globular clusters. These stars will be further investigated in future stages of the project, by obtaining their detailed chemical abundances. 

Section G contains a mixture of spectra from subgiants (\logg~$\geq$ 3.3-3.5) that are extremely hot  (\teff~$\geq$ 30000 K) or metal-poor ([Fe/H]~$\leq$ -1.1). It is almost entirely populated by reference spectra of these types. Both types of spectra feature very few spectral lines. In the R2 region, most of their similarity comes from similar H$_\beta$ profiles. As a result, the t-SNE algorithm groups them closely together on the map. Additionally, our sample includes relatively few such stars, which further limits their separation in the projection. 

\subsection{Testing large catalogues of stellar parameters}

\begin{figure}[htbp]
  \centering
  \includegraphics[width=0.9\linewidth]{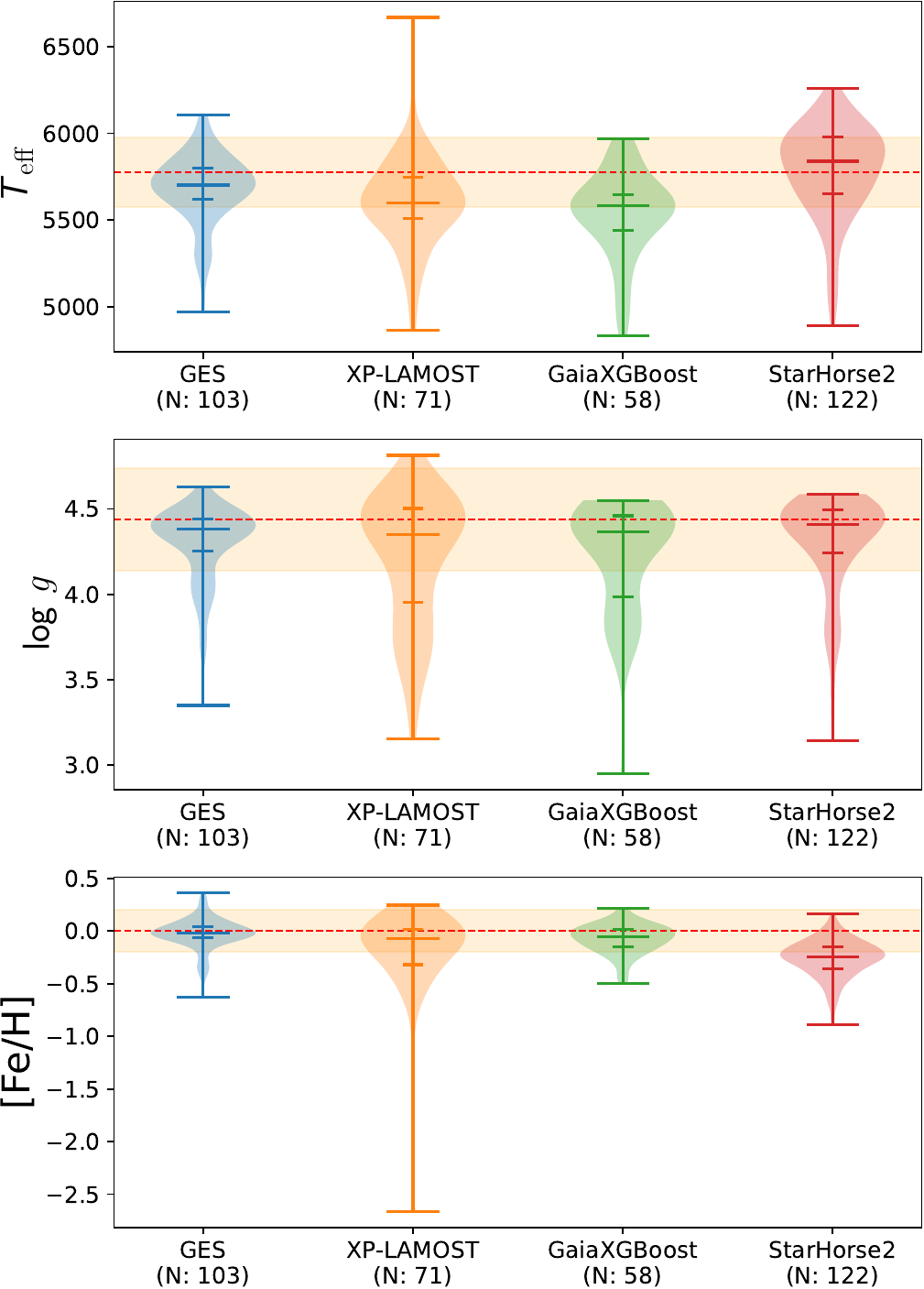}
  \caption{Violin plots showing the comparison between the stellar parameters of stars with spectra similar to the Sun in four catalogues.The red dashed line indicates the reference value of the parameter and the yellow coloured area is the region within 200 K in \teff (top panel), 0.3 dex in \logg (mid panel), and 0.2 dex in [Fe/H] (bottom panel) from the solar parameters.}
  \label{fig:violin_sunganym}
\end{figure}

In the context of a spectroscopic survey, one alternative use of our method of identifying similar spectra is to test the consistency of the results. When applying our method, stars with similar spectra should have similar atmospheric parameters, at least within the thresholds of 200 K in \teff, 0.3 dex in \logg, and 0.2 dex in [Fe/H]. Furthermore, comparison with reference stars can identify particularly problematic regions of the parameter space where analysis methods do not deliver accurate parameters.

Using the blind method to identify spectra similar to reference stars, we can test the catalogues introduced in \ref{sec:externalcats}. For each reference spectra, we calculated the bias and the interquartile range (IQR) of the parameters. The bias is the absolute deviation between the median value of a given stellar parameter and the corresponding value for the reference spectra. The IQR provides a measure of the dispersion of the values in a specific region of the parameter space. Figure \ref{fig:violin_sunganym} illustrates the case for a solar spectrum. Although the method does not select pure samples, as discussed before, Fig. \ref{fig:violin_sunganym} shows how the bias and dispersion change between catalogues. With such a diagnosis, it is possible to quickly select cases for detailed investigation with the aim of identifying the factors that limit the quality of the results.

\begin{figure*}
    \centering
    \includegraphics[width=\linewidth]{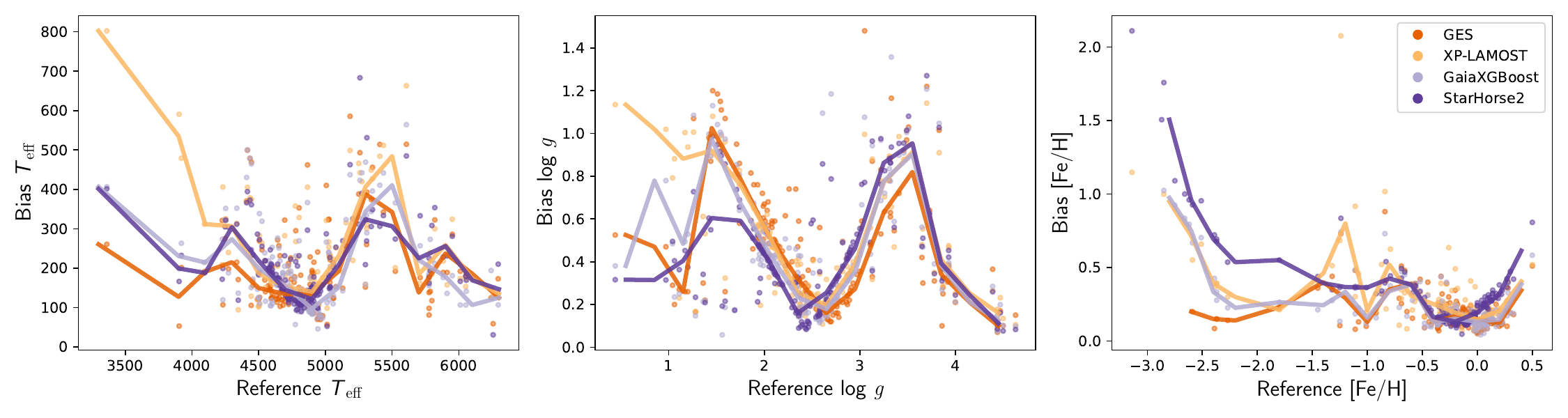}
    \includegraphics[width=\linewidth]{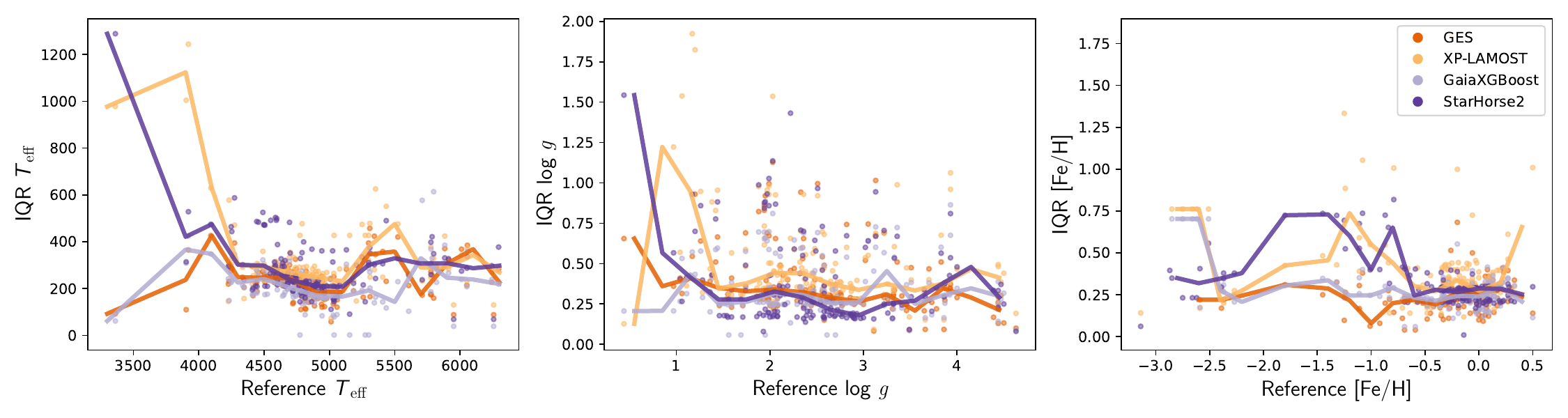}
    \caption{Comparison of four large catalogues of stellar parameters. Spectra similar to the reference stars were selected using our blind method. The median biases of these samples are plotted as a function of the parameters of the reference stars on the top row. The bottom row show the median interquartile range (IQR) as a function of the parameters. Low IQR indicates high precision and low bias indicates high accuracy.}
    \label{fig:cats_comparison}
\end{figure*}

Figure \ref{fig:cats_comparison} illustrates the overall comparison. Ideally, a high-quality catalogue should have both high accuracy and high precision, i.e. low bias and low IQR within the limits that the method can provide. In general, the biases behave similarly in all four catalogues. The values look particularly worrying for \teff~around 5500 K and for \logg~typical of subgiants (\logg~$\sim$ 3.5) and bright giants (\logg~$<$ 2.0). In this case, we cannot exclude problems in the adopted reference values. For metallicities, the biases increase for metal-poor stars with [Fe/H] $<$ $-$2.5 and metal-rich giants with [Fe/H] $>$+0.3.

The plots of IQR ranges in Fig. \ref{fig:cats_comparison} show where each catalogue has issues with precision, i.e. where parameters of stars with similar spectra tend to be different. Unlike the bias case, the precision evaluation is independent of the reference values. Regimes where the catalogues have low precision include cool stars (\teff~$<$ 4000 K), bright giants (\logg~$<$ 1.5), and metal-poor stars (in particular for StarHorse2 and XP-LAMOST, already below [Fe/H] $<$ $-$0.5). The bias and IQR values are tabulated in the Appendix \ref{app:tab}.

\section{Conclusions}\label{sec:conclusion}

This work is the first in a series of papers related to the CHESS pipeline and our project to analyse the UVES spectra of FGK-type stars with high precision and high accuracy. For that, the pipeline will rely on the differential analysis method using a series of reference stars that have accurate parameters distributed over the parameter space. In this paper, we present and discuss our method for finding similar stars based purely on their spectra. The performance of the method was tested and calibrated using catalogues of atmospheric parameters available in the literature.

Our main findings can be summarised as follows.

\begin{itemize}

    \item Identifying the correct astrophysical sources that correspond to each UVES spectrum proved to be a significant challenge due to inconsistencies in the metadata of the files.
 
    \item Because of that, we strongly recommend that observatories and data archives implement standardised formats for recording object coordinates and identifiers, in a way that can ensure the long-term utility of their datasets.

    \item We used the dimensionality reduction algorithm t-SNE to create 2D projection maps of our sample using their spectra as the multi-dimensional data. We tested the method using six spectral regions and found that the morphologies of the maps are different, but their overall characteristics are the same. Giants and dwarfs are well separated, with dwarfs distributed according to a gradient of temperature and giants according to a gradient of metallicities. Smaller groups of stars can also be identified, including a group of Cepheids, one of bright giants, and one of peculiar globular cluster stars.
    
\item To mitigate class imbalance when identifying similar stars, we applied t-SNE in two steps: the first iteration separated the main groups of interest in our sample (giants, dwarfs, and metal-poor stars), and the second iteration was applied within these groups, using a smaller value for the perplexity, to create tighter groups in the projected space.

    \item Using the GES catalogue of stellar parameters, we calibrated a metric focused on spectra similarities that allows the identification of similar spectra surrounding a given reference in the projection map. The stars identified in this way have parameters within 200 K in \teff, 0.3 dex in \logg, and 0.2 dex in [Fe/H]. The similarity threshold can be adjusted depending on the balance between completeness and purity needed in the selection.

\item Tests using our own differential analysis showed that with our choice of spectral metric, we can achieve completeness between 74-98 \% and typical purity between 39-54\% (with one case of very low purity, around 8\%).

\item Our method can also be used to test the precision and accuracy of large catalogues of stellar parameters. Testing GES, XP-LAMOST, GaiaXGBoost, and StarHorse2, we found a general strong and worrying bias in \logg~values for subgiants (\logg $\sim$ 3.5) and for stars with \teff~$\sim$ 5500 K. However, we cannot exclude the fact that the issue lies with the choice of reference stars. The biases are also large for cool stars (\teff $<$ 4000 K) and for metal-poor stars ([Fe/H] $<$ $-$0.5). 

\item We found the precision of the catalogues to decrease for cool stars (\teff~$<$ 4000 K), bright giants (\logg~$<$ 1.5), and metal-poor stars (in particular for StarHorse2 and XP-LAMOST, already below [Fe/H] $<$ $-$0.5). 

\item This analysis to identify similar stars is a crucial step in the CHESS pipeline. It will enable efficient selection of target stars for precise differential analysis.
    
\end{itemize}

\begin{acknowledgements}
We thank the referee for the fast and constructive report. J.~E.~Mart\'{i}nez Fern\'andez, S. \"Ozdemir, and R. Smiljanic acknowledge support from the National Science Centre, Poland, project 2019/34/E/ST9/00133. MLLD acknowledges Agencia Nacional de Investigación y Desarollo (ANID), Chile, Fondecyt Postdoctorado Folio 3240344. MLLD also acknowledges ANID Basal Project FB210003.  Based on data obtained from the ESO Science Archive Facility with DOI: \url{https://doi.eso.org/10.18727/archive/50} \citep{UVESreduceddata}. This research has made use of: NASA’s Astrophysics Data System Bibliographic Services; the SIMBAD database, operated at CDS, Strasbourg, France; the VizieR catalogue access tool, CDS, Strasbourg, France. The original description of the VizieR service was published in \citet{vizier2000}. This work has made use of data from the European Space Agency (ESA) mission {\it Gaia} (\url{https://www.cosmos.esa.int/gaia}), processed by the {\it Gaia} Data Processing and Analysis Consortium (DPAC,
\url{https://www.cosmos.esa.int/web/gaia/dpac/consortium}). Funding for the DPAC has been provided by national institutions, in particular the institutions participating in the {\it Gaia} Multilateral Agreement.
\end{acknowledgements}

\bibliographystyle{aa}
\bibliography{chess} 

\begin{appendix}

\onecolumn

\section{Catalog comparison summary table}
\label{app:tab}

\begin{table}[!ht]
    \centering
    \label{tab:cat_comparison}
    \caption{Catalogue comparison data from the Fig. \ref{fig:cats_comparison}. For each catalogue, the columns list values of "Bias | IQR". Each bin has a range of 200 K, 0.3 dex and 0.2 dex for \teff, \logg\ and [Fe/H], respectively.}
    \begin{tabular}{c c c c c}
    \hline\hline
        Bin Center (\teff) & GES & XP-LAMOST & GaiaXGBoost & StarHorse2 \\
    \hline
        3300 & 261 | 92 & 802 | 977 & 407 | 61 & 401 | 1288 \\ 
        3500 & - & - & - & - \\ 
        3700 & - & - & - & - \\ 
        3900 & 127 | 237 & 535 | 1124 & 231 | 366 & 199 | 421 \\ 
        4100 & 191 | 426 & 312 | 629 & 214 | 348 & 188 | 474 \\ 
        4300 & 215 | 248 & 307 | 304 & 274 | 225 & 304 | 302 \\ 
        4500 & 150 | 246 & 187 | 290 & 200 | 241 & 219 | 297 \\ 
        4700 & 134 | 244 & 153 | 268 & 154 | 203 & 145 | 237 \\ 
        4900 & 134 | 183 & 140 | 249 & 98 | 170 & 122 | 211 \\ 
        5100 & 193 | 186 & 239 | 231 & 169 | 165 & 208 | 207 \\
        5300 & 388 | 335 & 406 | 376 & 344 | 163 & 324 | 300 \\
        5500 & 343 | 358 & 484 | 476 & 411 | 143 & 306 | 329 \\
        5700 & 137 | 244 & 198 | 324 & 215 | 396 & 223 | 340 \\
        5900 & 243 | 332 & 253 | 285 & 173 | 223 & 271 | 309 \\
        6100 & 186 | 309 & 150 | 302 & 123 | 283 & 167 | 281 \\
        6300 & 99 | 189 & 135 | 228 & 125 | 181 & 161 | 306 \\ 
    \hline
    \end{tabular}
\end{table}

\begin{table}[!ht]
    \centering
    \begin{tabular}{c c c c c}
    \hline\hline
        Bin Center (\logg) & GES & XP-LAMOST & GaiaXGBoost & StarHorse2 \\
    \hline
        0.55 & 0.53 | 0.66 & 1.13 | 0.13 & 0.38 | 0.21 & 0.32 | 1.54 \\ 
        0.85 & 0.47 | 0.36 & 1.02 | 1.22 & 0.78 | 0.21 & 0.31 | 0.57 \\ 
        1.15 & 0.26 | 0.41 & 0.88 | 0.95 & 0.48 | 0.43 & 0.4 | 0.41 \\ 
        1.45 & 1.03 | 0.35 & 0.92 | 0.35 & 0.97 | 0.25 & 0.6 | 0.28 \\ 
        1.75 & 0.79 | 0.33 & 0.76 | 0.38 & 0.67 | 0.26 & 0.59 | 0.28 \\ 
        2.05 & 0.52 | 0.34 & 0.52 | 0.44 & 0.44 | 0.32 & 0.4 | 0.33 \\ 
        2.35 & 0.32 | 0.32 & 0.25 | 0.44 & 0.23 | 0.31 & 0.16 | 0.29 \\ 
        2.65 & 0.16 | 0.29 & 0.2 | 0.37 & 0.18 | 0.26 & 0.26 | 0.21 \\ 
        2.95 & 0.27 | 0.27 & 0.39 | 0.34 & 0.35 | 0.26 & 0.46 | 0.18 \\ 
        3.25 & 0.63 | 0.31 & 0.77 | 0.37 & 0.78 | 0.45 & 0.86 | 0.25 \\ 
        3.55 & 0.82 | 0.17 & 0.92 | 0.33 & 0.91 | 0.21 & 0.95 | 0.27 \\ 
        3.85 & 0.35 | 0.33 & 0.4 | 0.37 & 0.42 | 0.3 & 0.42 | 0.39 \\ 
        4.15 & 0.22 | 0.28 & 0.25 | 0.43 & 0.19 | 0.34 & 0.19 | 0.45 \\ 
        4.45 & 0.12 | 0.27 & 0.17 | 0.37 & 0.14 | 0.25 & 0.11 | 0.32 \\ 
    \hline
    \end{tabular}
\end{table}

\begin{table}[!ht]
    \centering
    \begin{tabular}{c c c c c}
    \hline\hline
        Bin Center ([Fe/H]) & GES & XP-LAMOST & GaiaXGBoost & StarHorse2 \\
    \hline
        -2.80 & - & 0.95 | 0.76 & 0.97 | 0.7 & 1.51 | 0.35 \\ 
        -2.60 & 0.2 | 0.22 & 0.71 | 0.76 & 0.75 | 0.7 & 0.96 | 0.32 \\ 
        -2.40 & 0.15 | 0.22 & 0.39 | 0.22 & 0.33 | 0.28 & 0.69 | 0.35 \\ 
        -2.20 & 0.14 | 0.25 & 0.3 | 0.27 & 0.23 | 0.21 & 0.54 | 0.38 \\ 
        -2.00 & - & - & - & - \\ 
        -1.80 & 0.23 | 0.31 & 0.21 | 0.43 & 0.26 | 0.31 & 0.55 | 0.72 \\ 
        -1.60 & - & - & - & - \\ 
        -1.40 & 0.4 | 0.29 & 0.46 | 0.46 & 0.24 | 0.34 & 0.39 | 0.73 \\ 
        -1.20 & 0.29 | 0.22 & 0.8 | 0.74 & 0.34 | 0.25 & 0.37 | 0.6 \\ 
        -1.00 & 0.13 | 0.08 & 0.21 | 0.55 & 0.16 | 0.25 & 0.36 | 0.4 \\ 
        -0.80 & 0.35 | 0.2 & 0.52 | 0.44 & 0.34 | 0.29 & 0.42 | 0.65 \\ 
        -0.60 & 0.39 | 0.22 & 0.31 | 0.3 & 0.38 | 0.23 & 0.38 | 0.25 \\ 
        -0.40 & 0.13 | 0.2 & 0.25 | 0.27 & 0.13 | 0.21 & 0.16 | 0.28 \\ 
        -0.20 & 0.17 | 0.26 & 0.19 | 0.34 & 0.15 | 0.24 & 0.13 | 0.27 \\ 
        0.00 & 0.15 | 0.26 & 0.14 | 0.27 & 0.13 | 0.24 & 0.19 | 0.29 \\ 
        0.20 & 0.13 | 0.29 & 0.2 | 0.32 & 0.15 | 0.28 & 0.32 | 0.28 \\ 
        0.40 & 0.35 | 0.24 & 0.41 | 0.65 & 0.39 | 0.21 & 0.61 | 0.25 \\ 
    \hline

    \hline
    \end{tabular}
\end{table}

\section{Projection maps for all the spectral regions with data from all the catalogues}
\label{app:A}

\begin{figure*}[htbp]
    \centering
    \begin{tabular}{ccc}
        \includegraphics[width=0.3\textwidth]{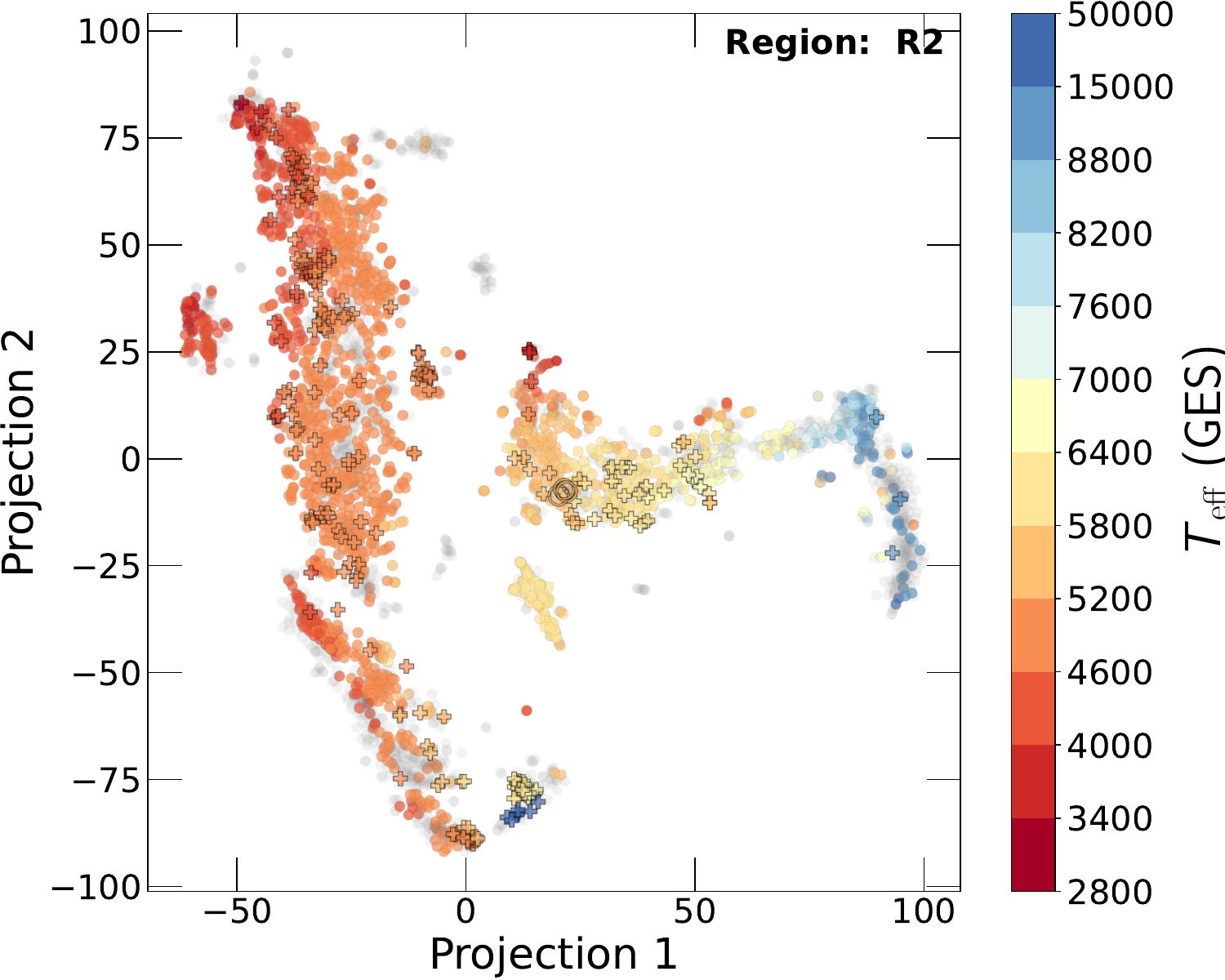} & \includegraphics[width=0.3\textwidth]{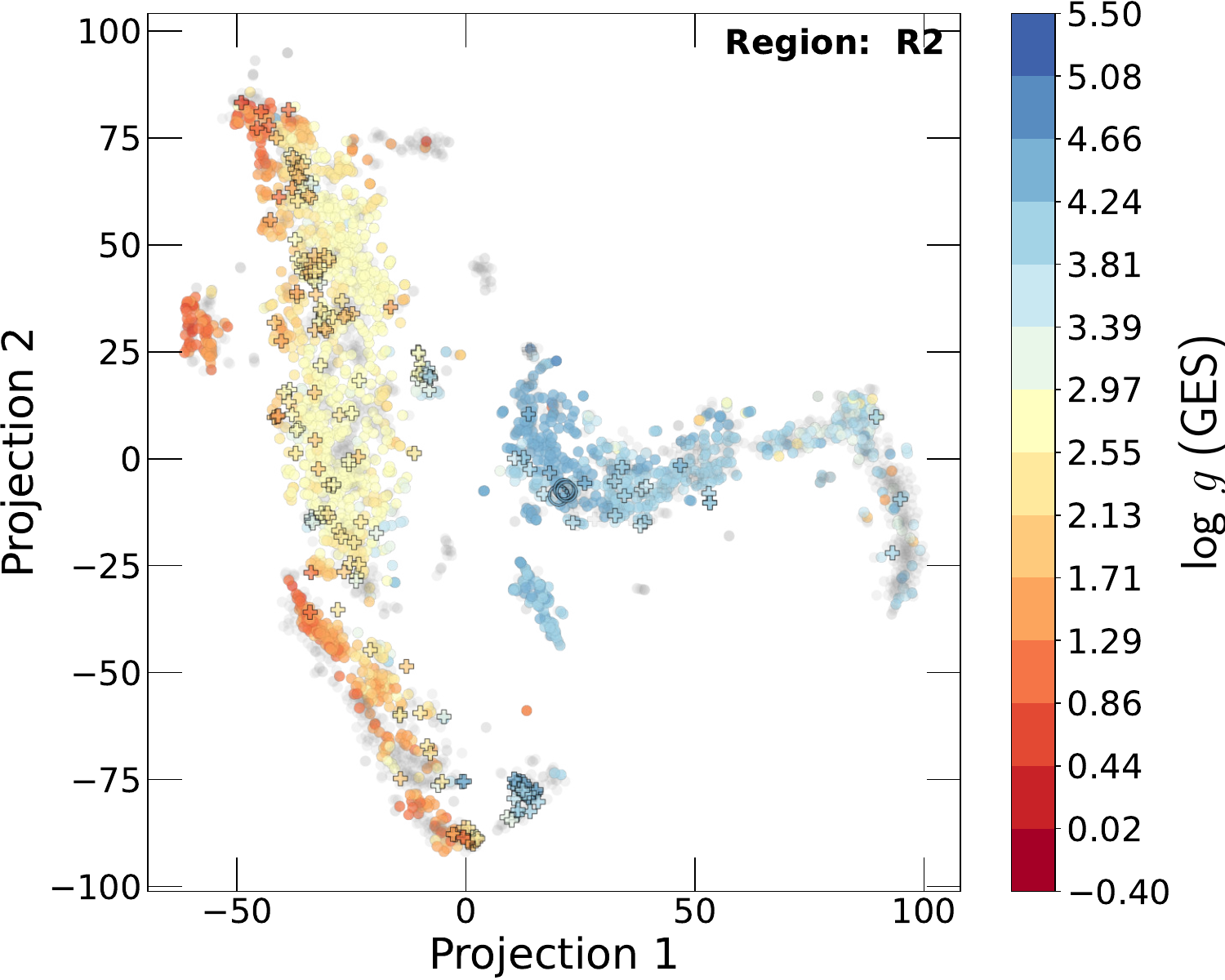} & \includegraphics[width=0.3\textwidth]{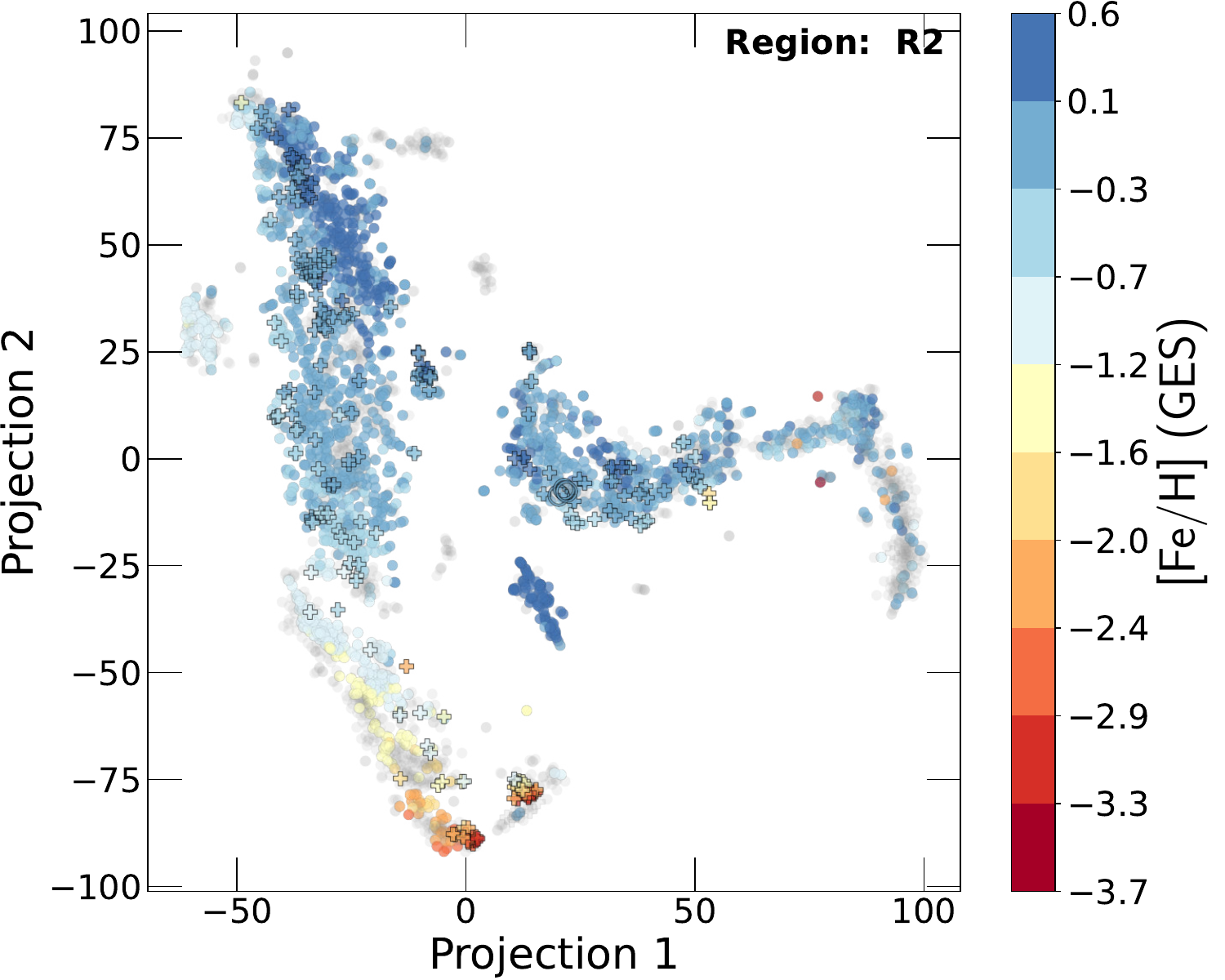} \\
        \includegraphics[width=0.3\textwidth]{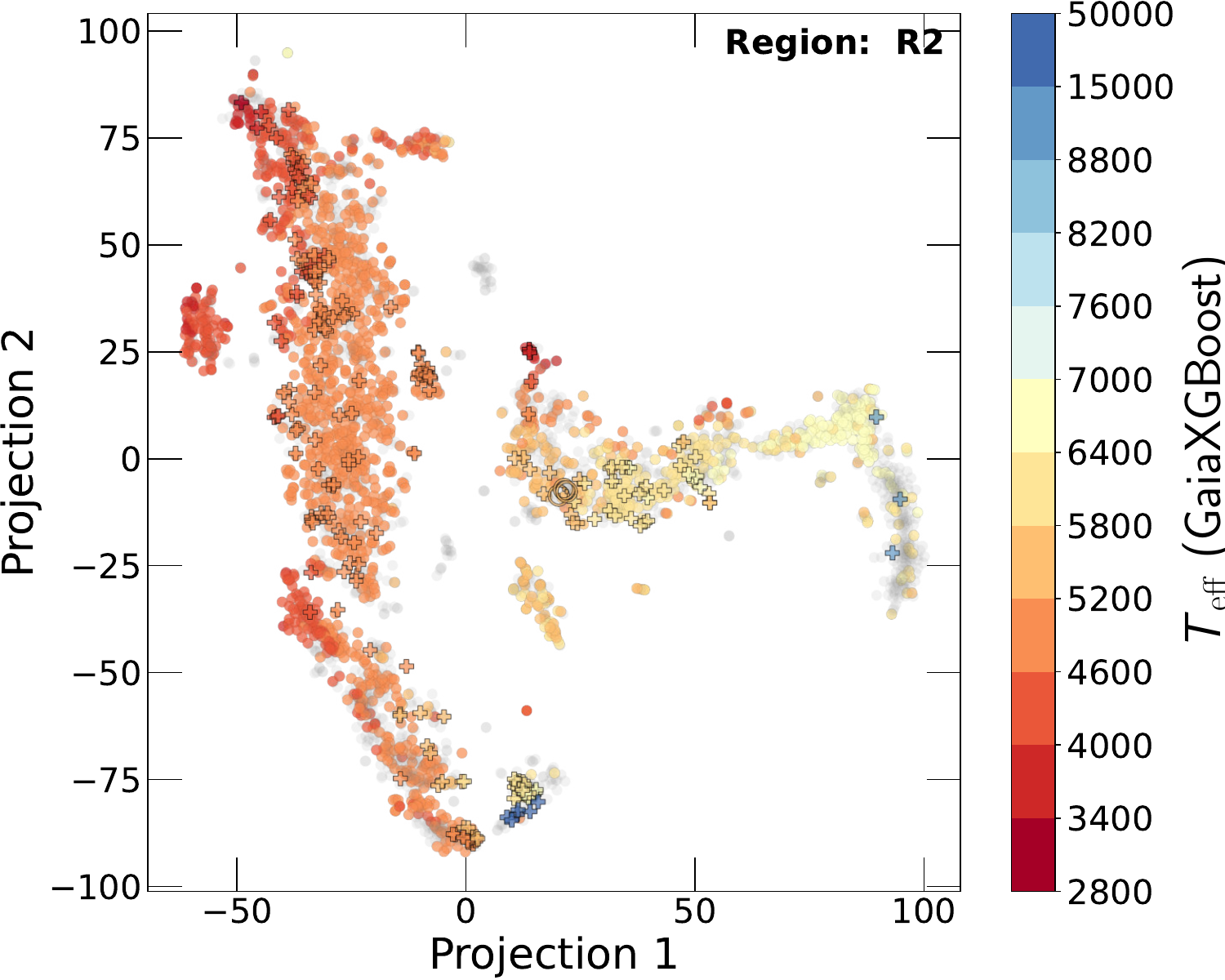} & \includegraphics[width=0.3\textwidth]{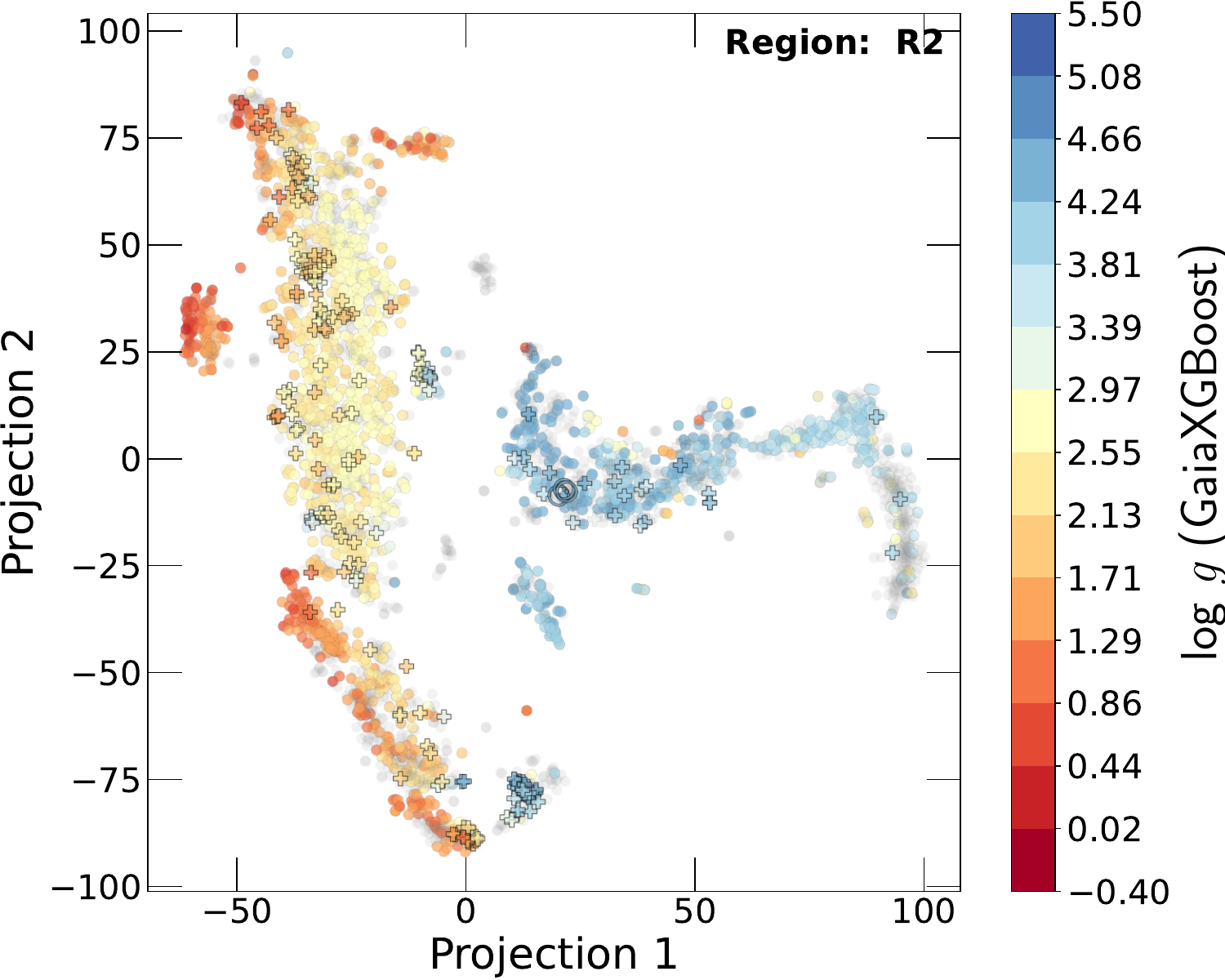} & \includegraphics[width=0.3\textwidth]{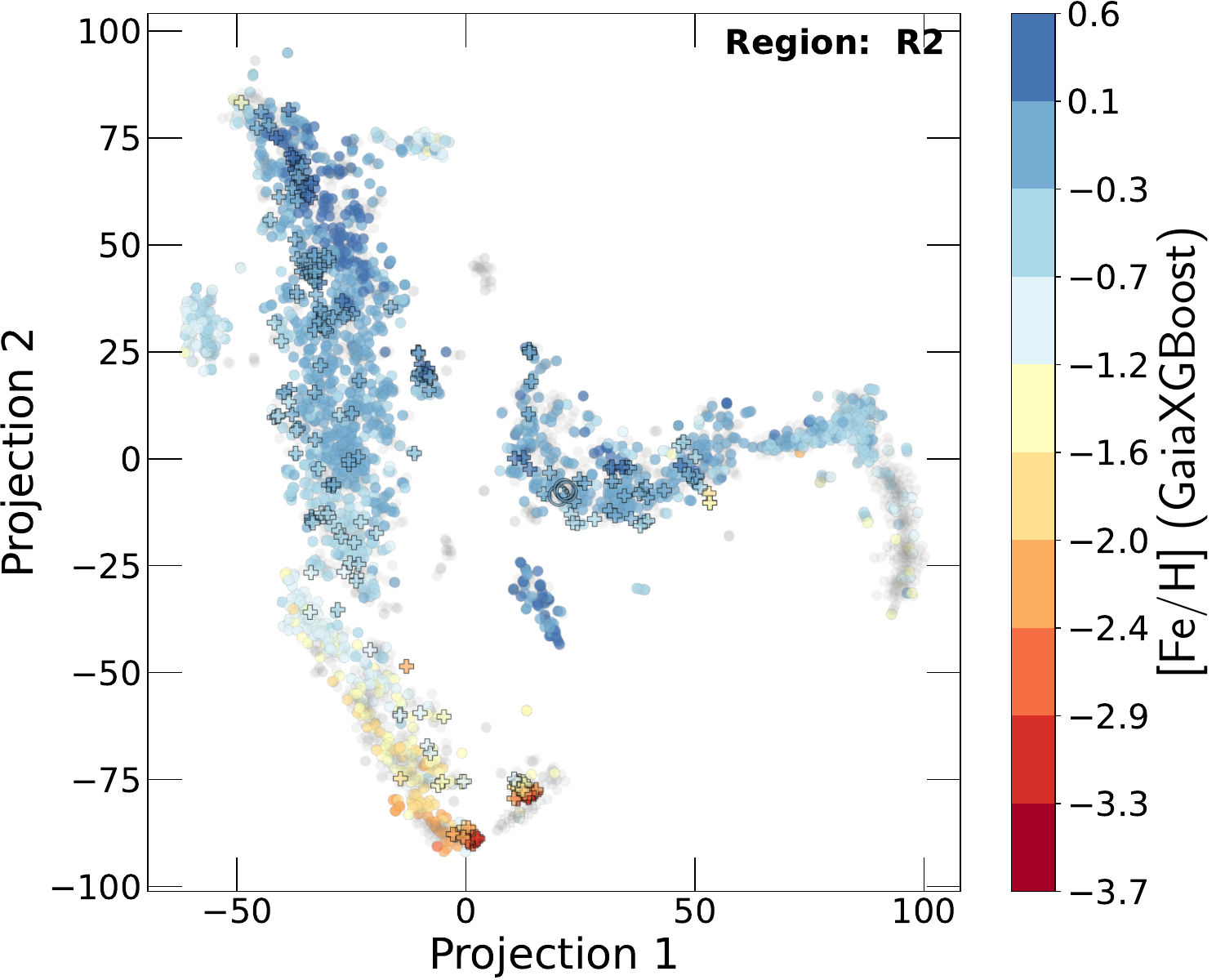} \\
        \includegraphics[width=0.3\textwidth]{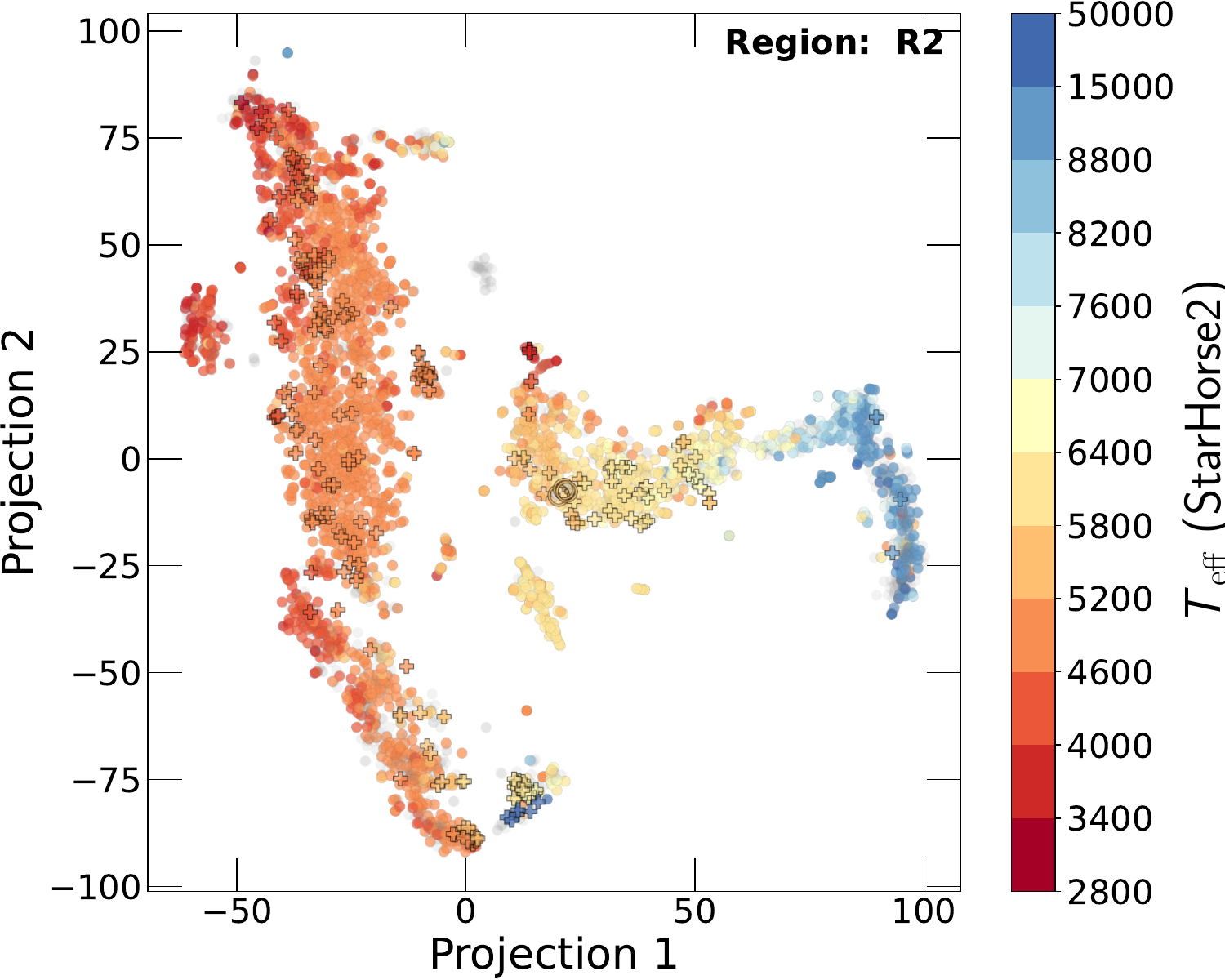} &
        \includegraphics[width=0.3\textwidth]{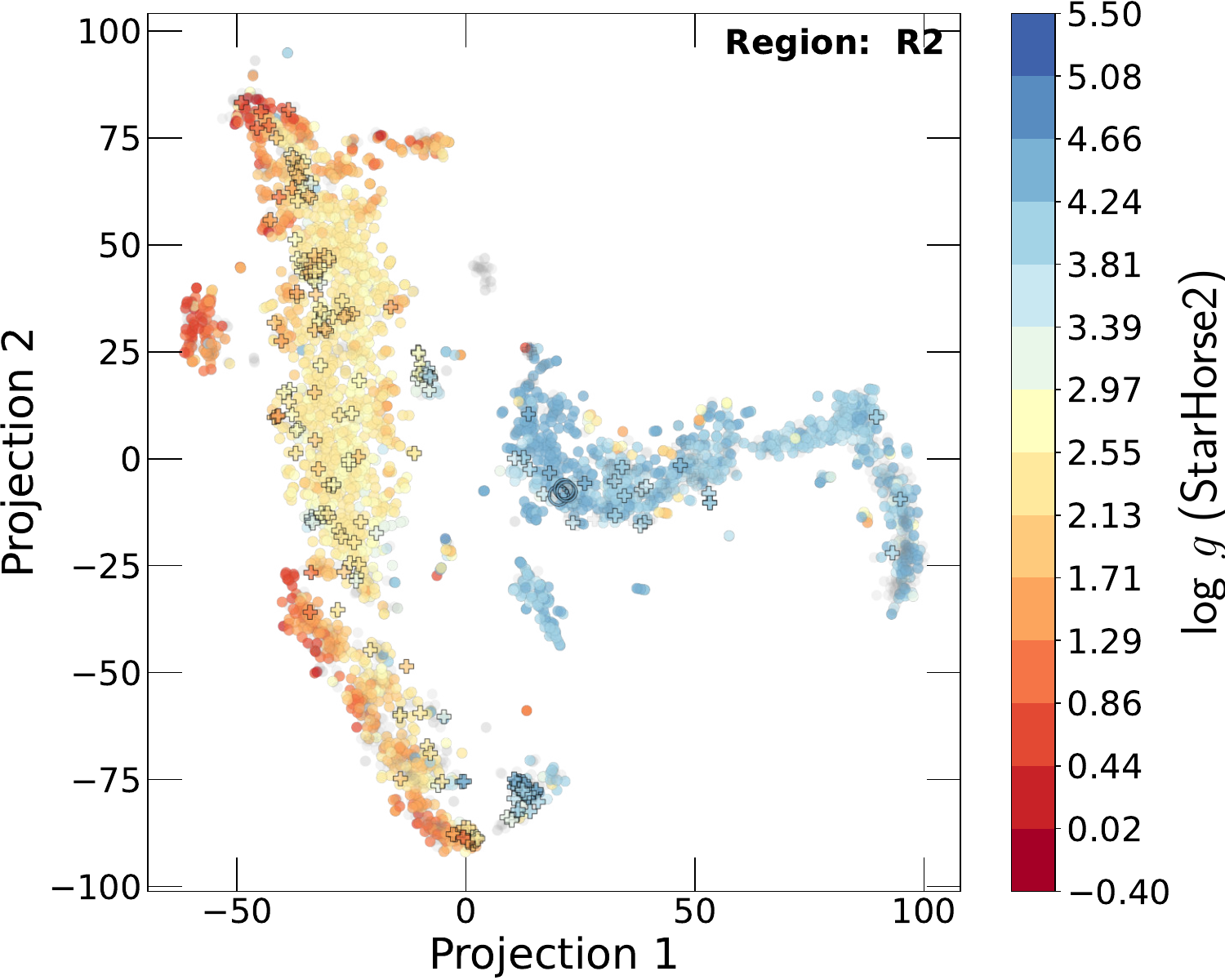} &
        \includegraphics[width=0.3\textwidth]{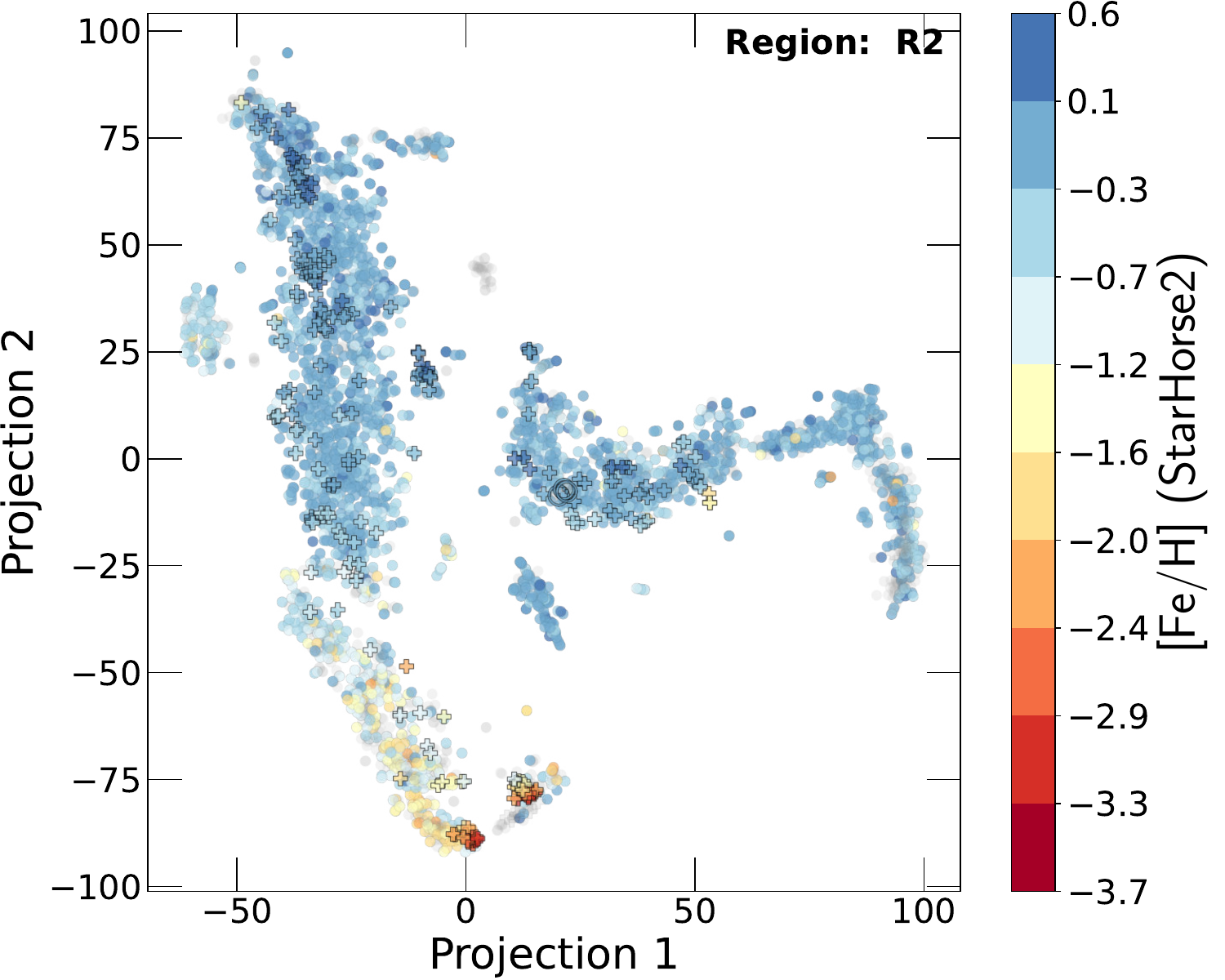} \\
        \includegraphics[width=0.3\textwidth]{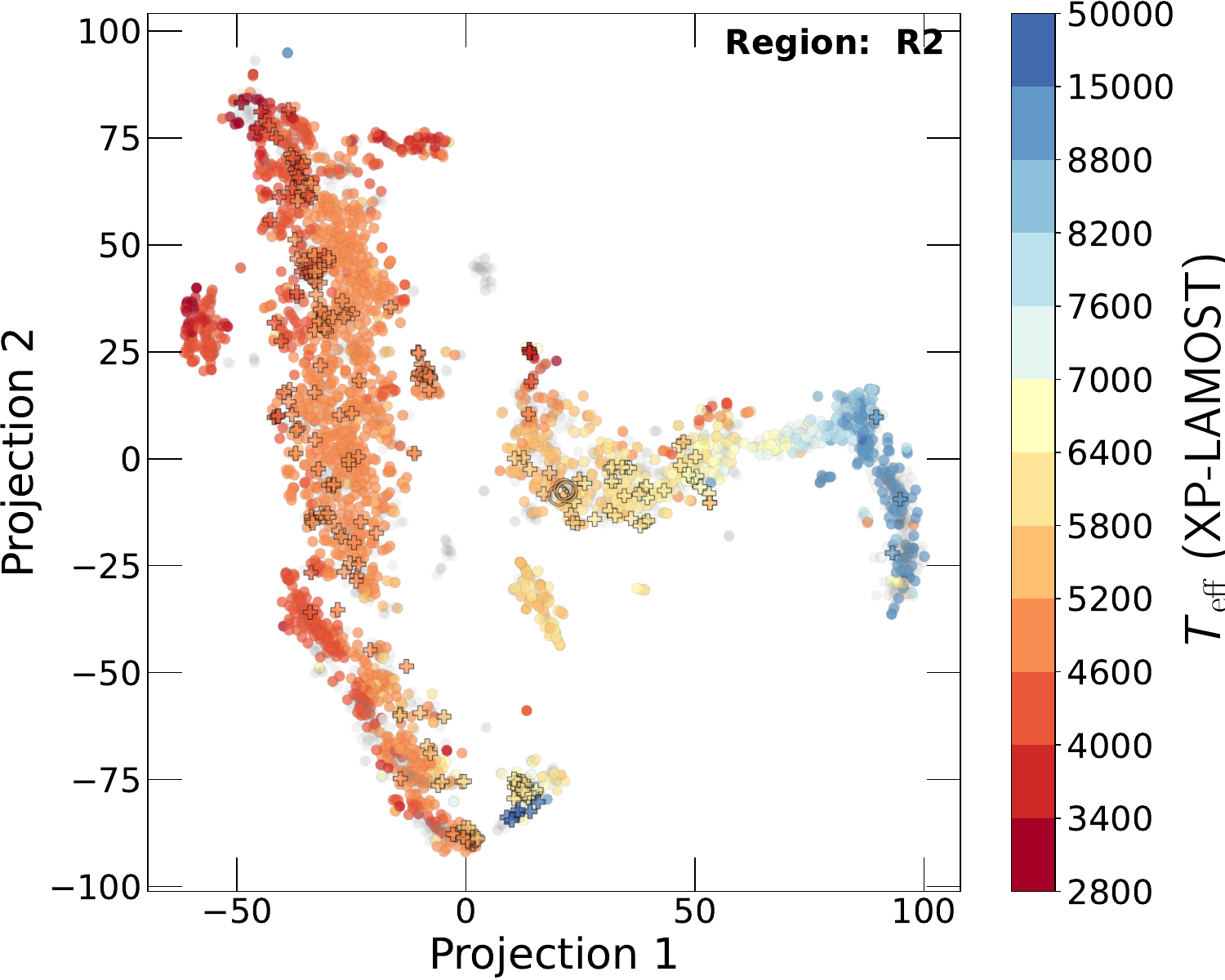} &
        \includegraphics[width=0.3\textwidth]{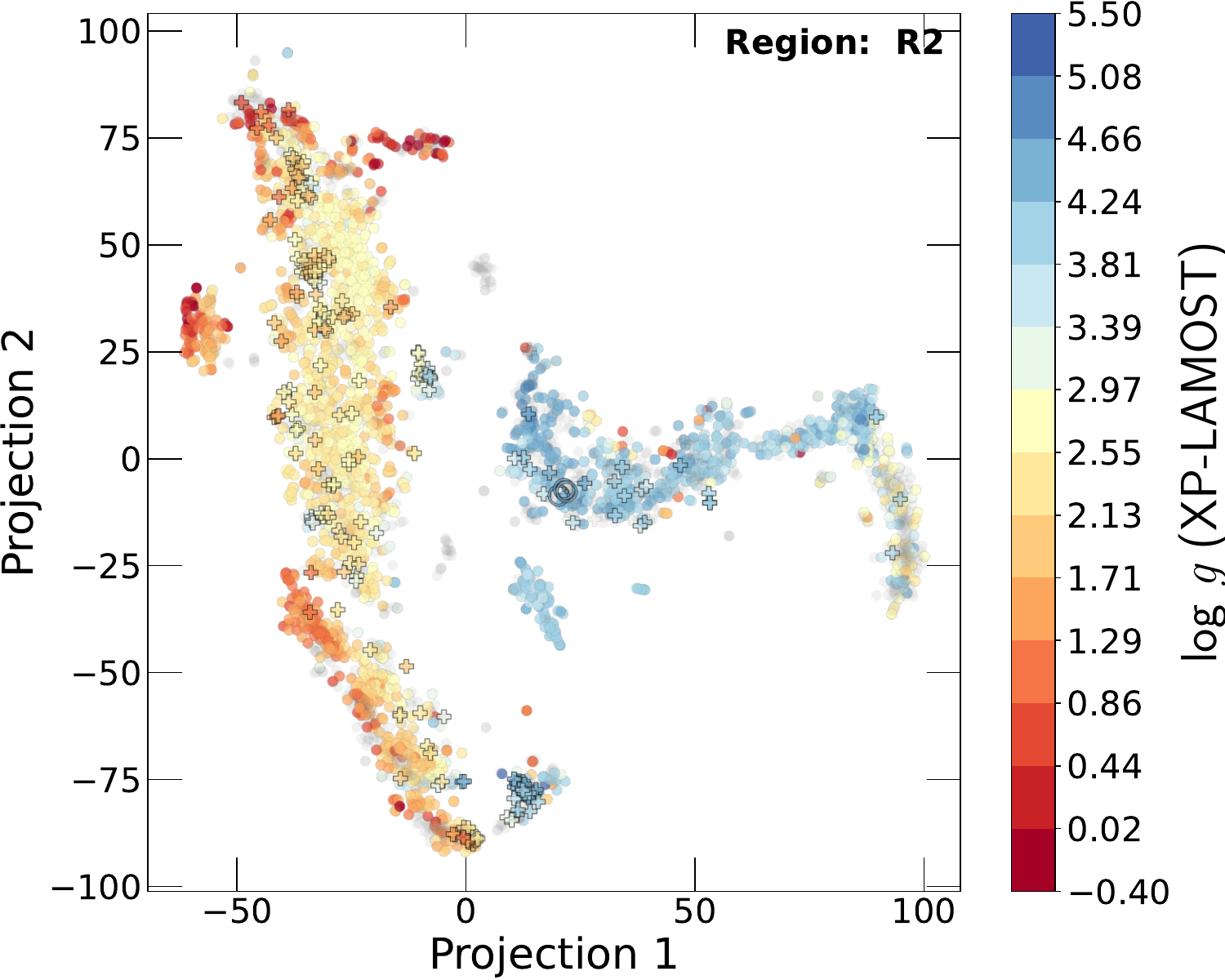} &
        \includegraphics[width=0.3\textwidth]{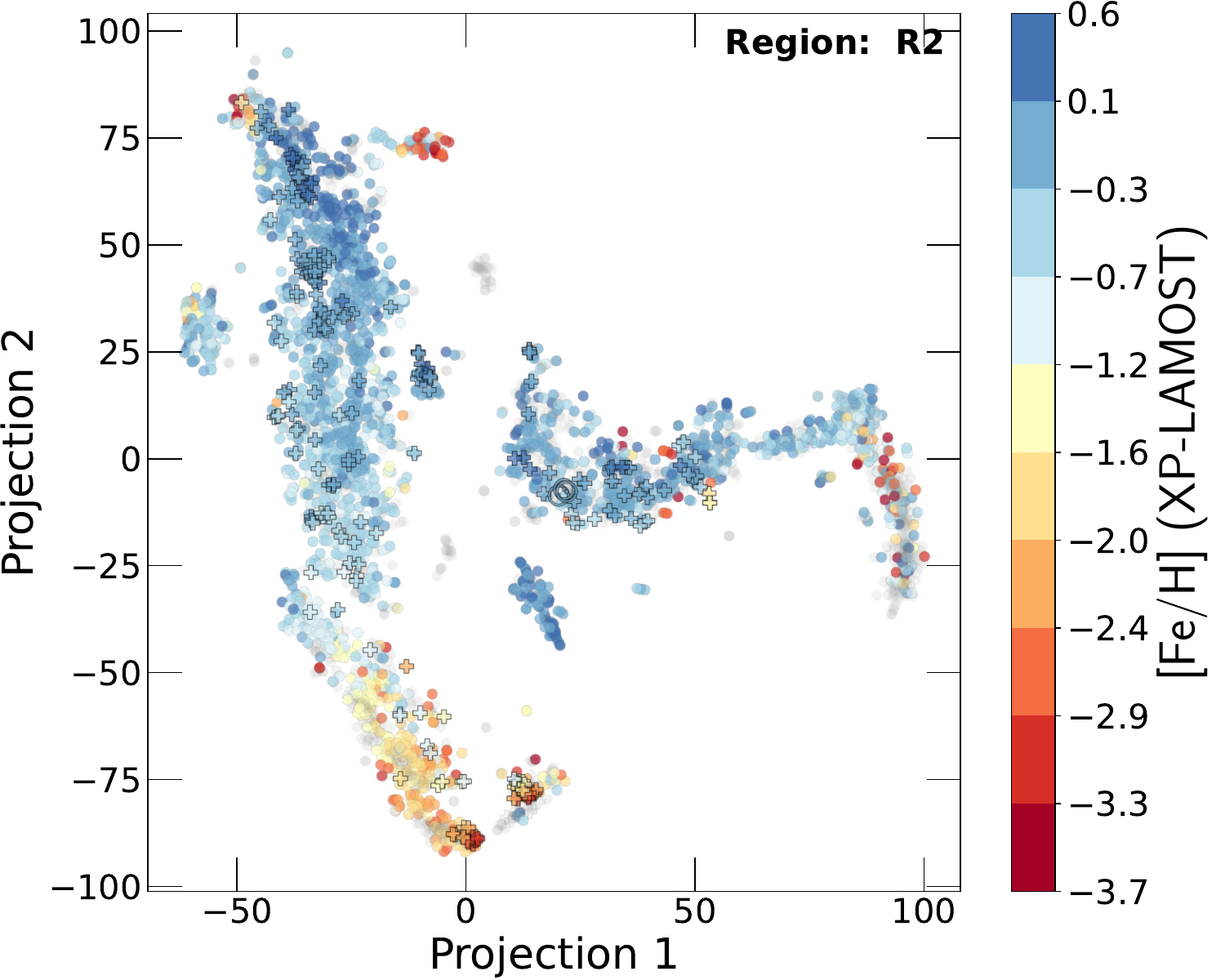} \\
    \end{tabular}
    \caption{Projection maps for the region R2 coloured with the atmospheric parameters of four different catalogues.}
\end{figure*}

\begin{figure*}[htbp]
    \centering
    \begin{tabular}{ccc}
        \includegraphics[width=0.3\textwidth]{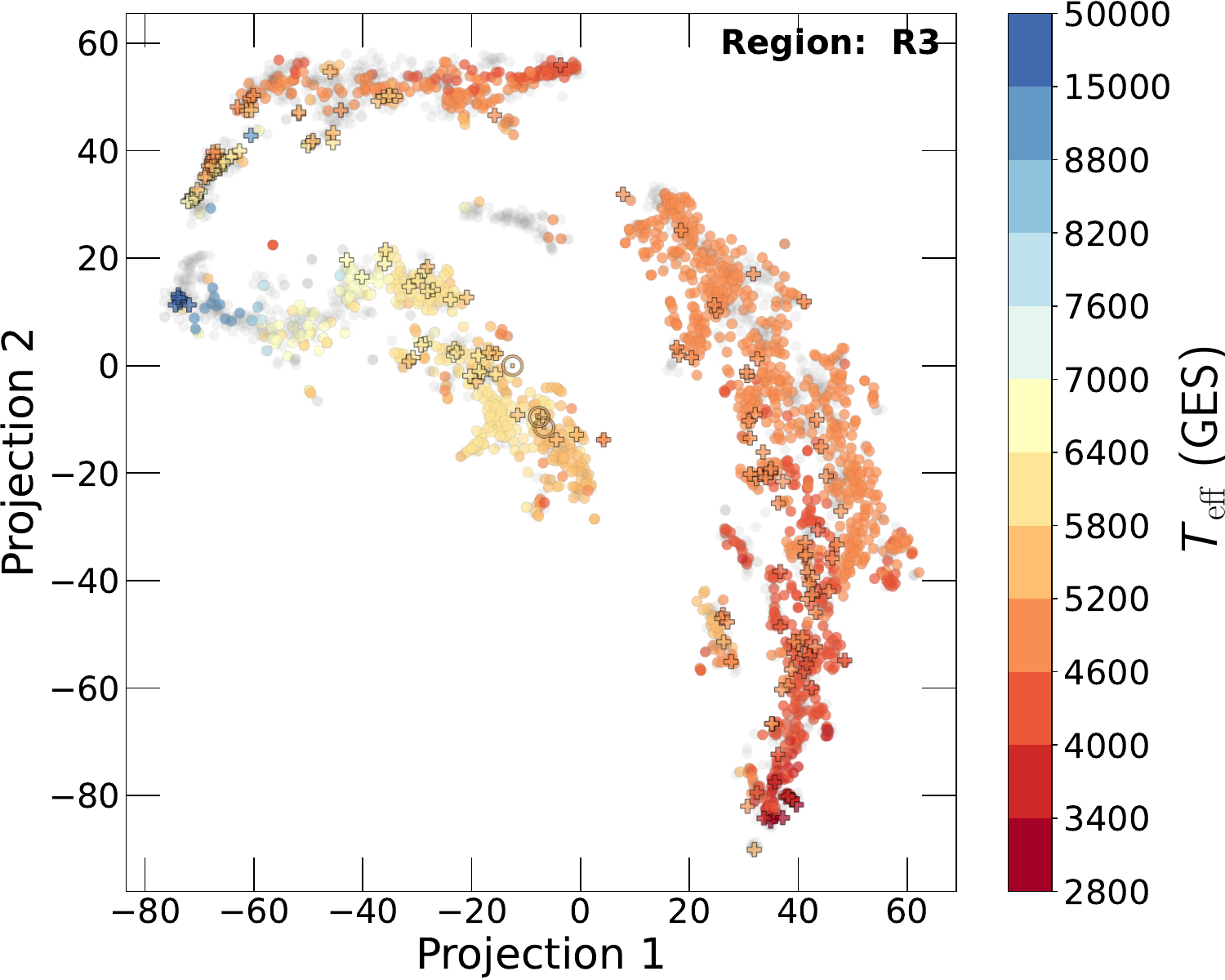} & \includegraphics[width=0.3\textwidth]{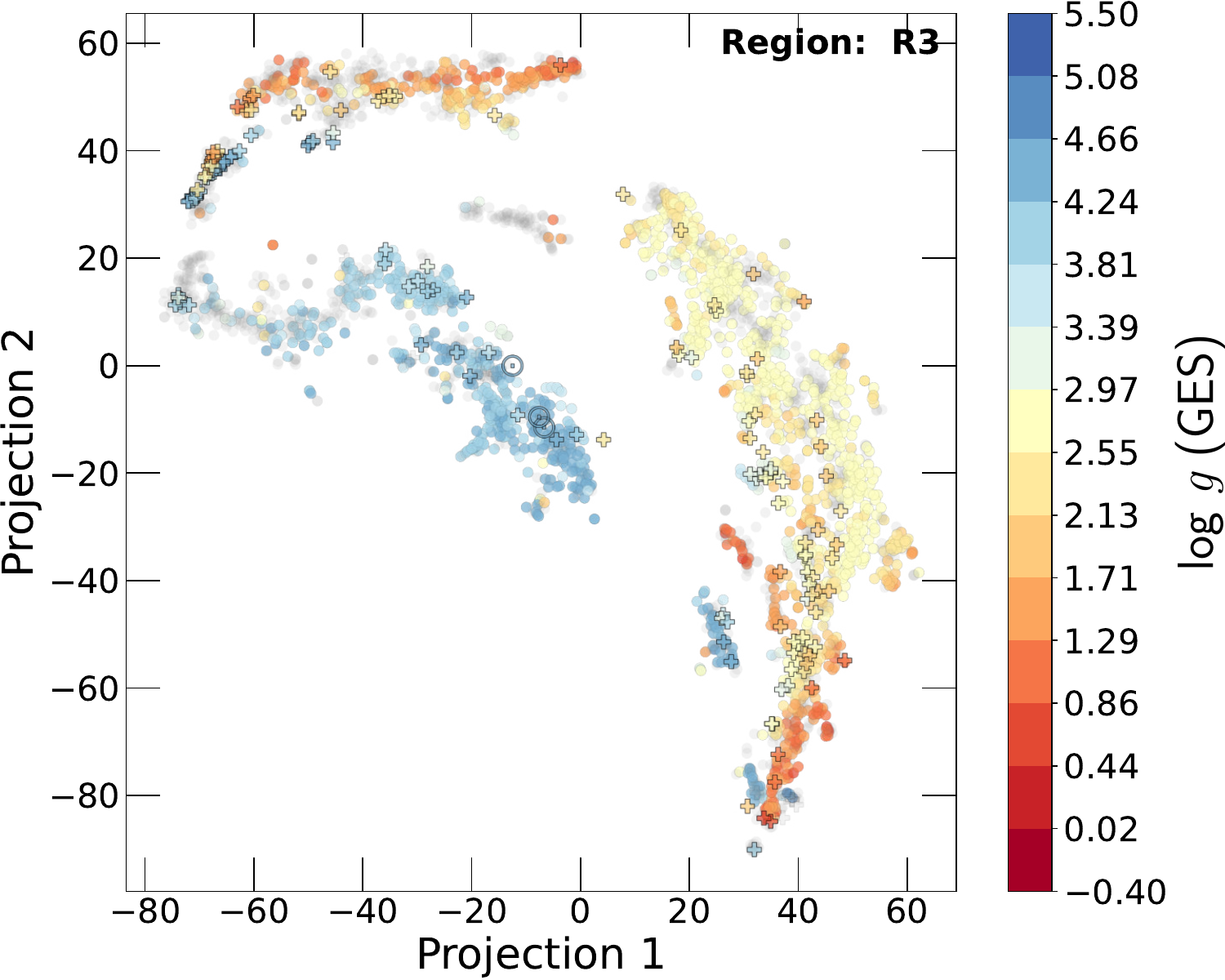} & \includegraphics[width=0.3\textwidth]{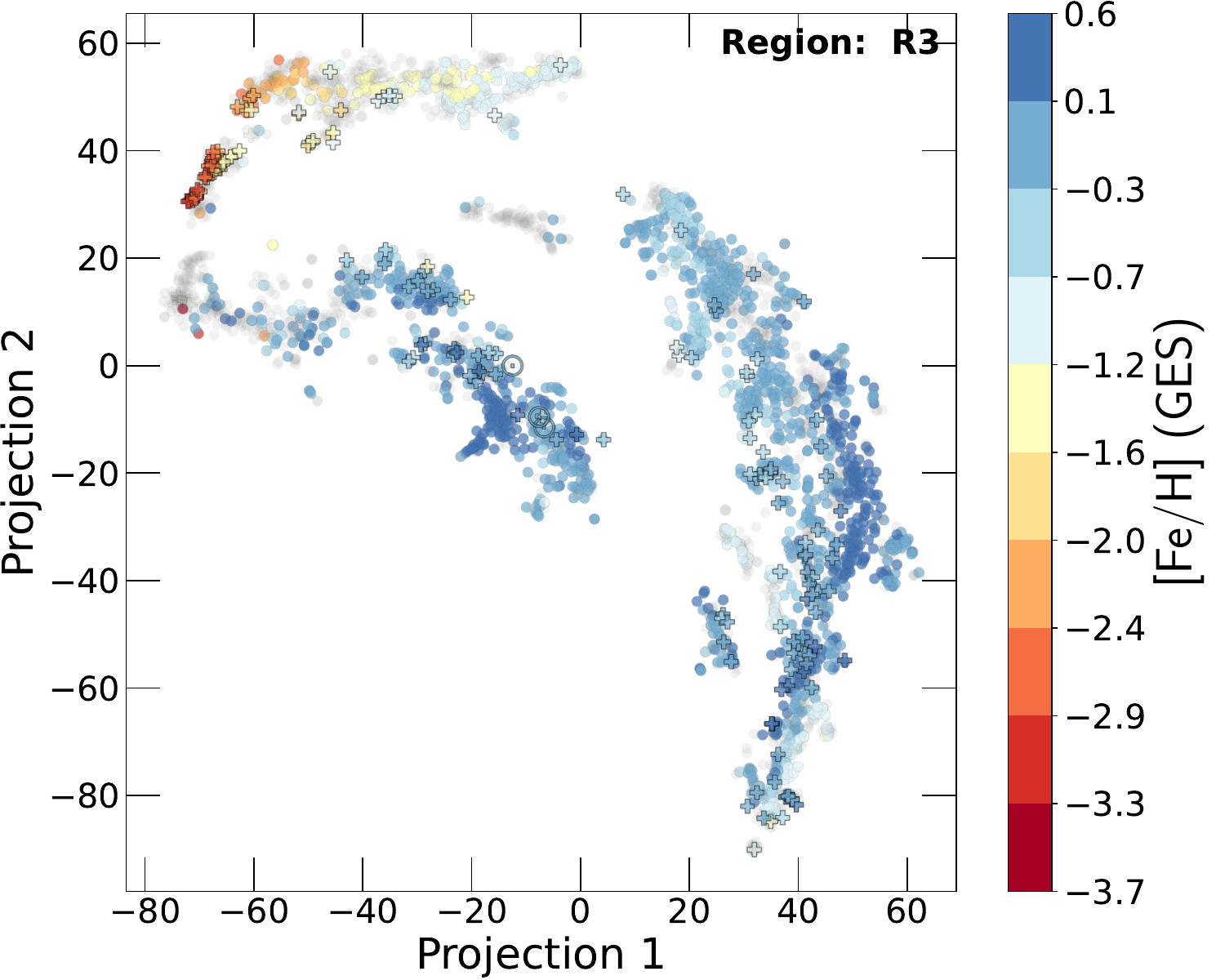} \\
        \includegraphics[width=0.3\textwidth]{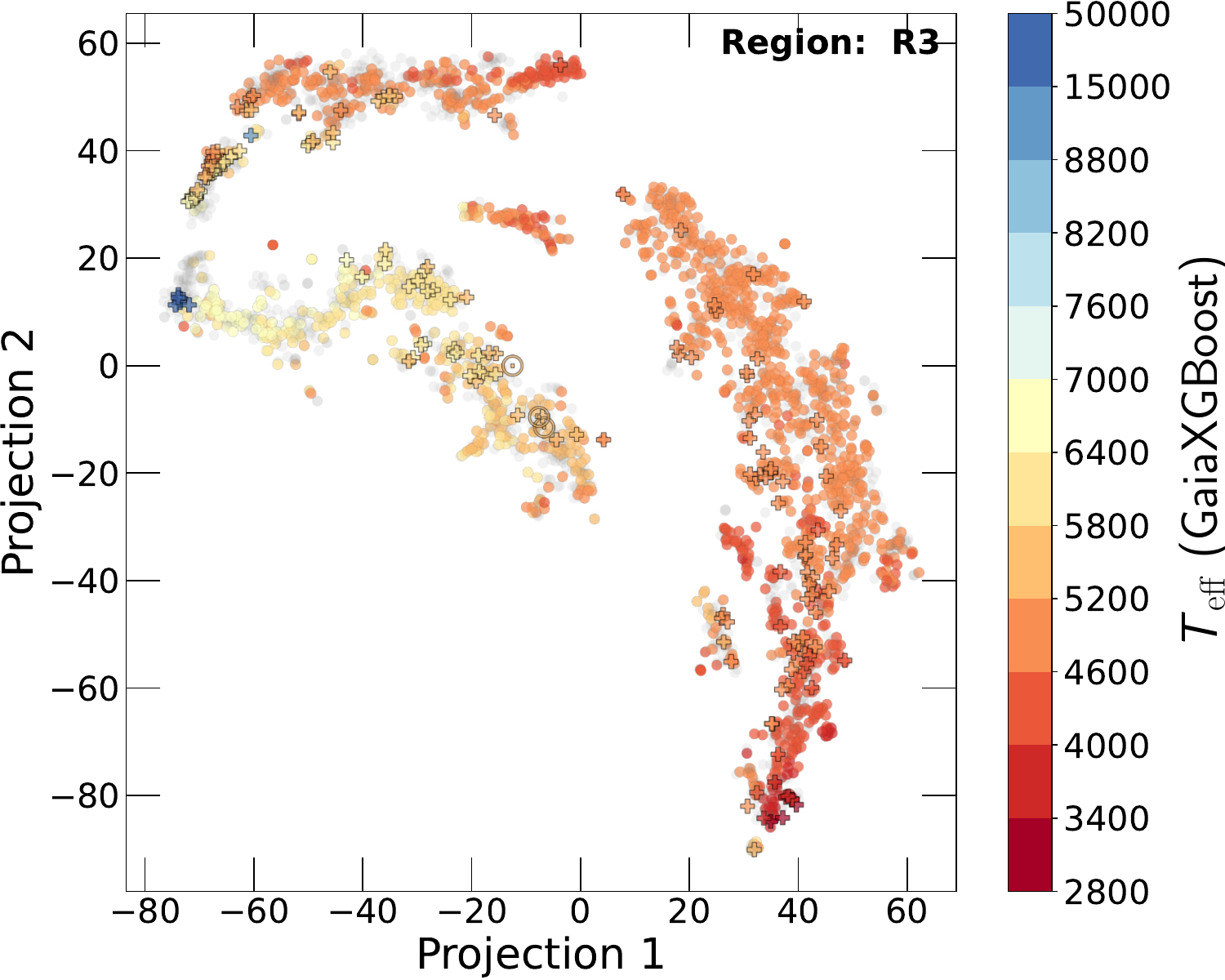} & \includegraphics[width=0.3\textwidth]{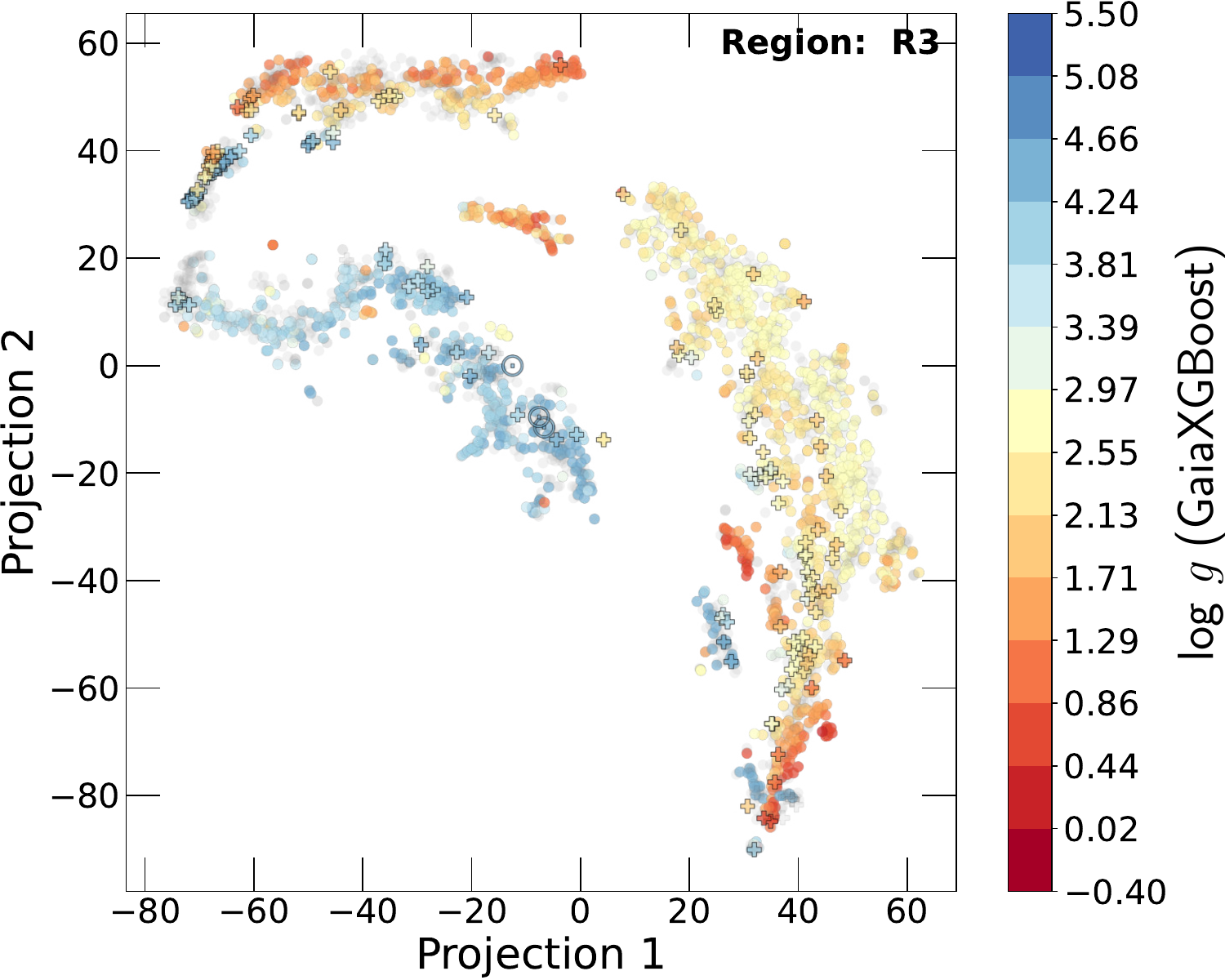} & \includegraphics[width=0.3\textwidth]{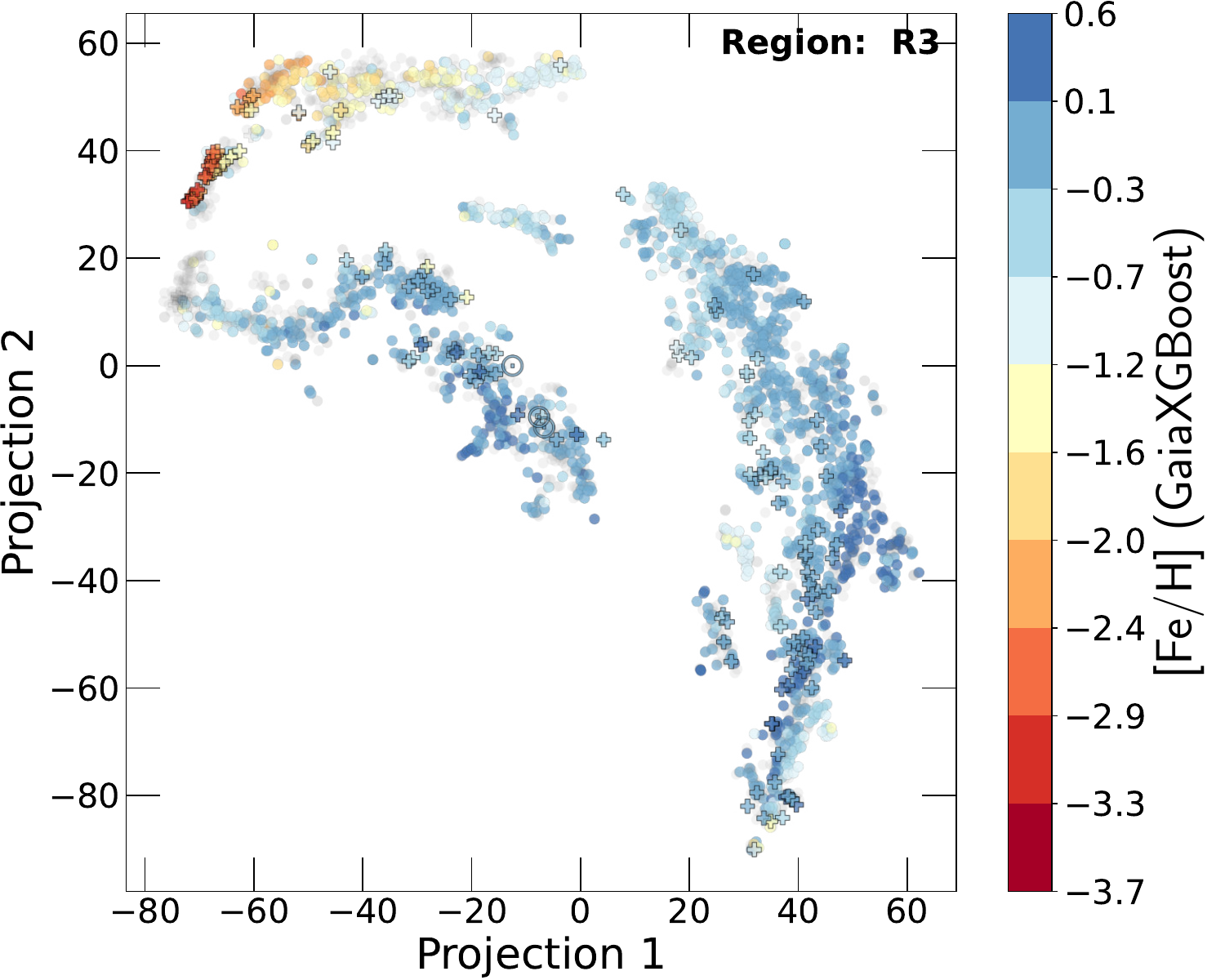} \\
        \includegraphics[width=0.3\textwidth]{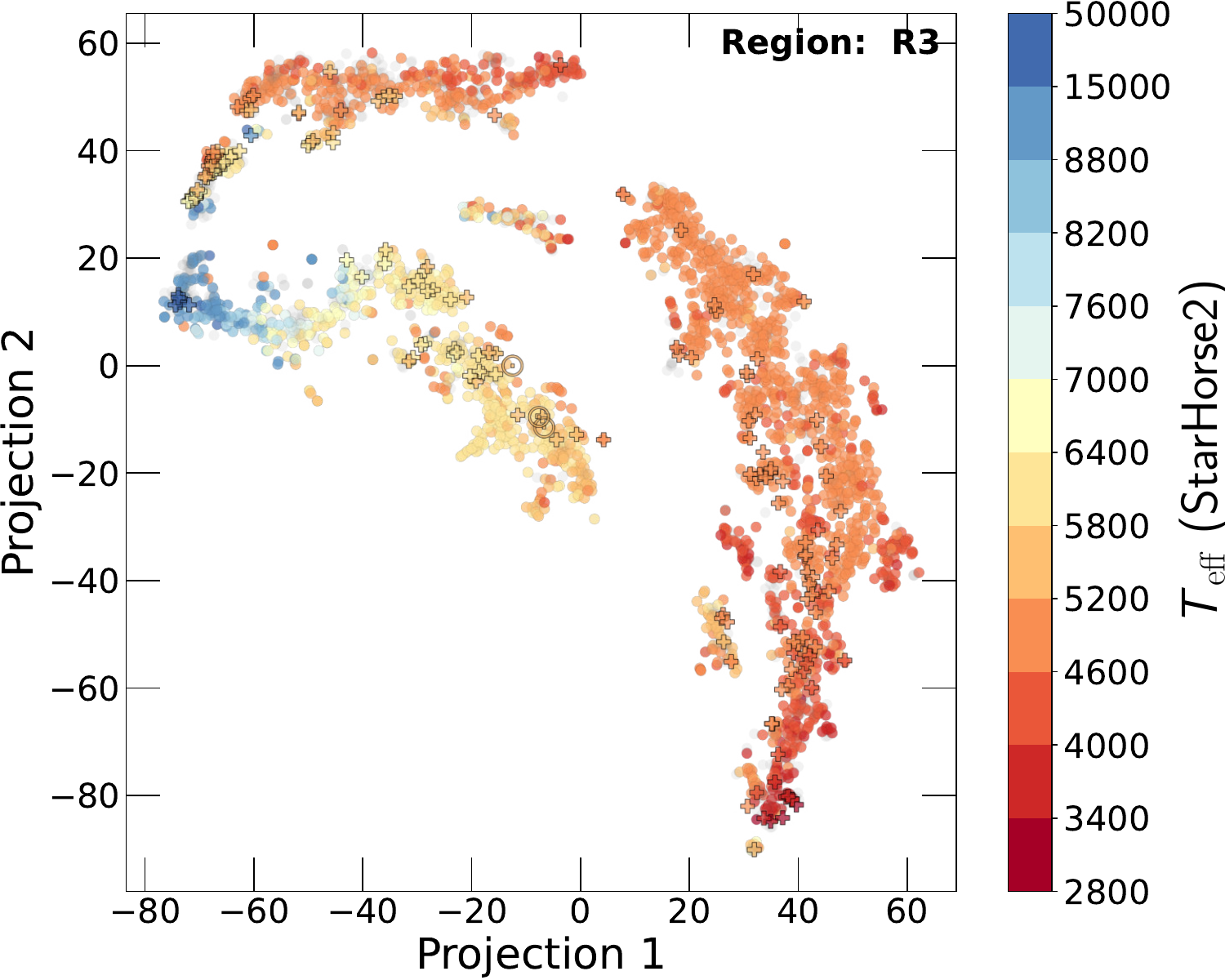} &
        \includegraphics[width=0.3\textwidth]{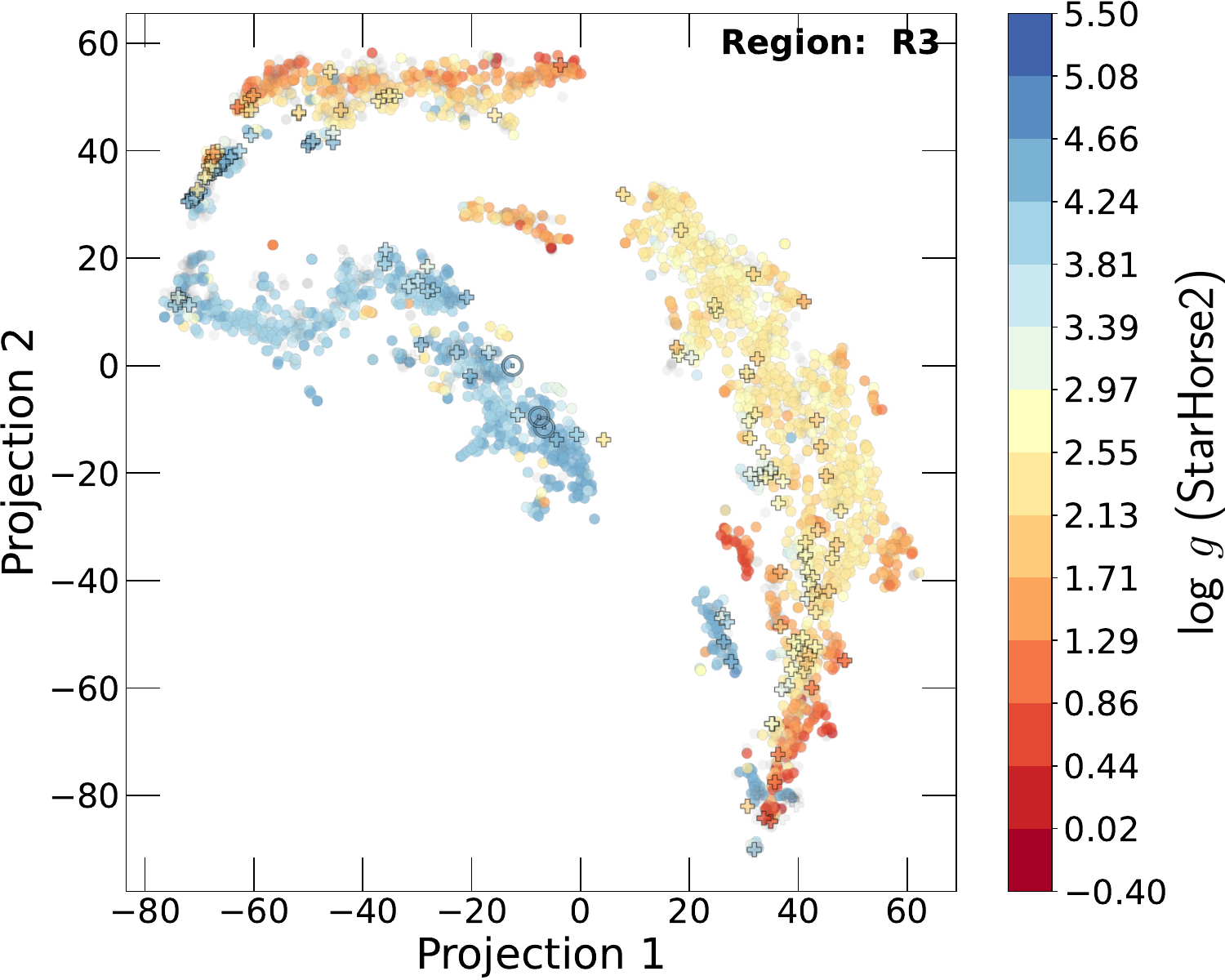} &
        \includegraphics[width=0.3\textwidth]{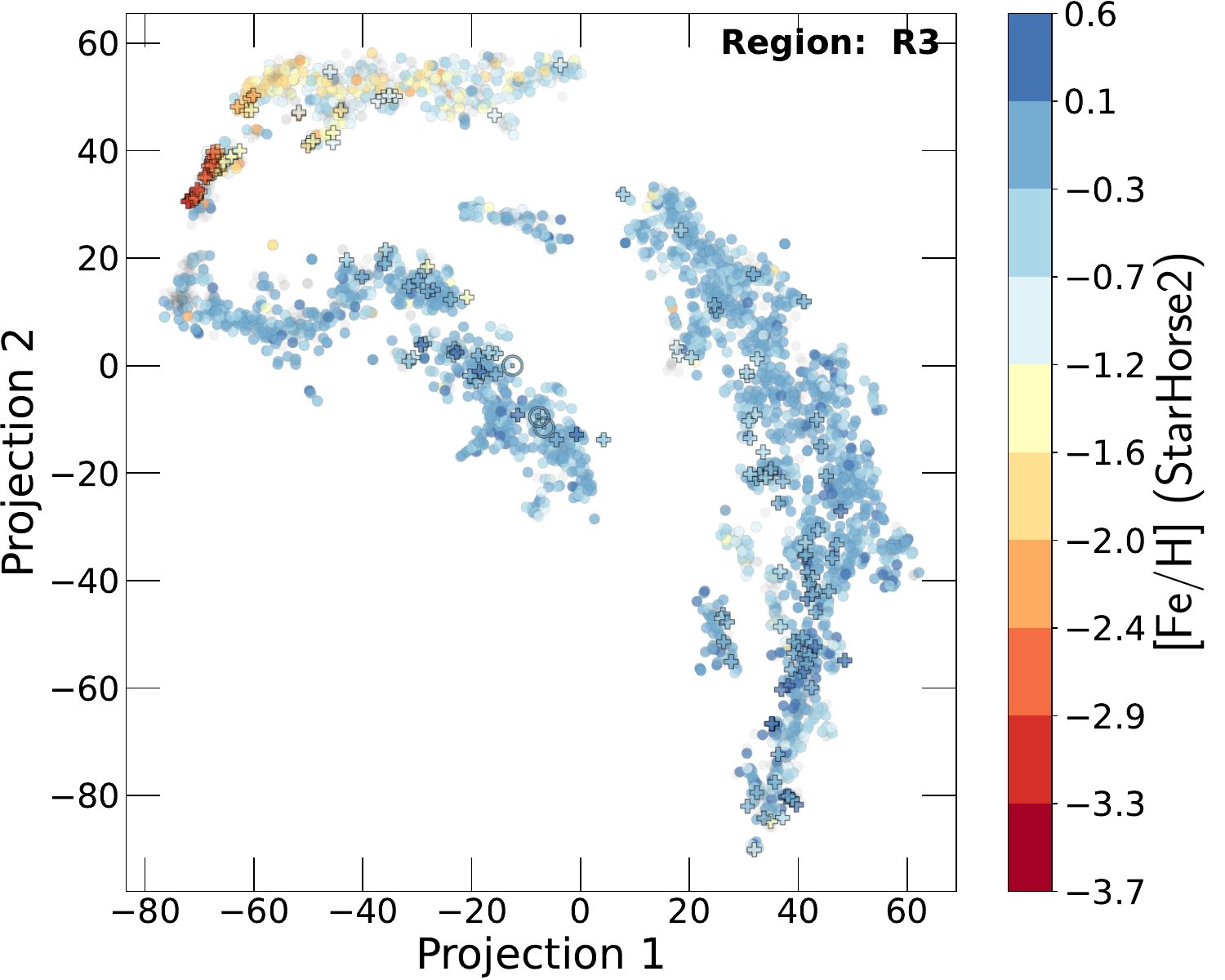} \\
        \includegraphics[width=0.3\textwidth]{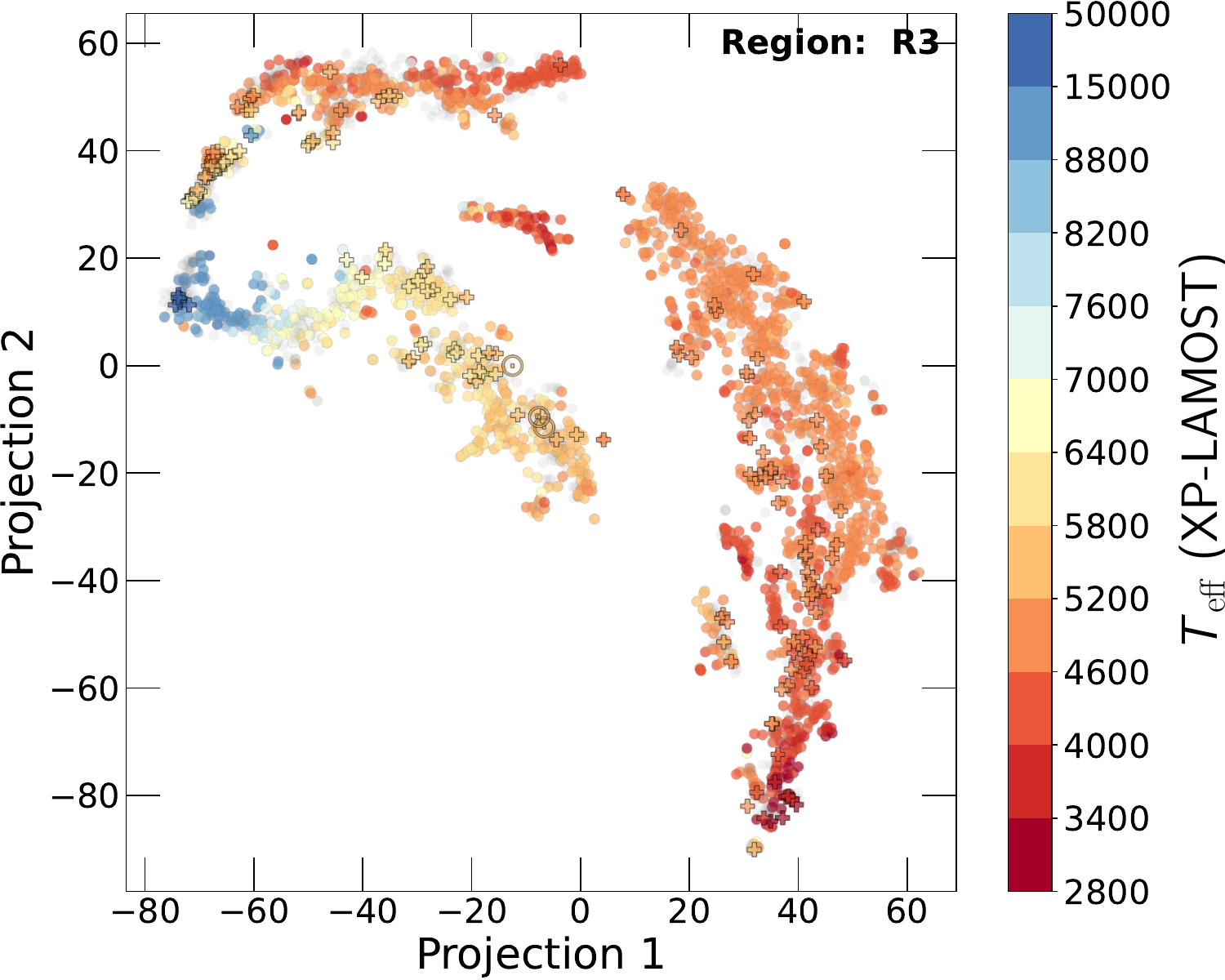} &
        \includegraphics[width=0.3\textwidth]{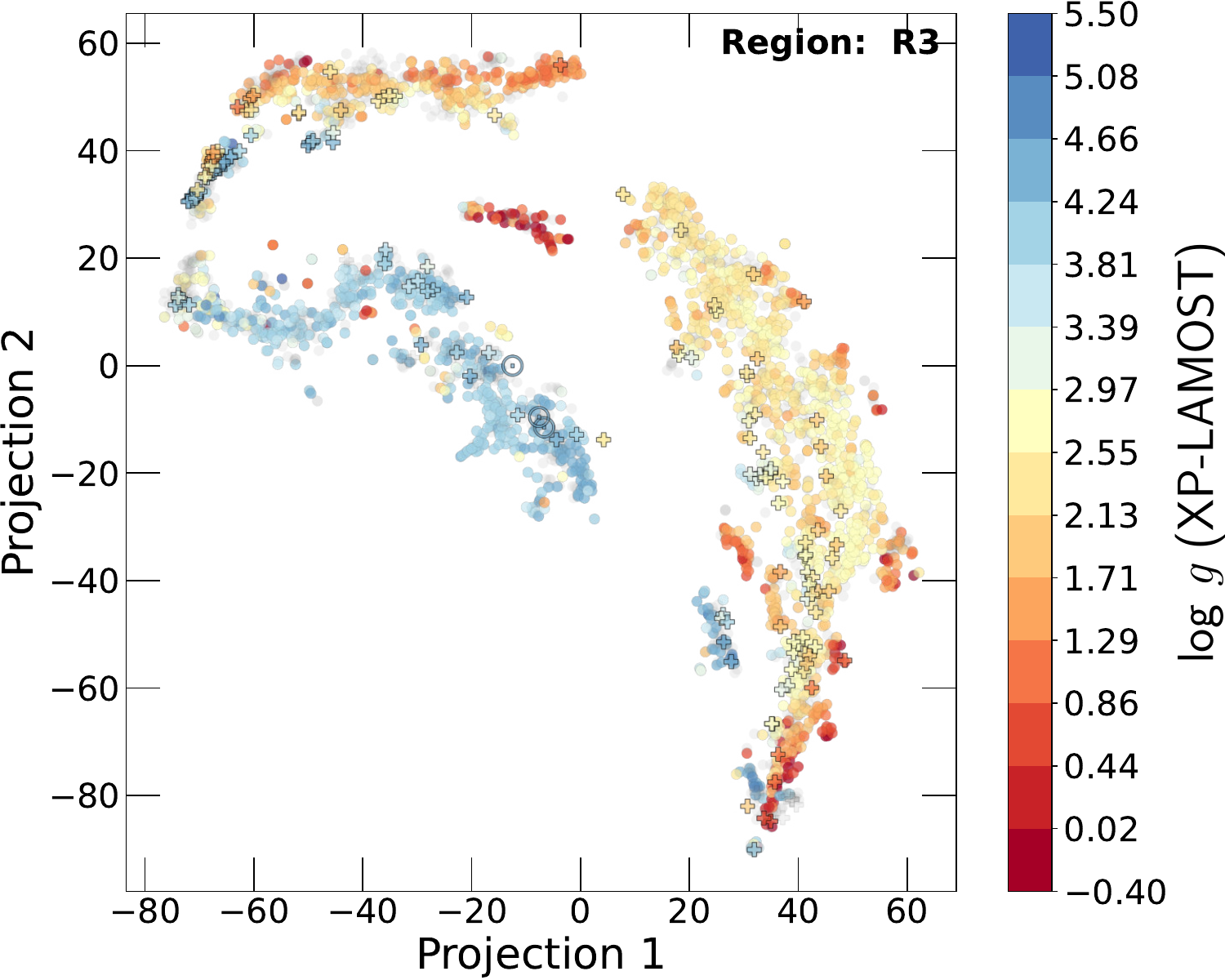} &
        \includegraphics[width=0.3\textwidth]{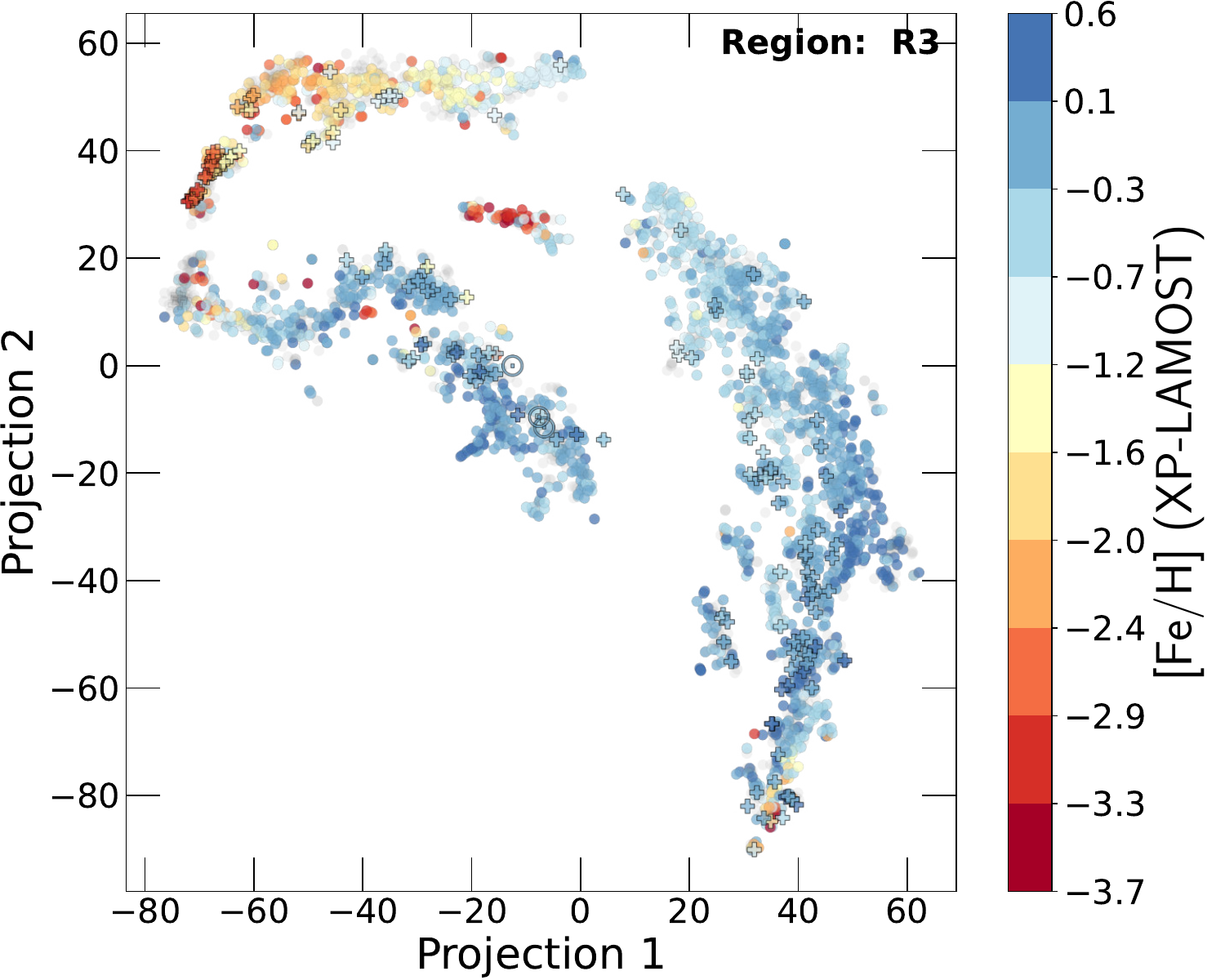} \\
    \end{tabular}
    \caption{Projection maps for the region R3 coloured with the atmospheric parameters of four different catalogues.}
\end{figure*}

\begin{figure*}[htbp]
    \centering
    \begin{tabular}{ccc}
        \includegraphics[width=0.3\textwidth]{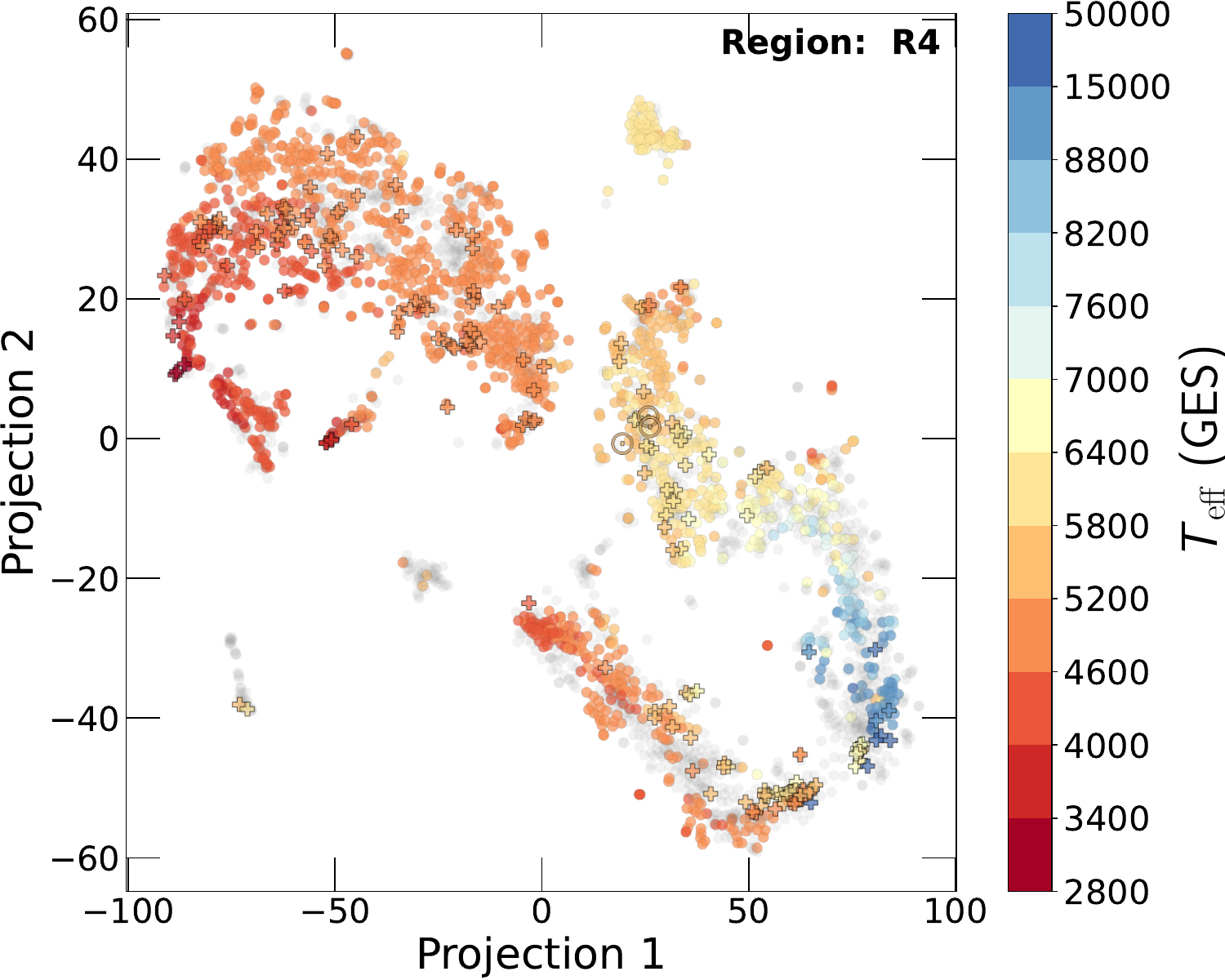} & \includegraphics[width=0.3\textwidth]{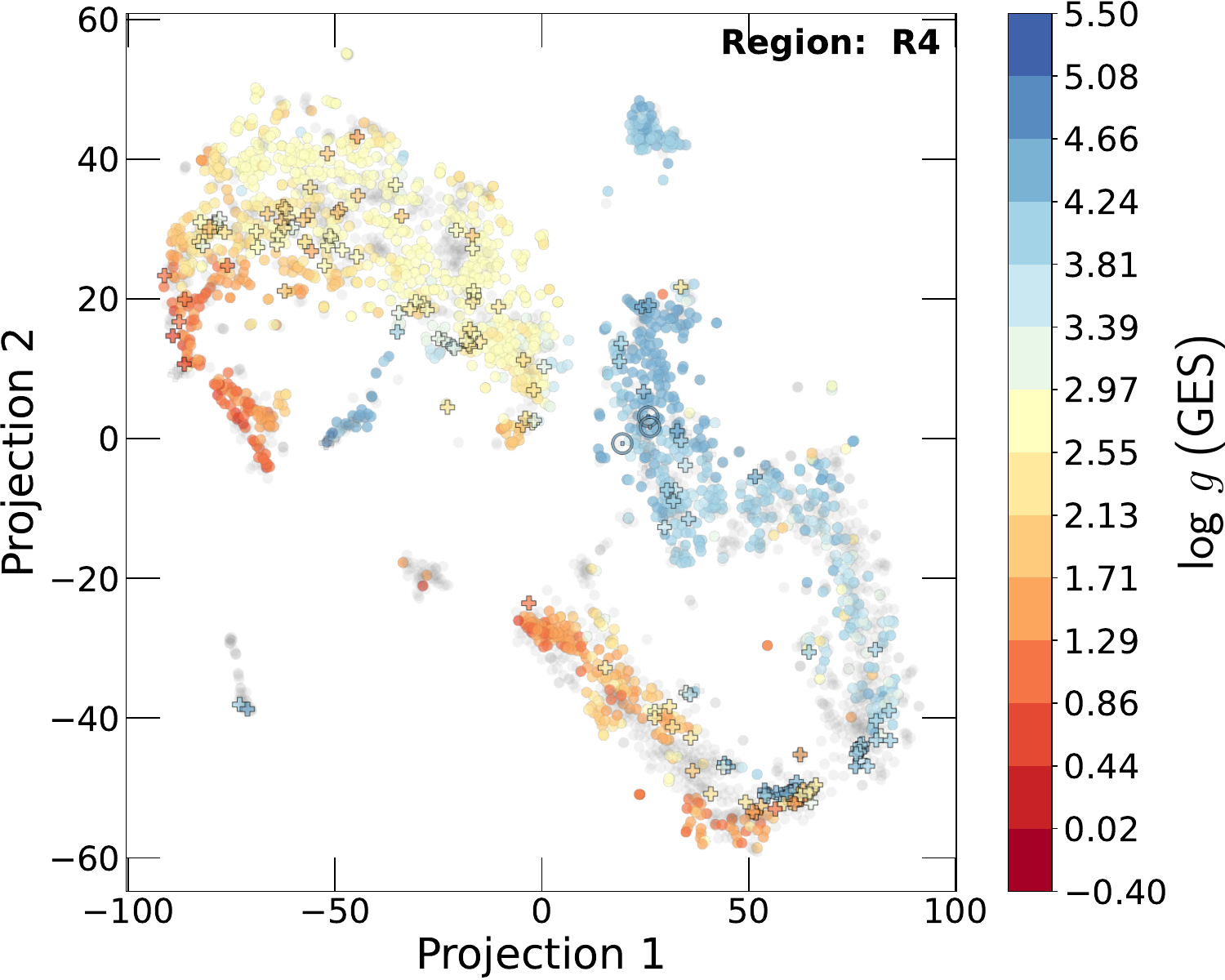} & \includegraphics[width=0.3\textwidth]{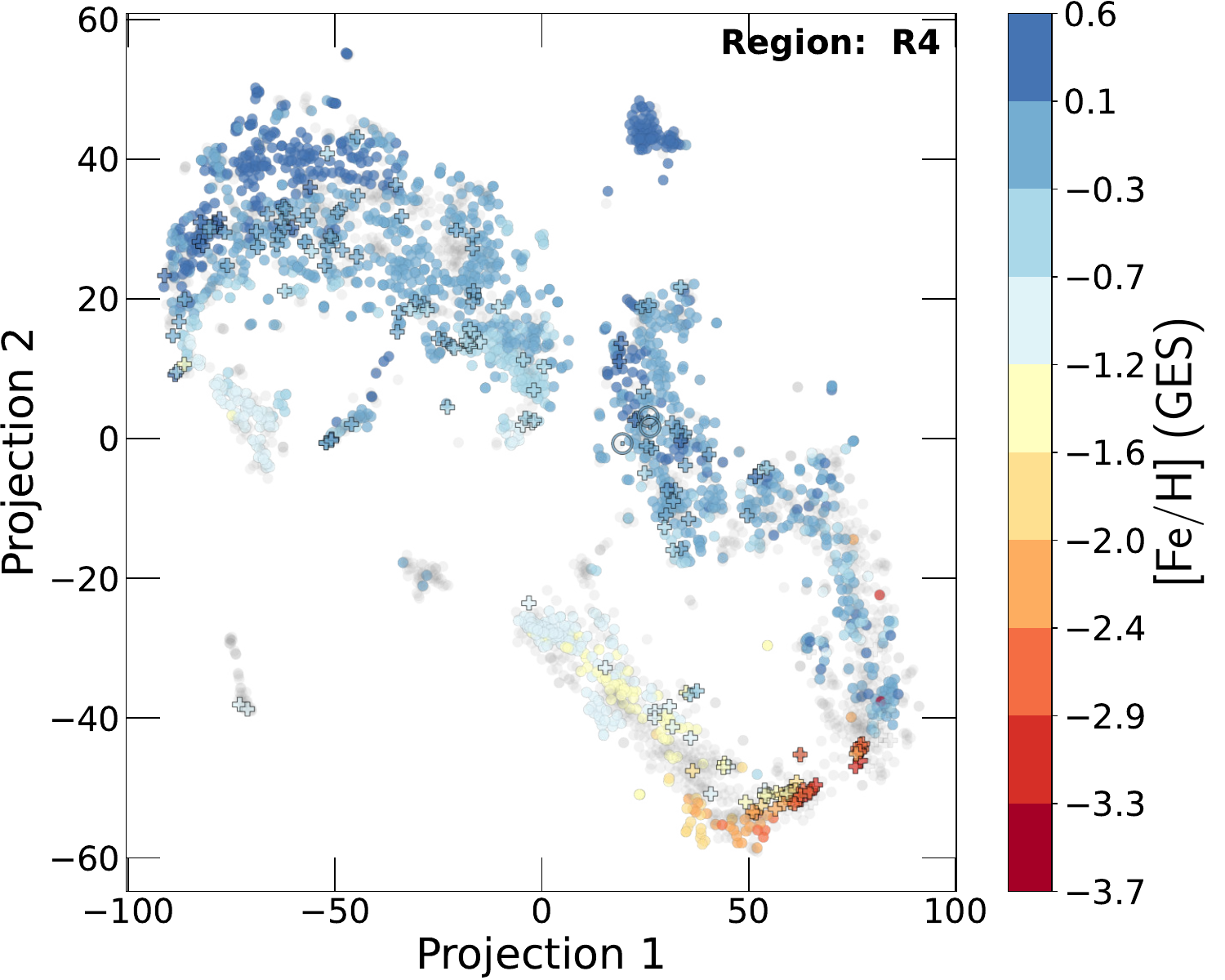} \\
        \includegraphics[width=0.3\textwidth]{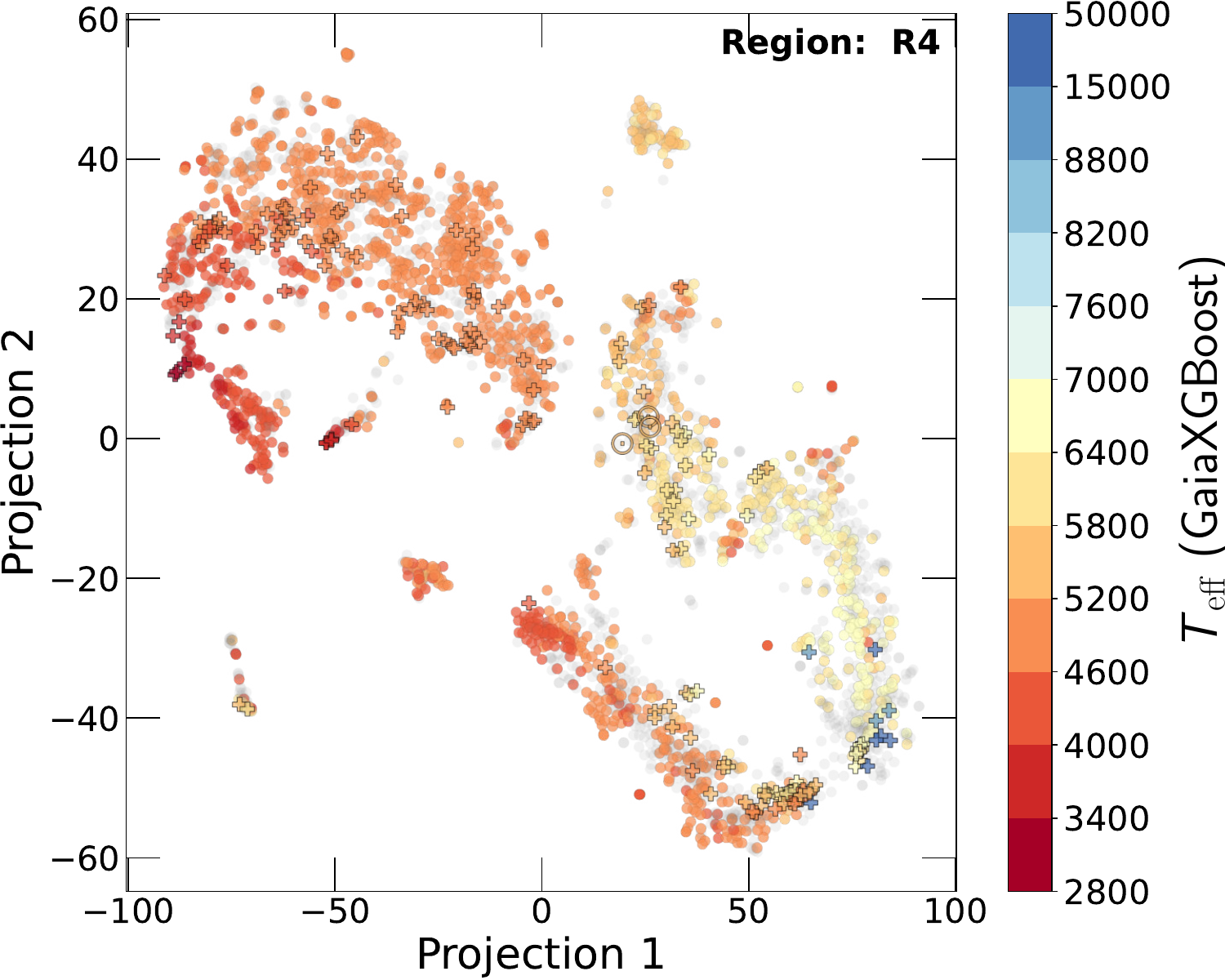} & \includegraphics[width=0.3\textwidth]{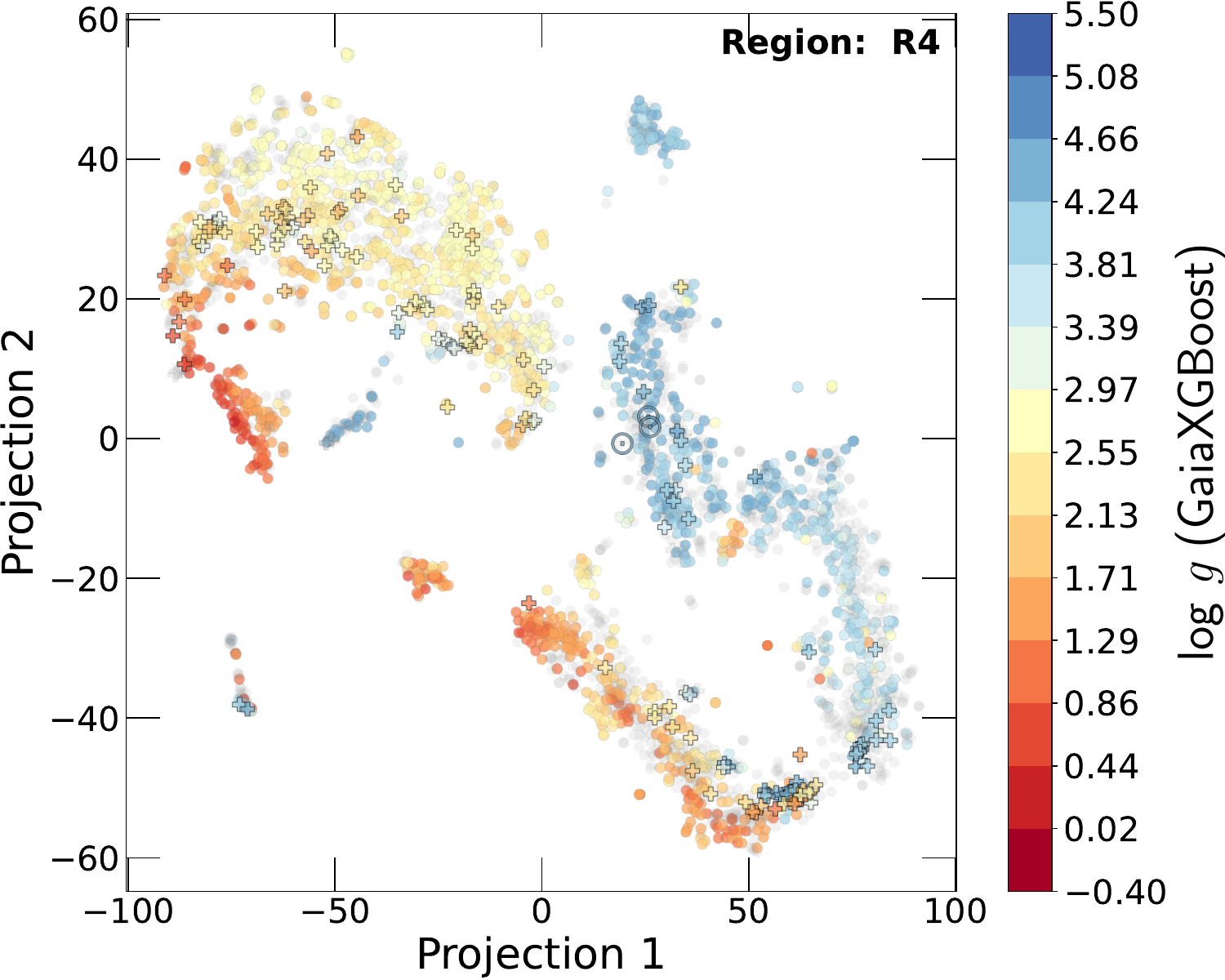} & \includegraphics[width=0.3\textwidth]{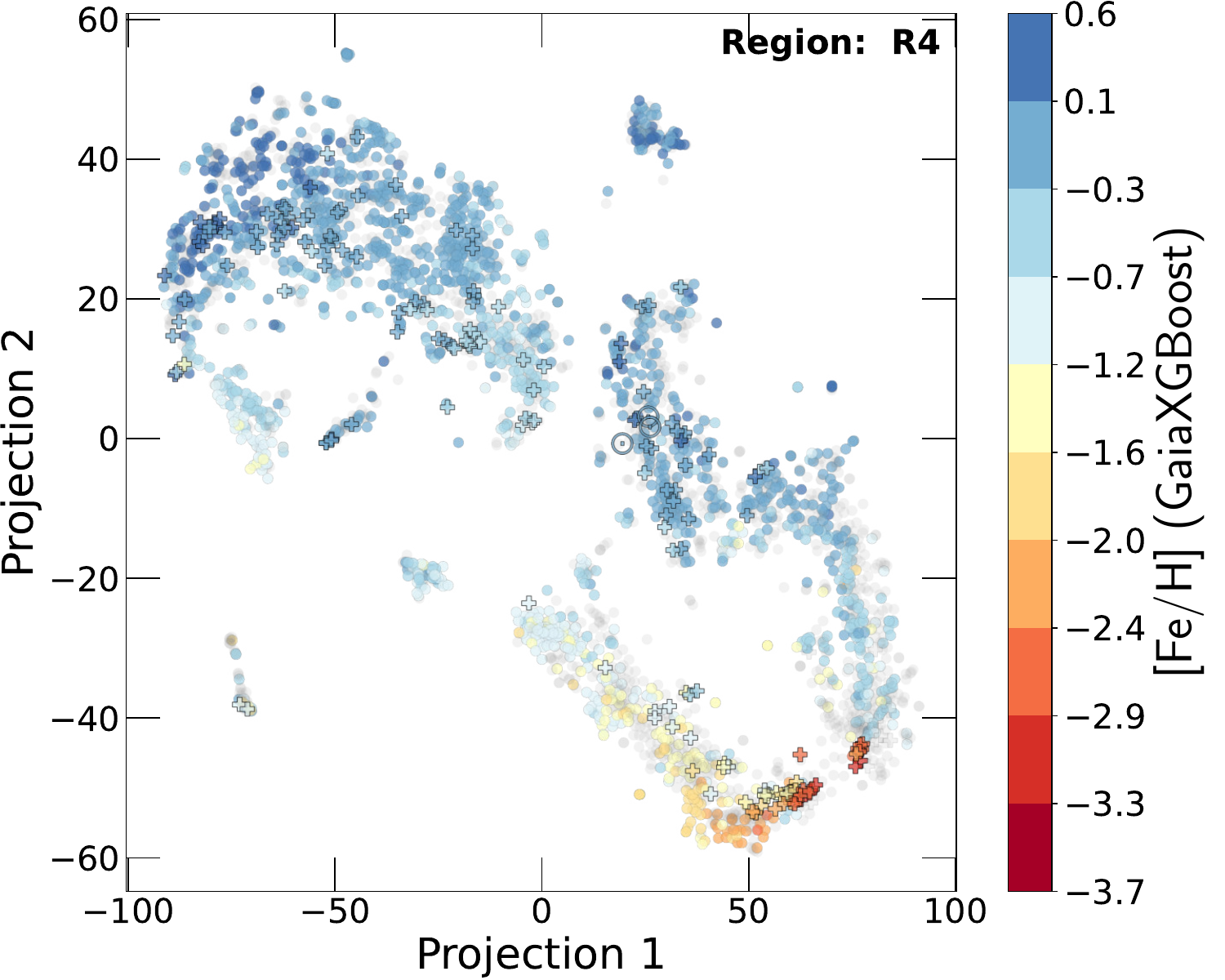} \\
        \includegraphics[width=0.3\textwidth]{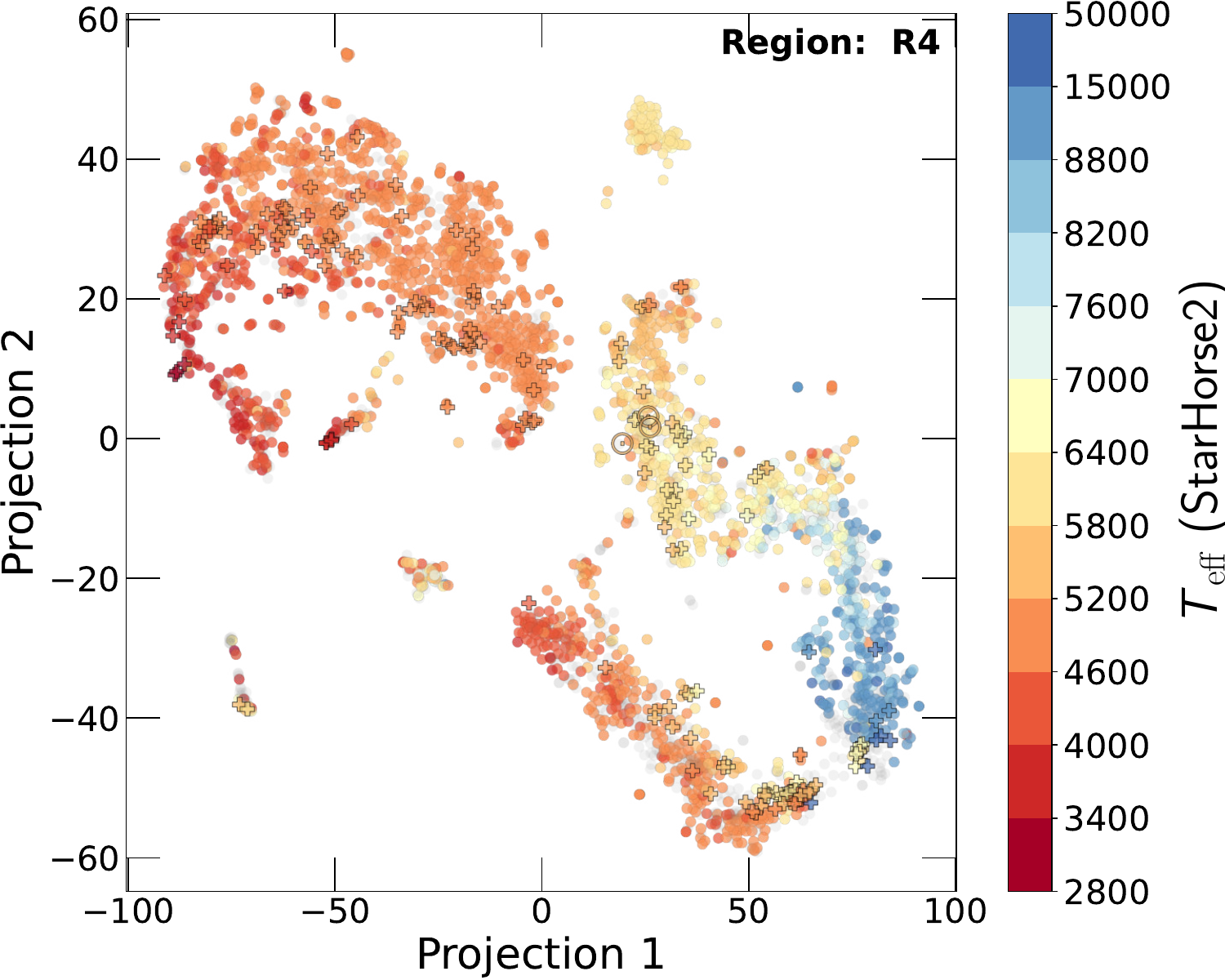} &
        \includegraphics[width=0.3\textwidth]{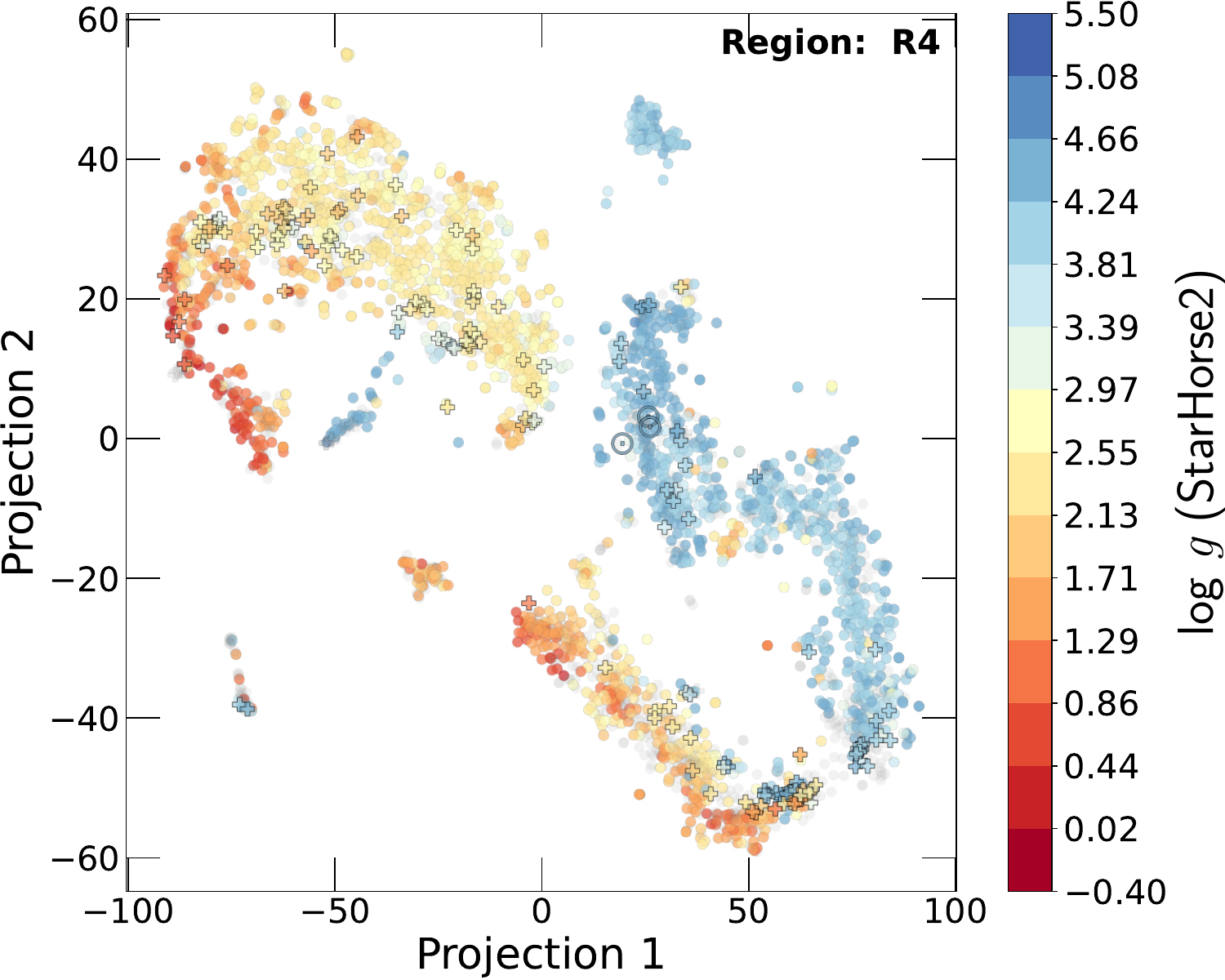} &
        \includegraphics[width=0.3\textwidth]{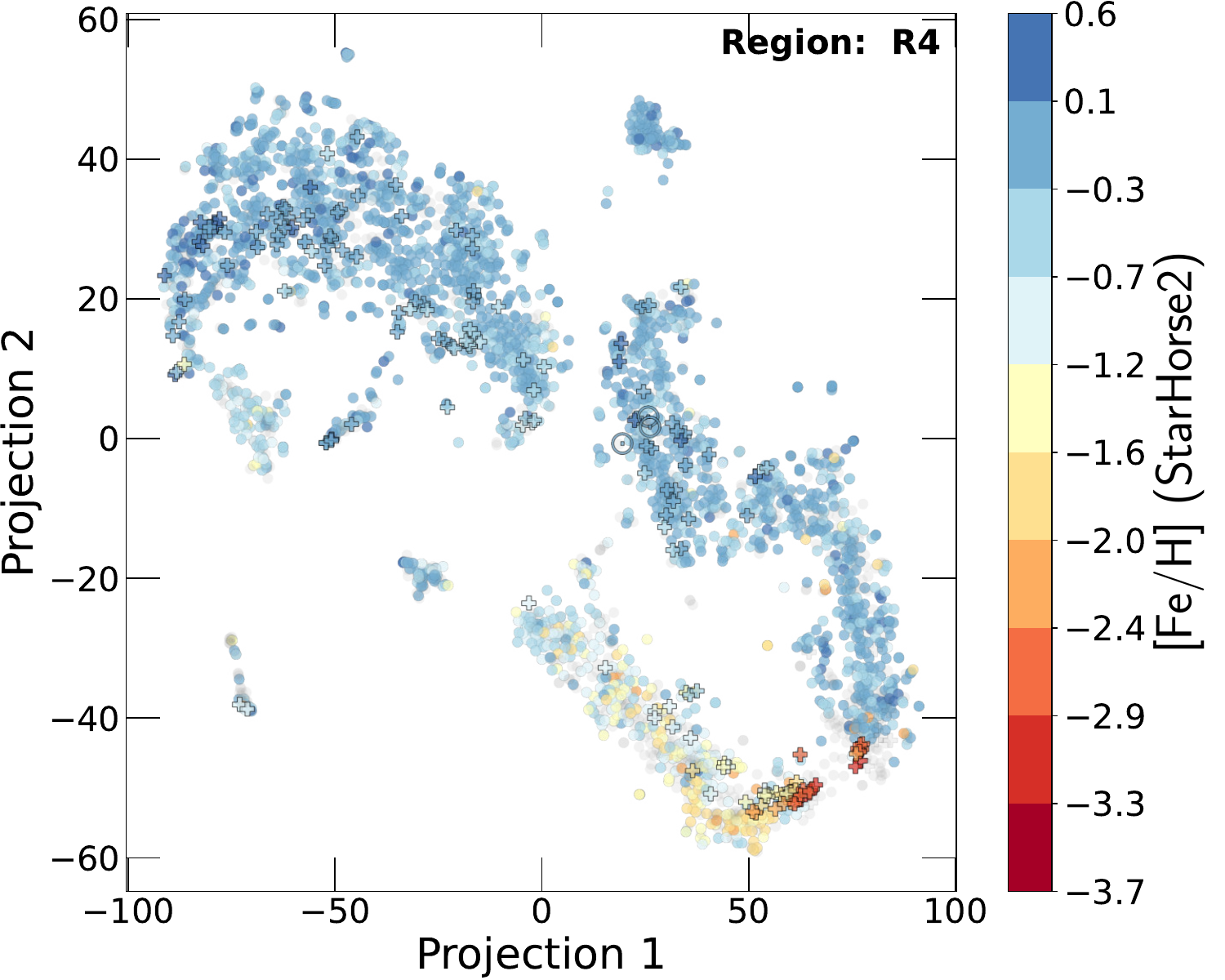} \\
        \includegraphics[width=0.3\textwidth]{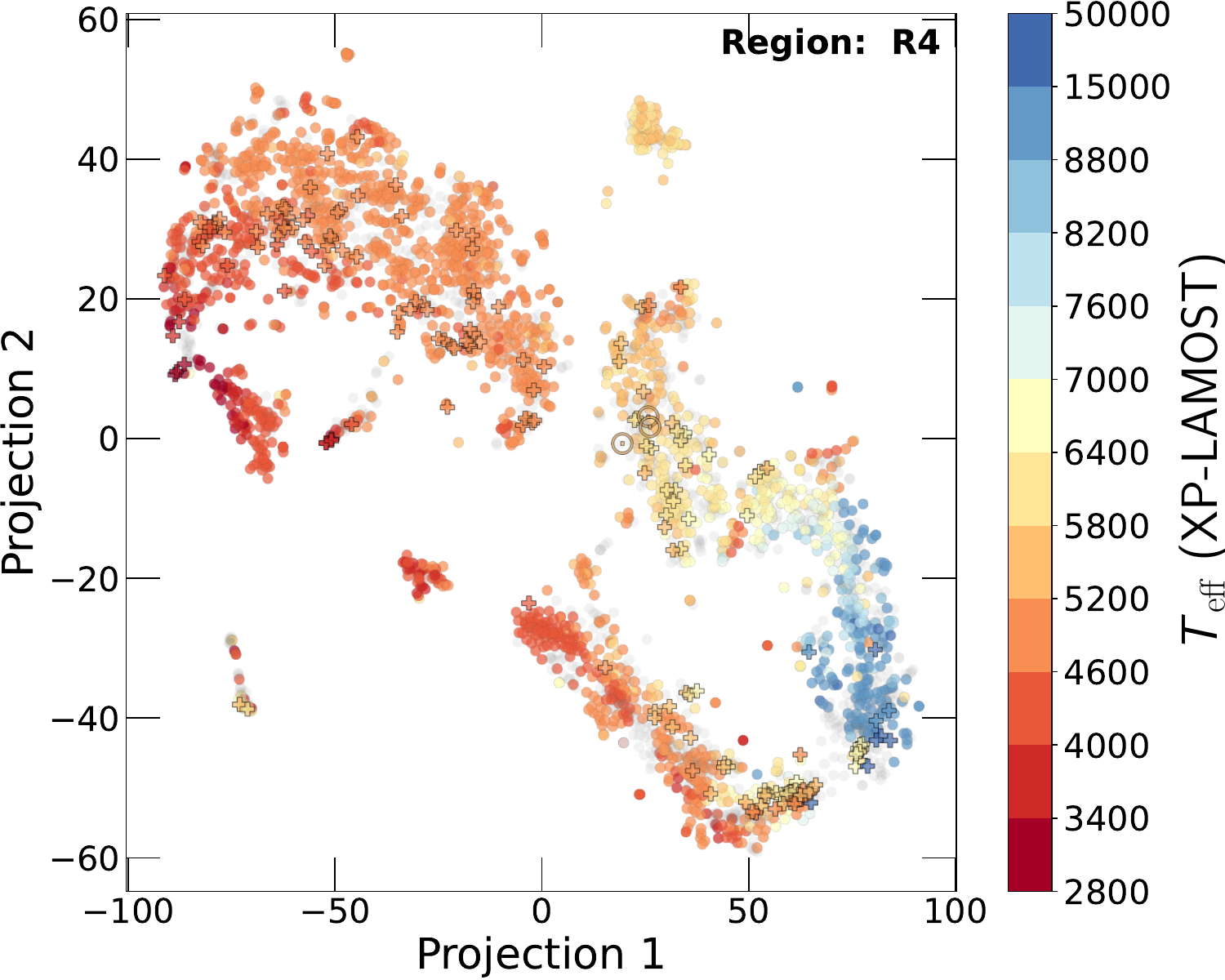} &
        \includegraphics[width=0.3\textwidth]{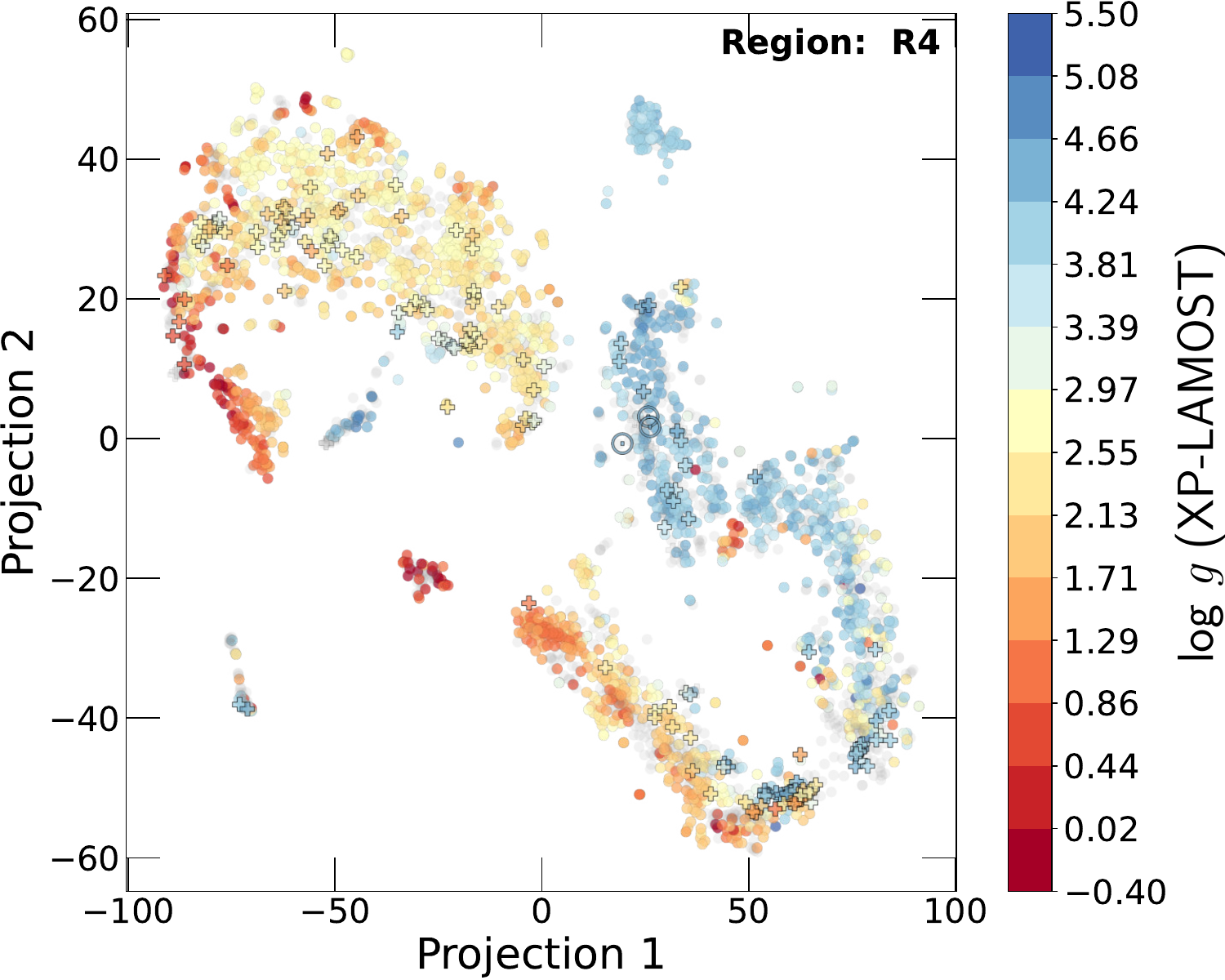} &
        \includegraphics[width=0.3\textwidth]{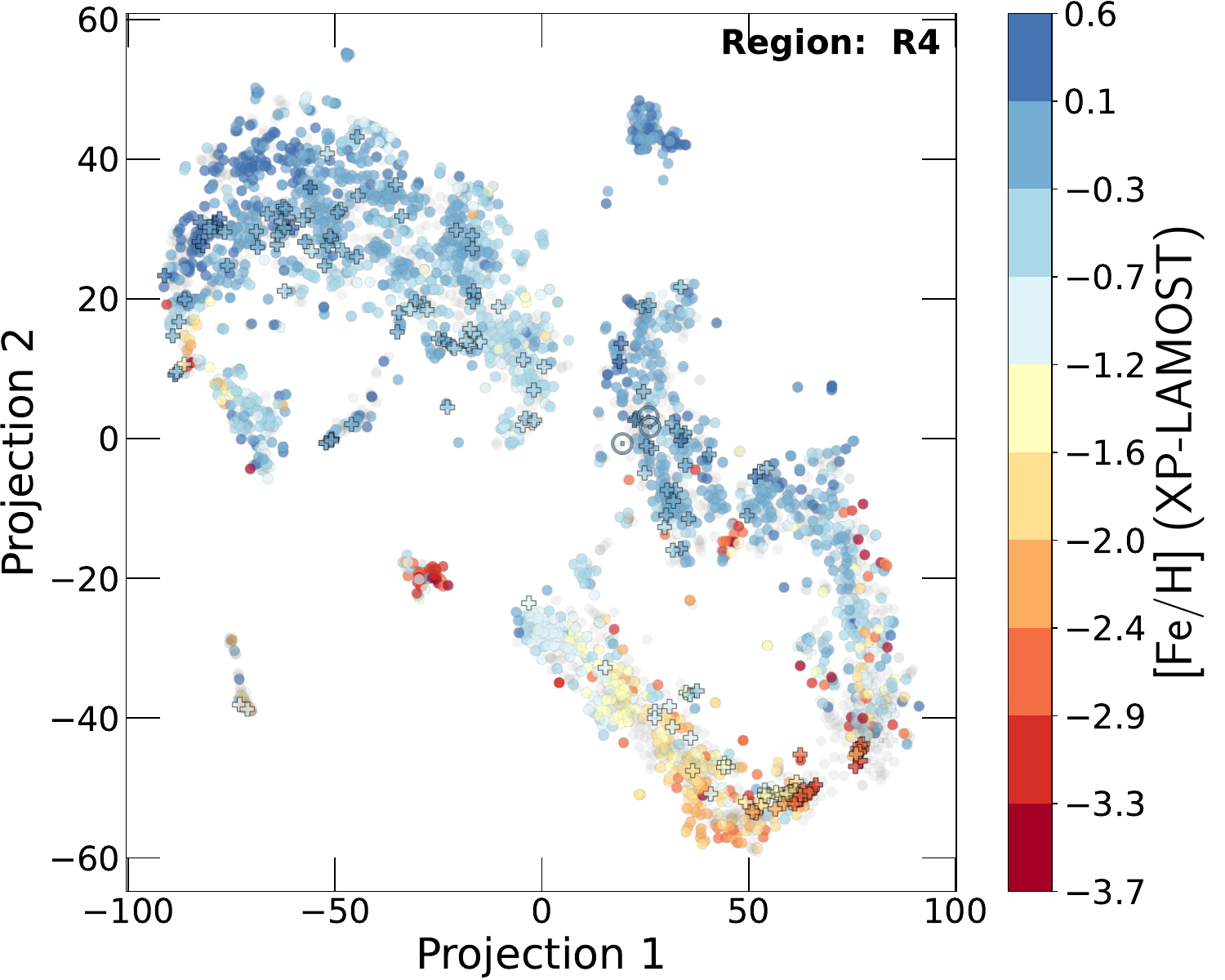} \\
    \end{tabular}
    \caption{Projection maps for the region R4 coloured with the atmospheric parameters of four different catalogues.}
\end{figure*}

\begin{figure*}[htbp]
    \centering
    \begin{tabular}{ccc}
        \includegraphics[width=0.3\textwidth]{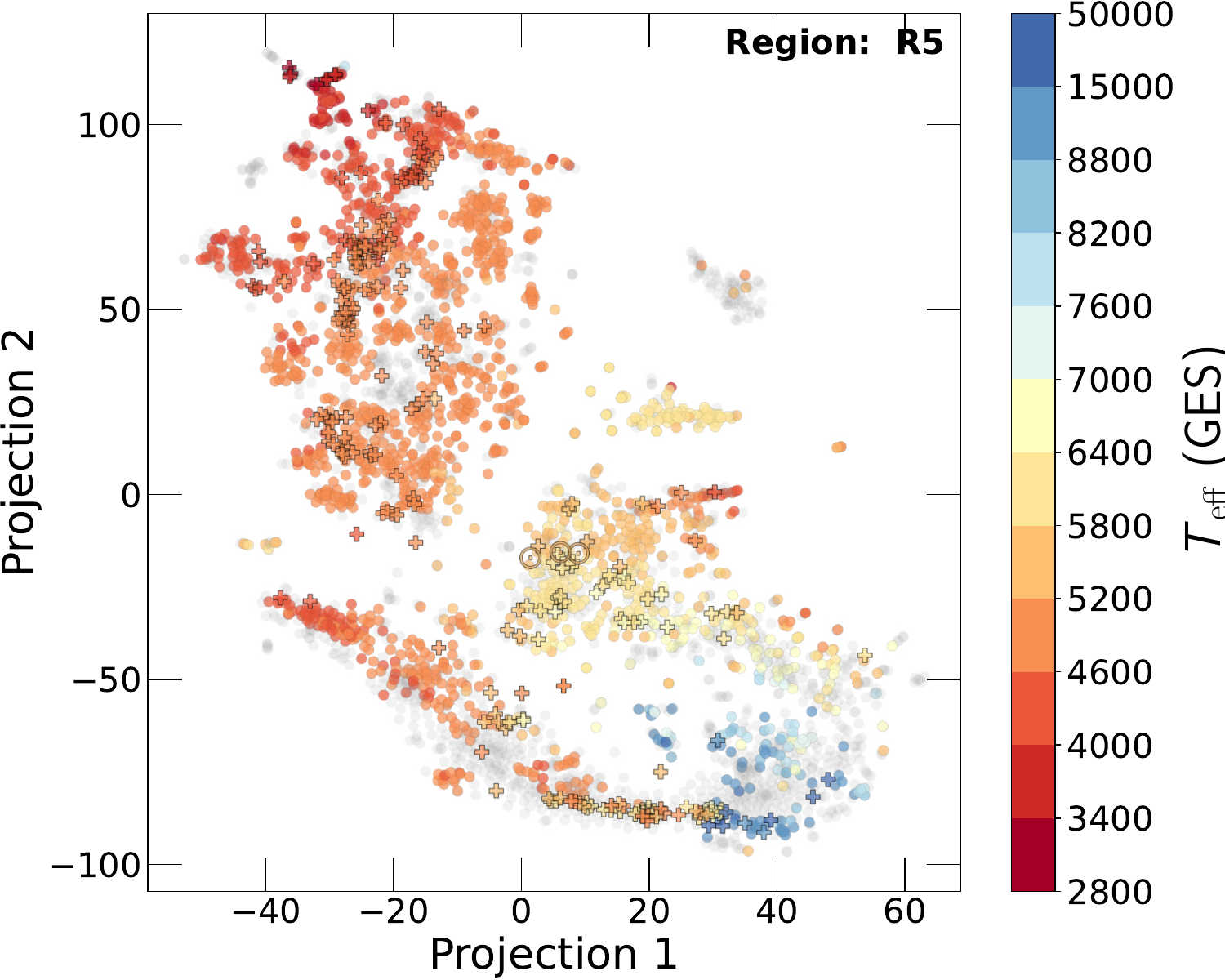} & \includegraphics[width=0.3\textwidth]{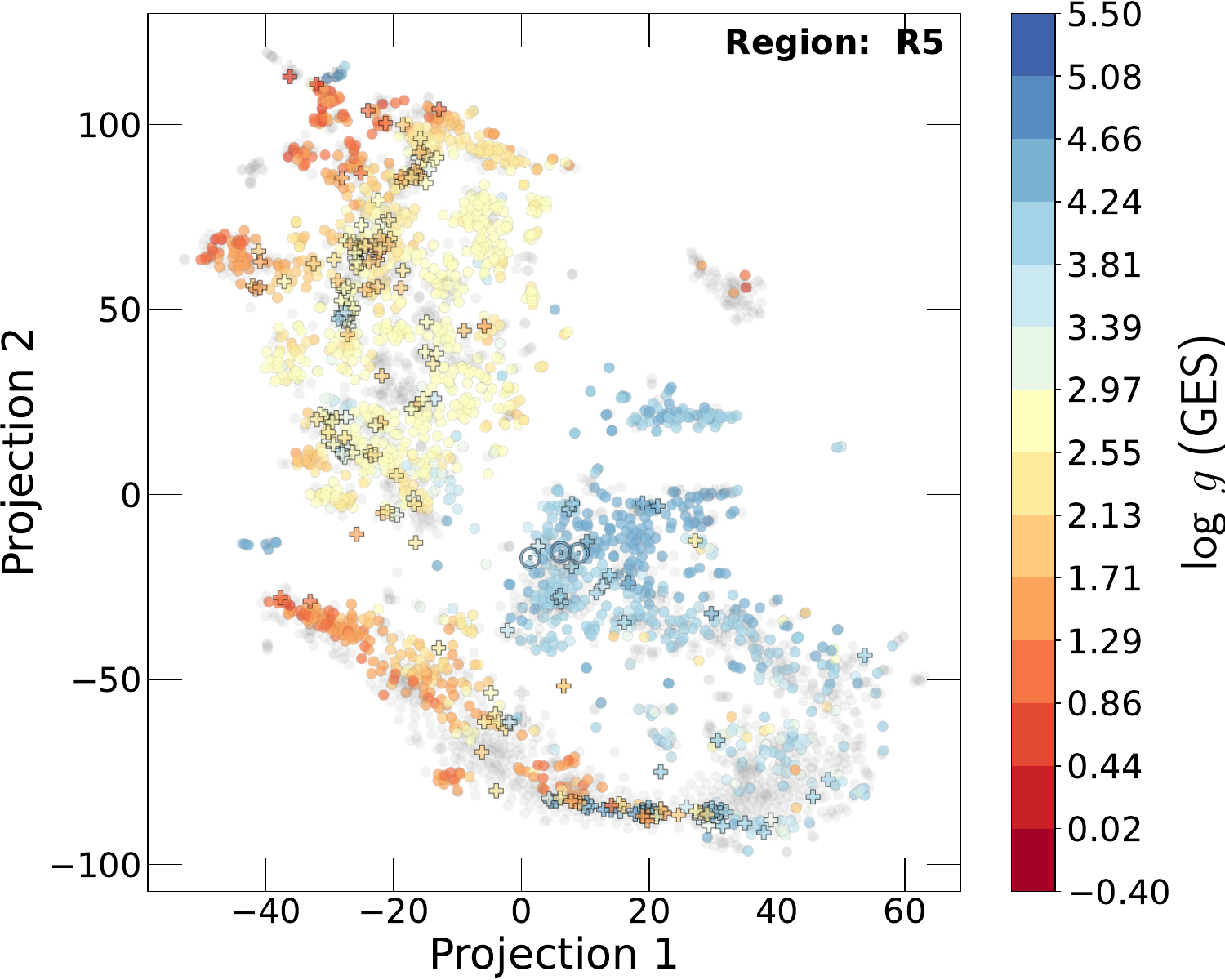} & \includegraphics[width=0.3\textwidth]{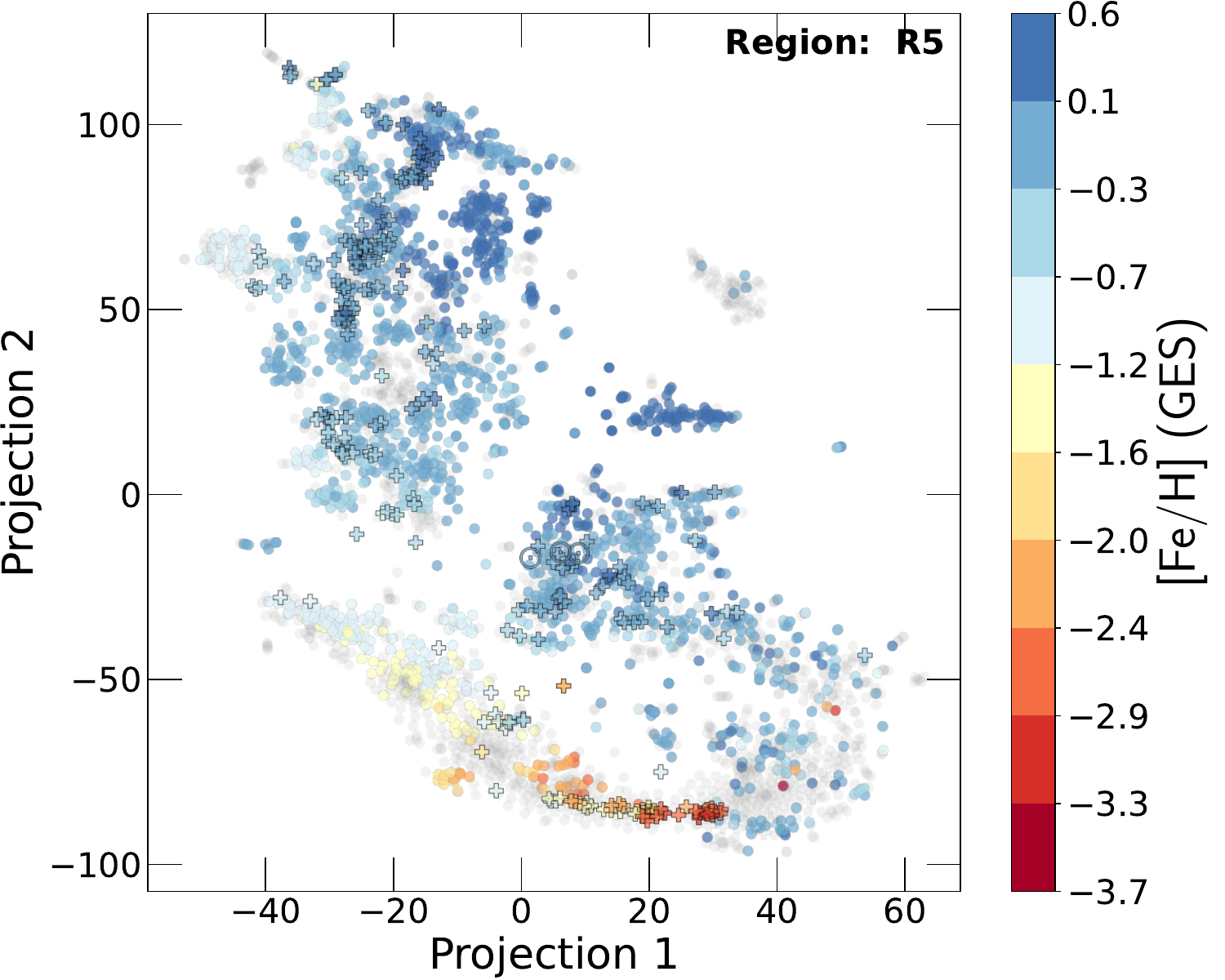} \\
        \includegraphics[width=0.3\textwidth]{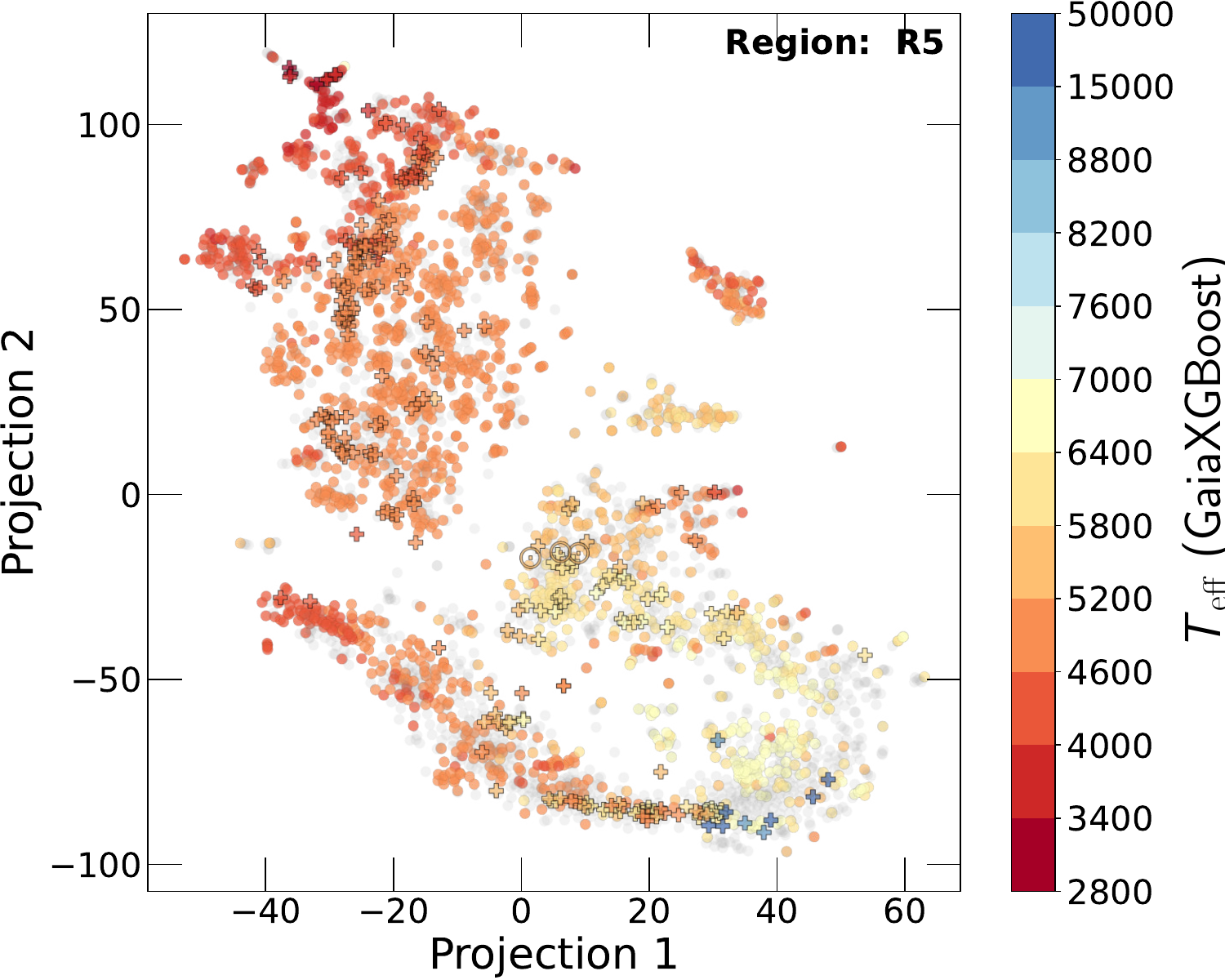} & \includegraphics[width=0.3\textwidth]{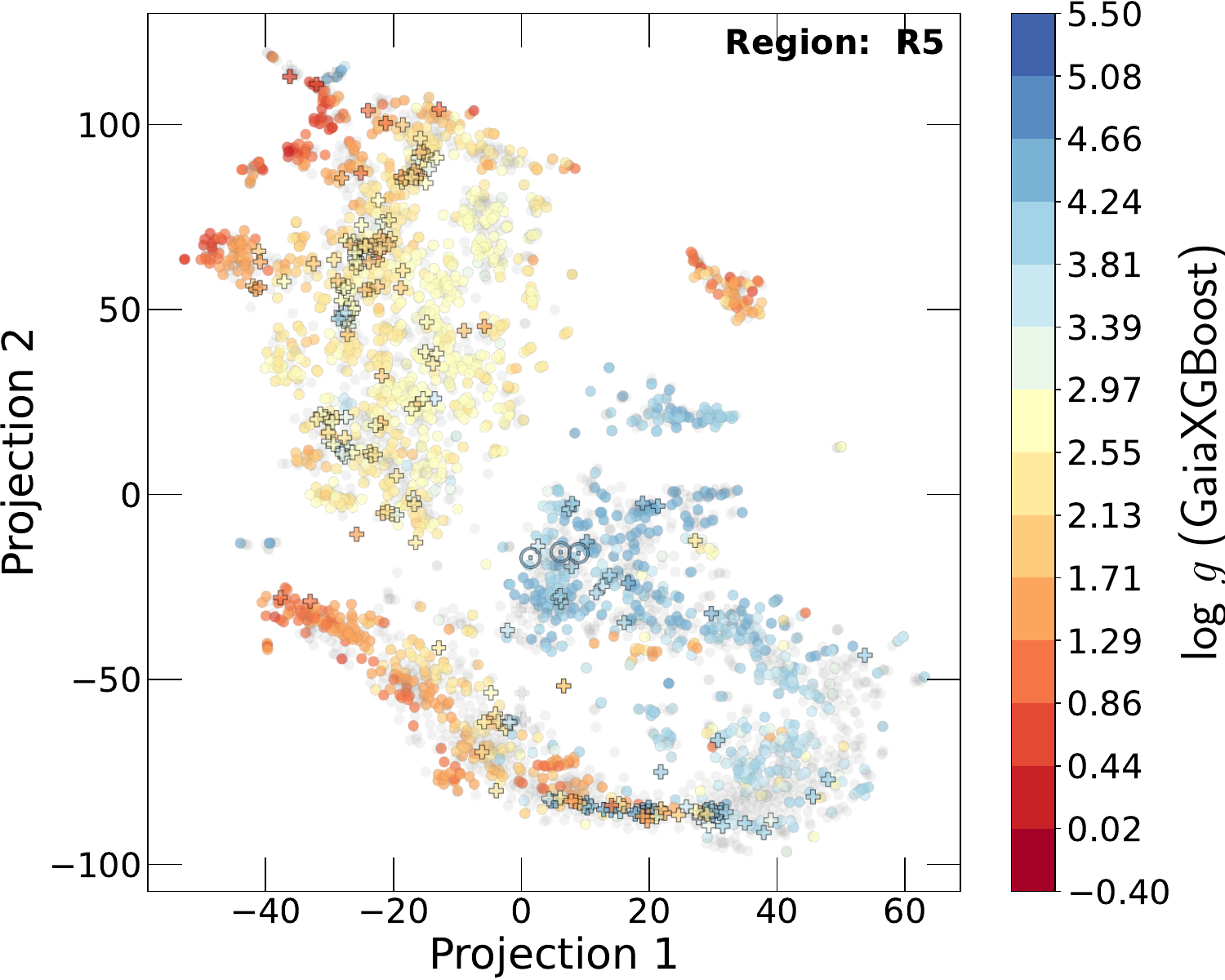} & \includegraphics[width=0.3\textwidth]{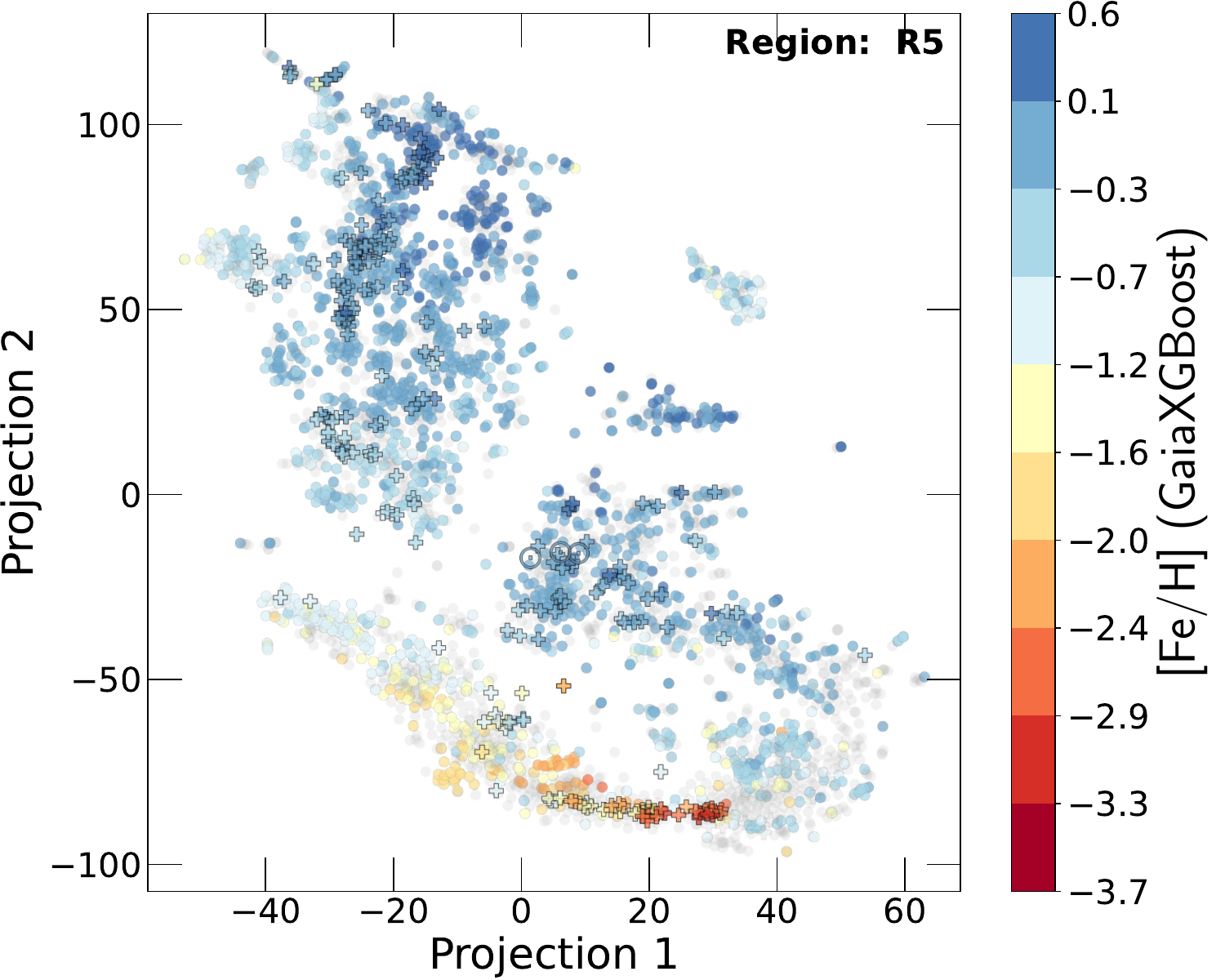} \\
        \includegraphics[width=0.3\textwidth]{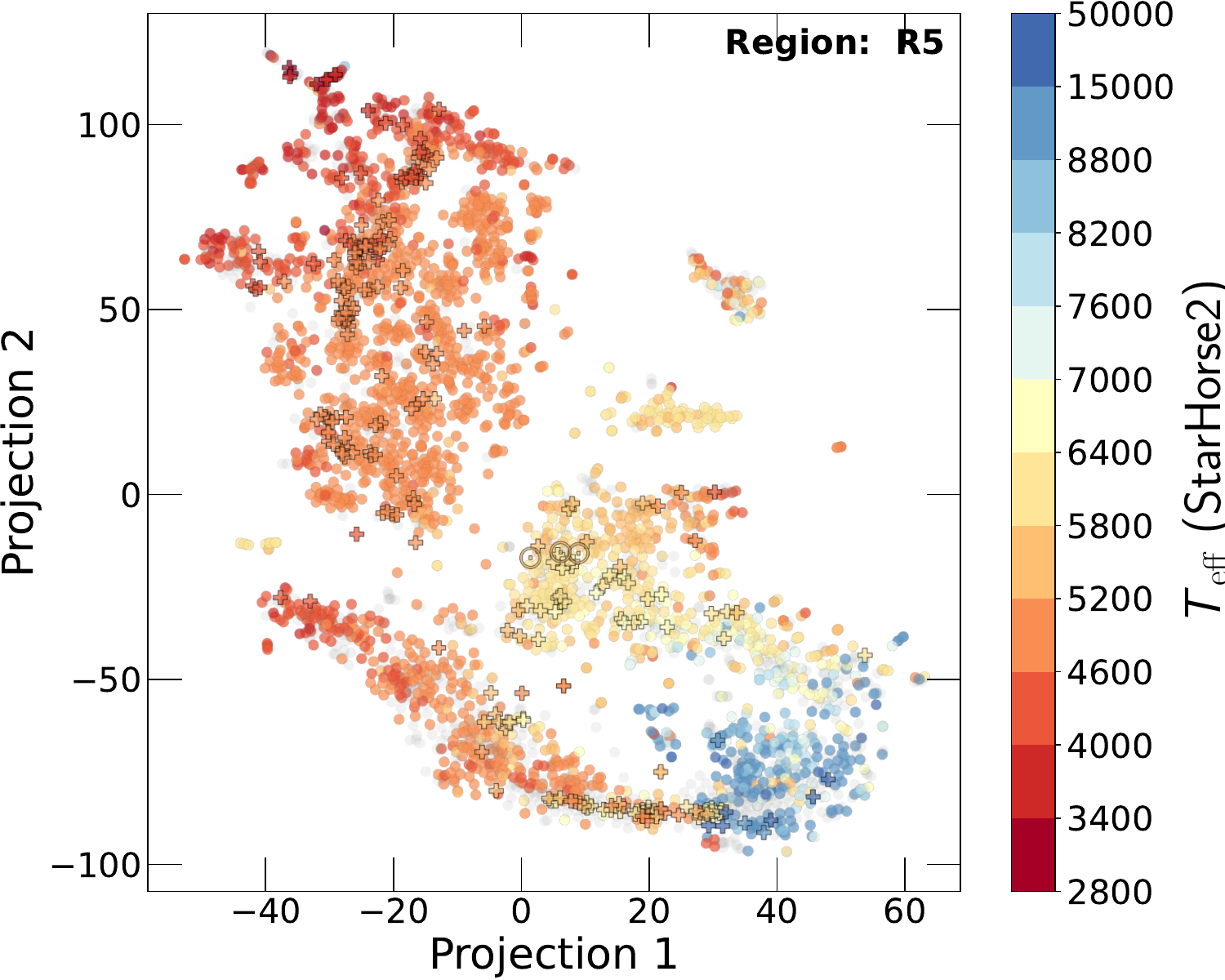} &
        \includegraphics[width=0.3\textwidth]{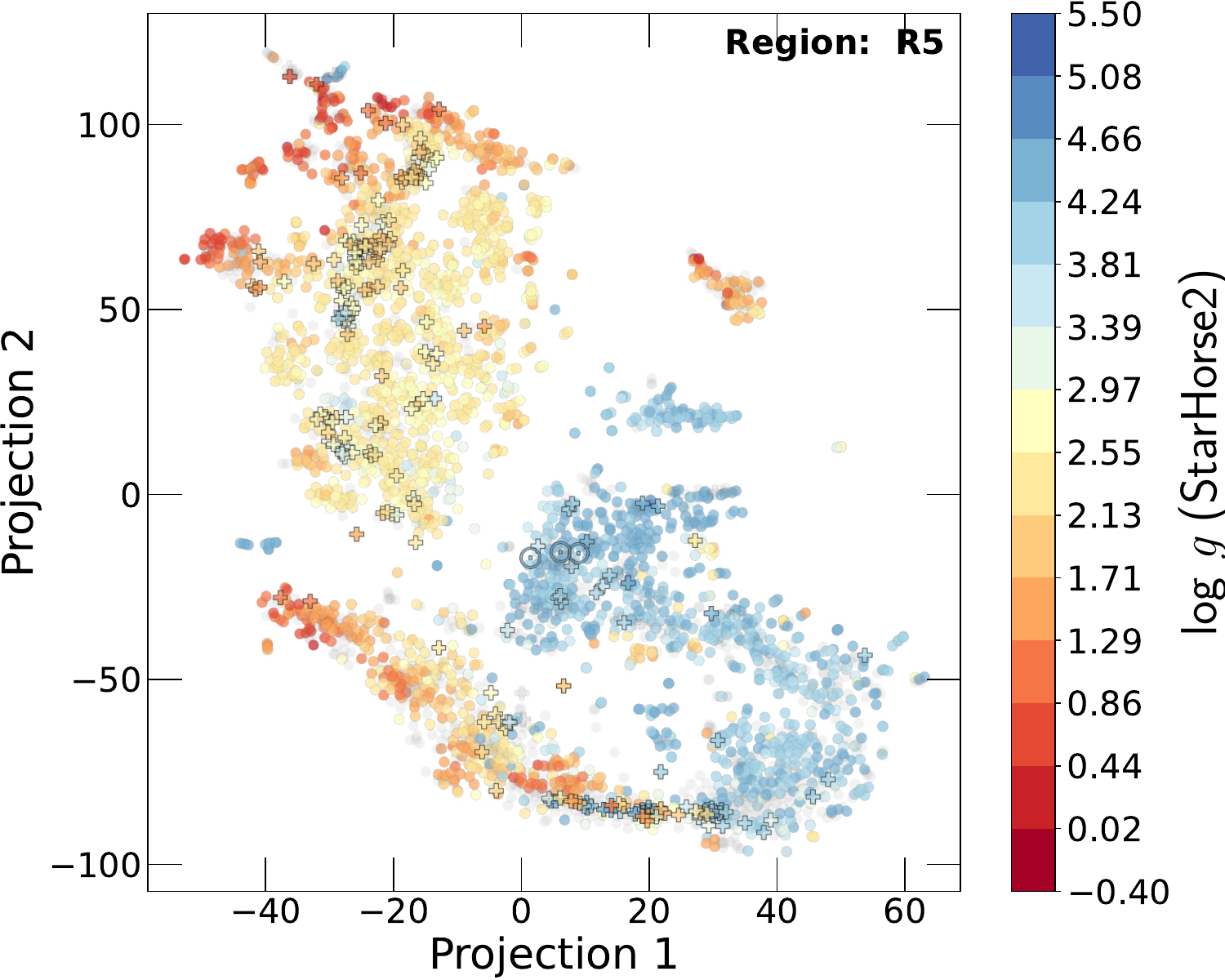} &
        \includegraphics[width=0.3\textwidth]{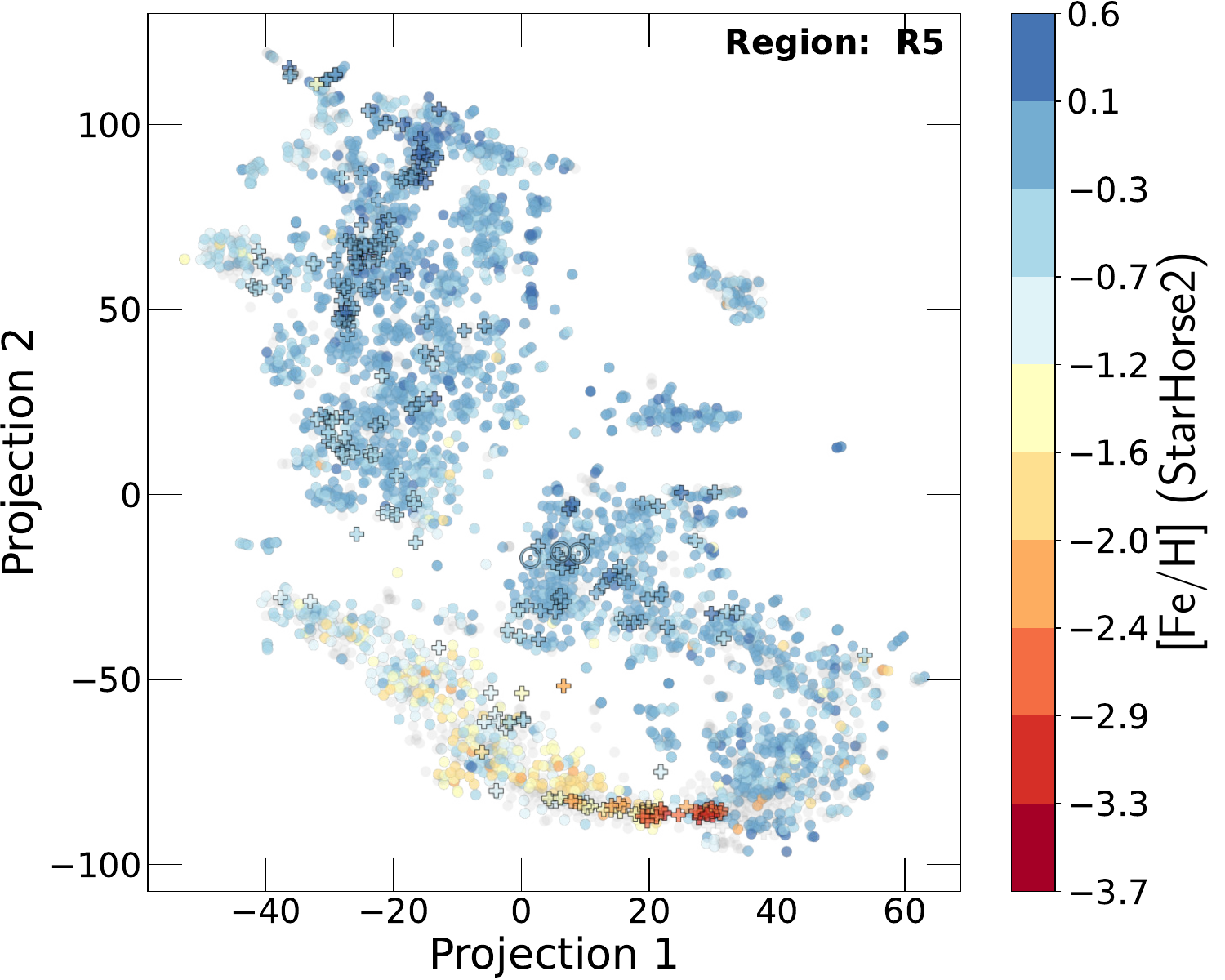} \\
        \includegraphics[width=0.3\textwidth]{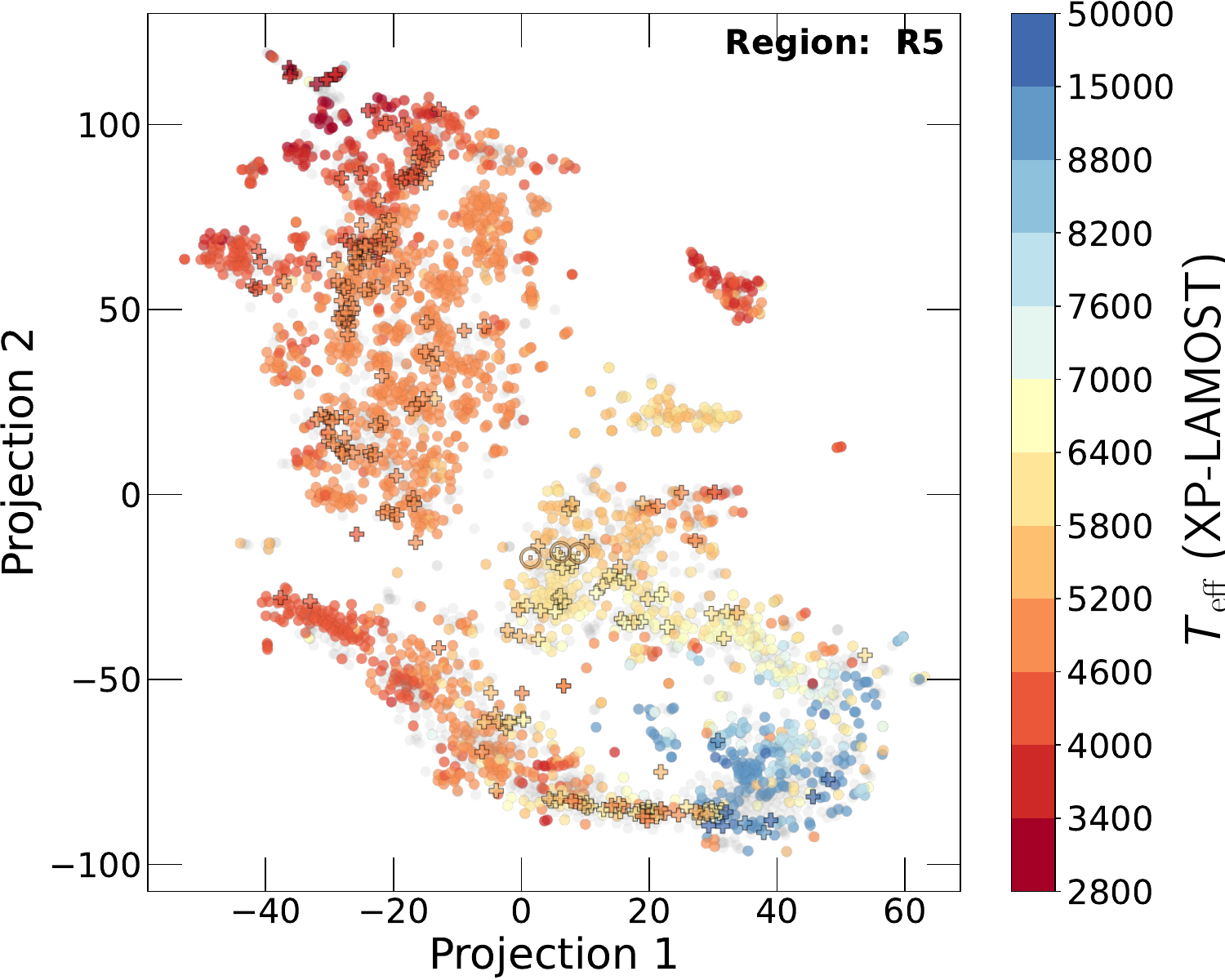} &
        \includegraphics[width=0.3\textwidth]{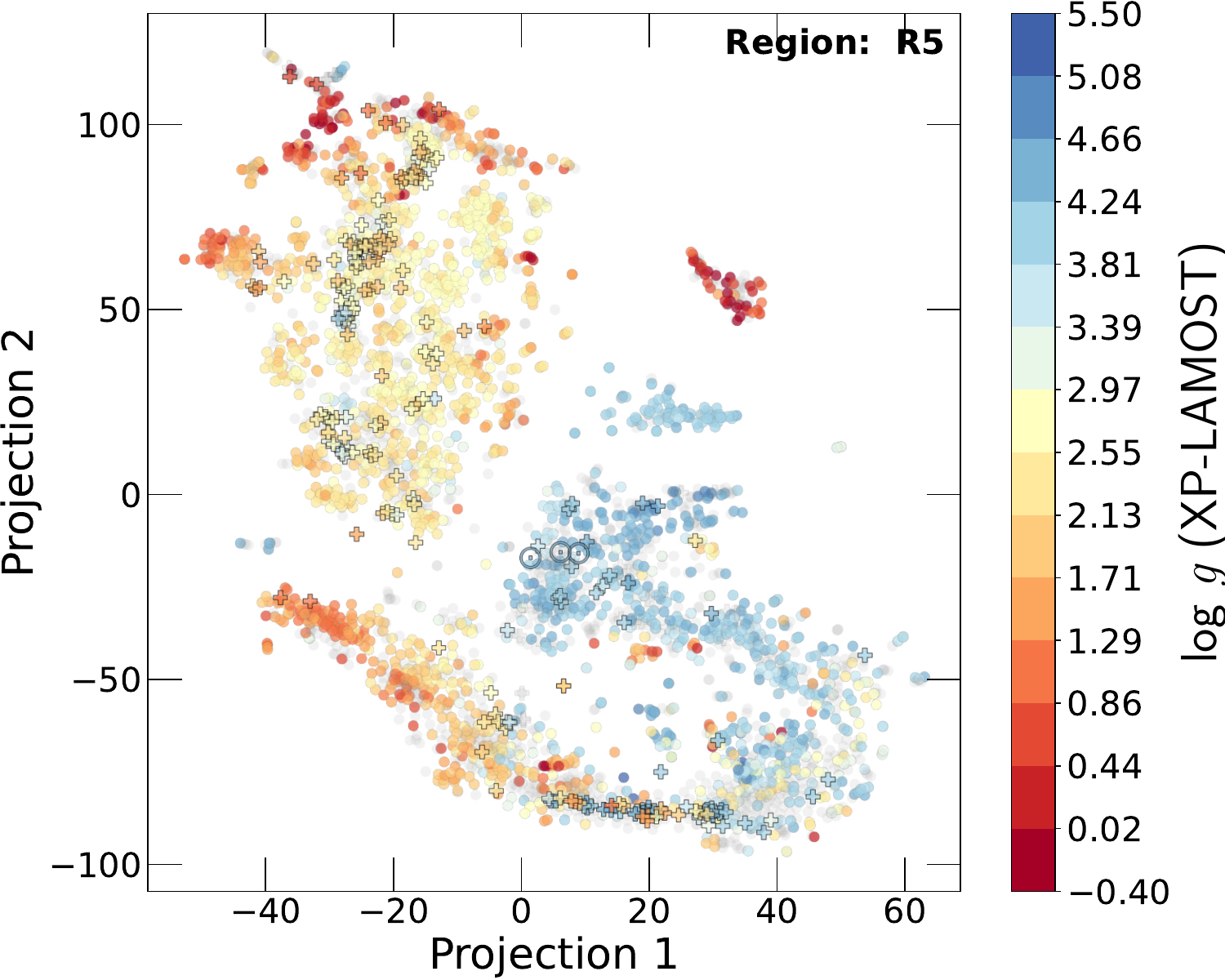} &
        \includegraphics[width=0.3\textwidth]{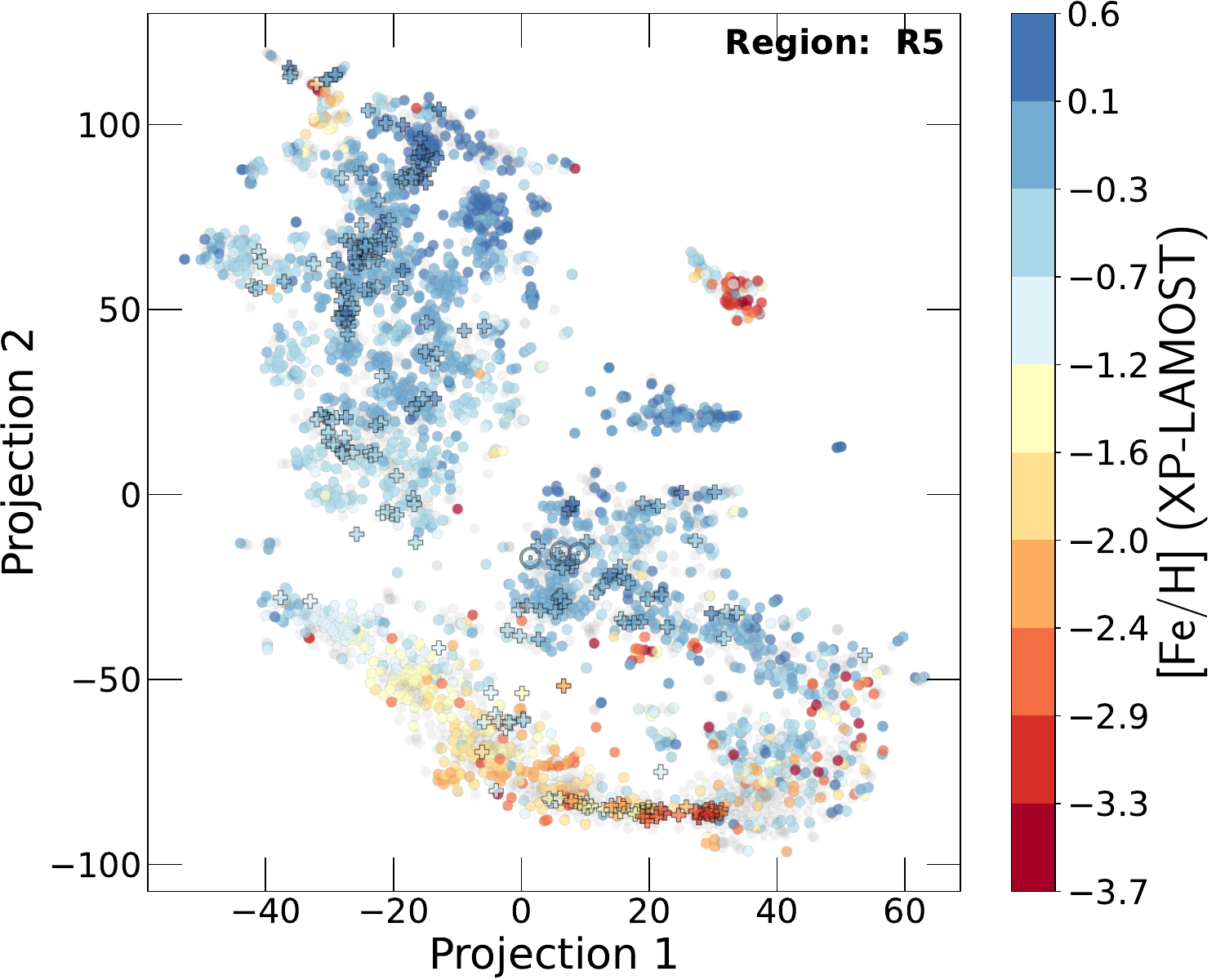} \\
    \end{tabular}
    \caption{Projection maps for the region R5 coloured with the atmospheric parameters of four different catalogues.}
\end{figure*}

\begin{figure*}[htbp]
    \centering
    \begin{tabular}{ccc}
        \includegraphics[width=0.3\textwidth]{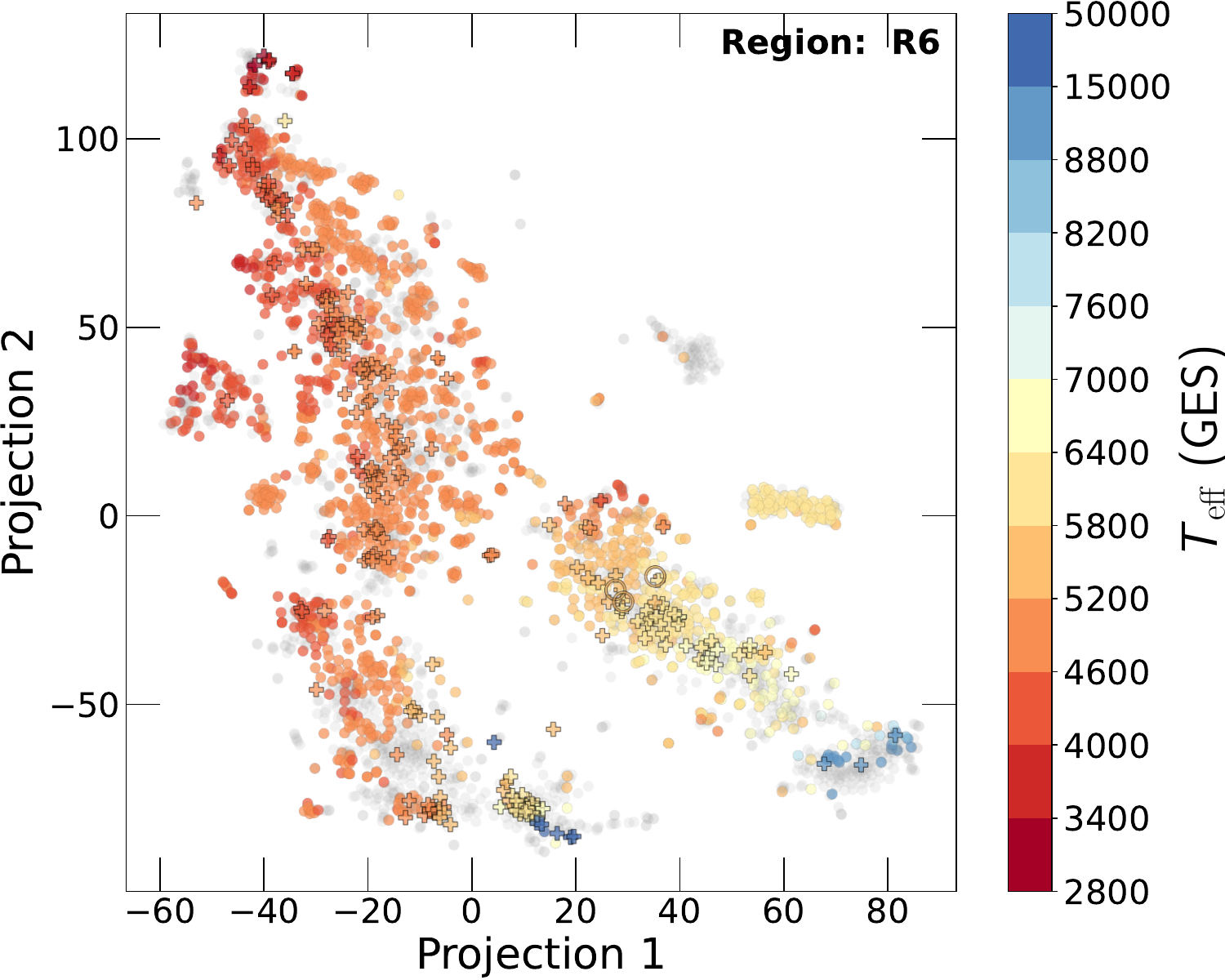} & \includegraphics[width=0.3\textwidth]{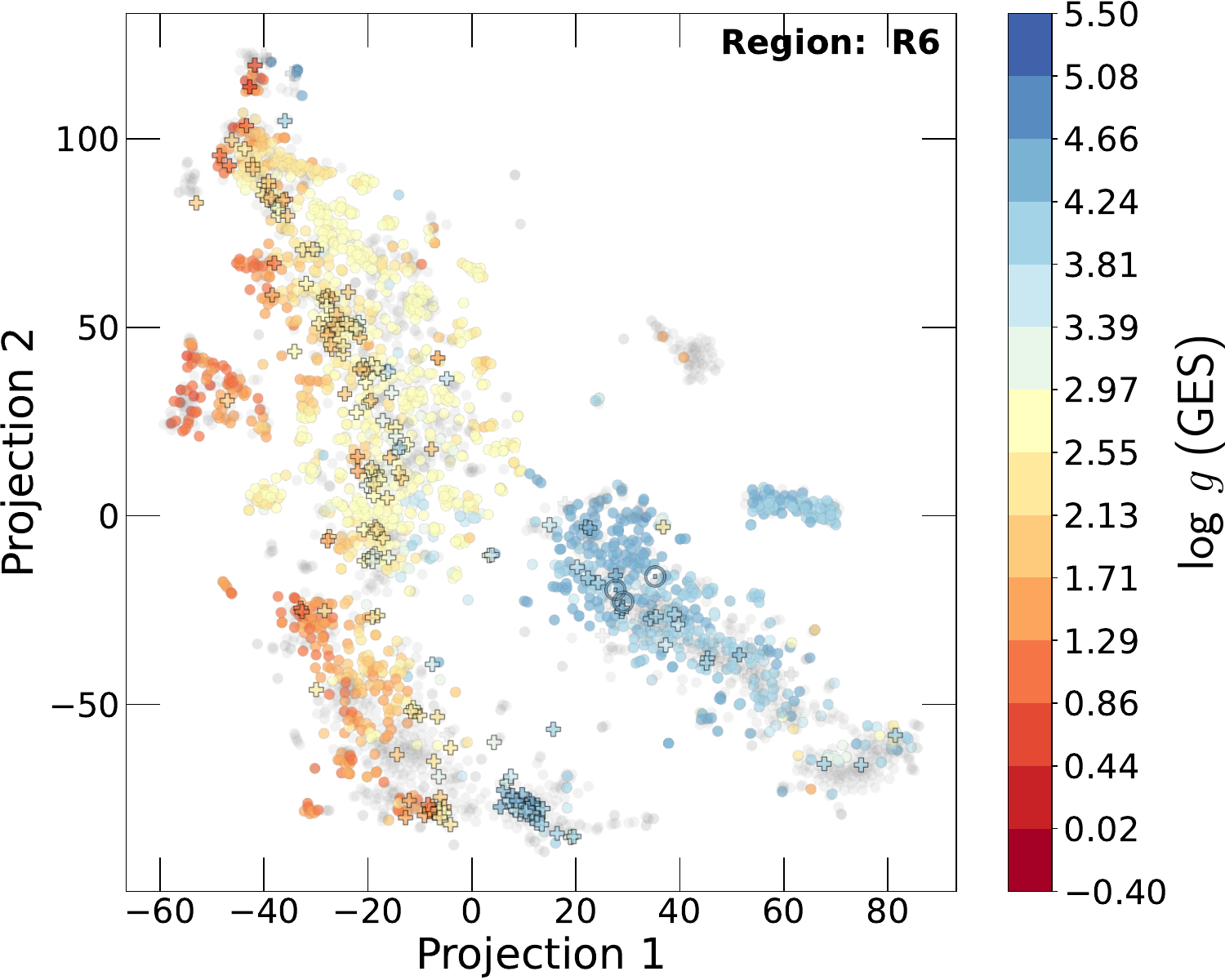} & \includegraphics[width=0.3\textwidth]{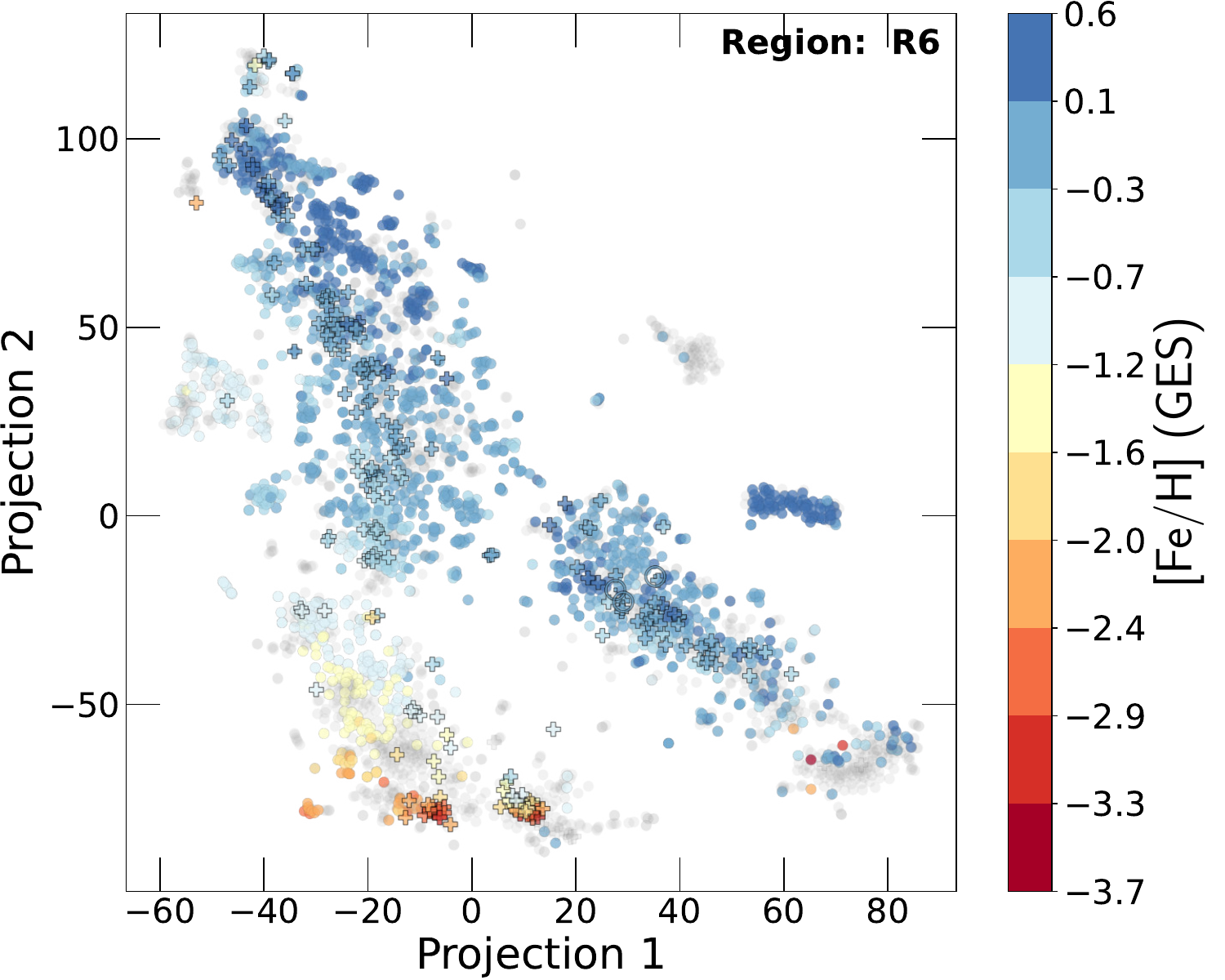} \\
        \includegraphics[width=0.3\textwidth]{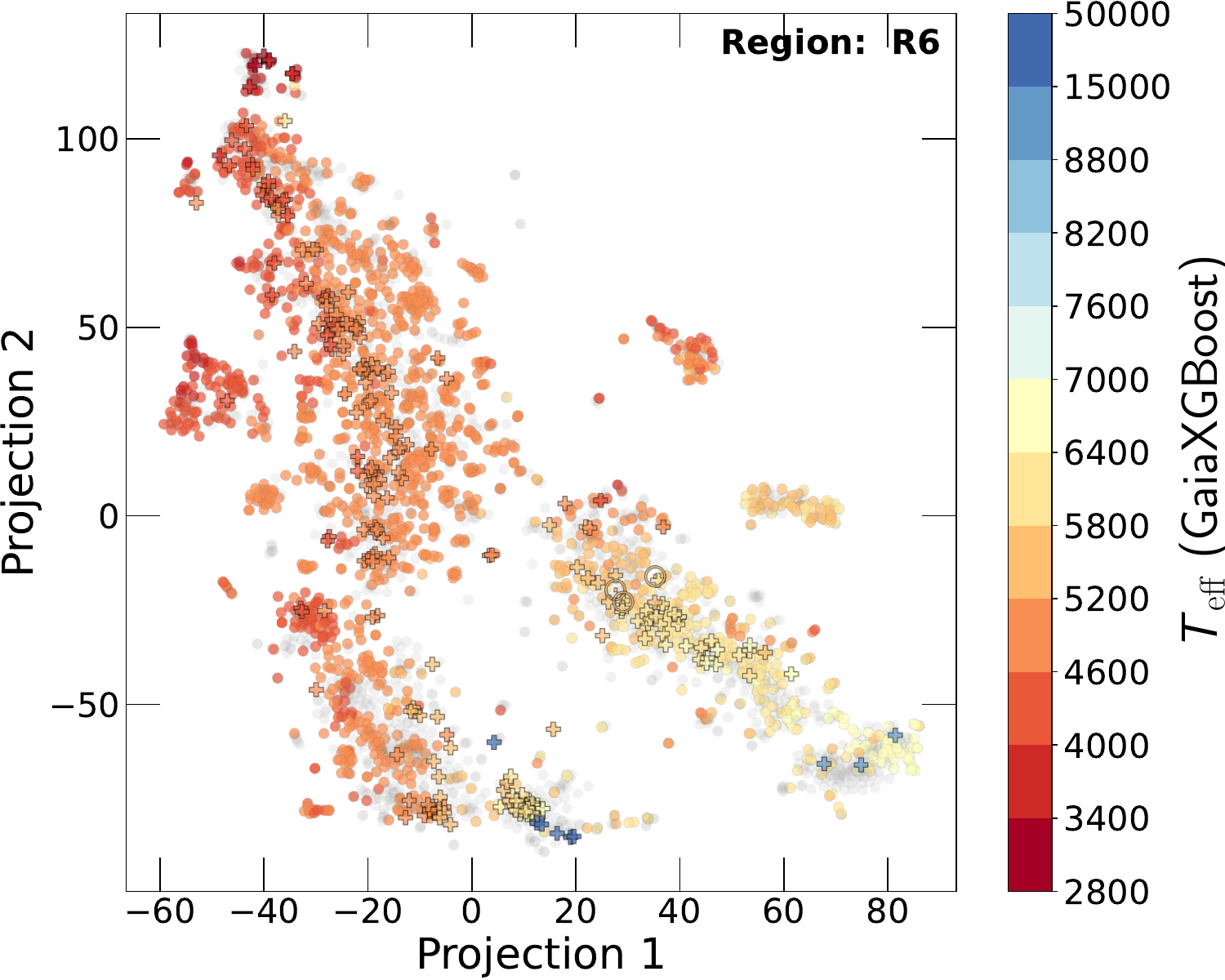} & \includegraphics[width=0.3\textwidth]{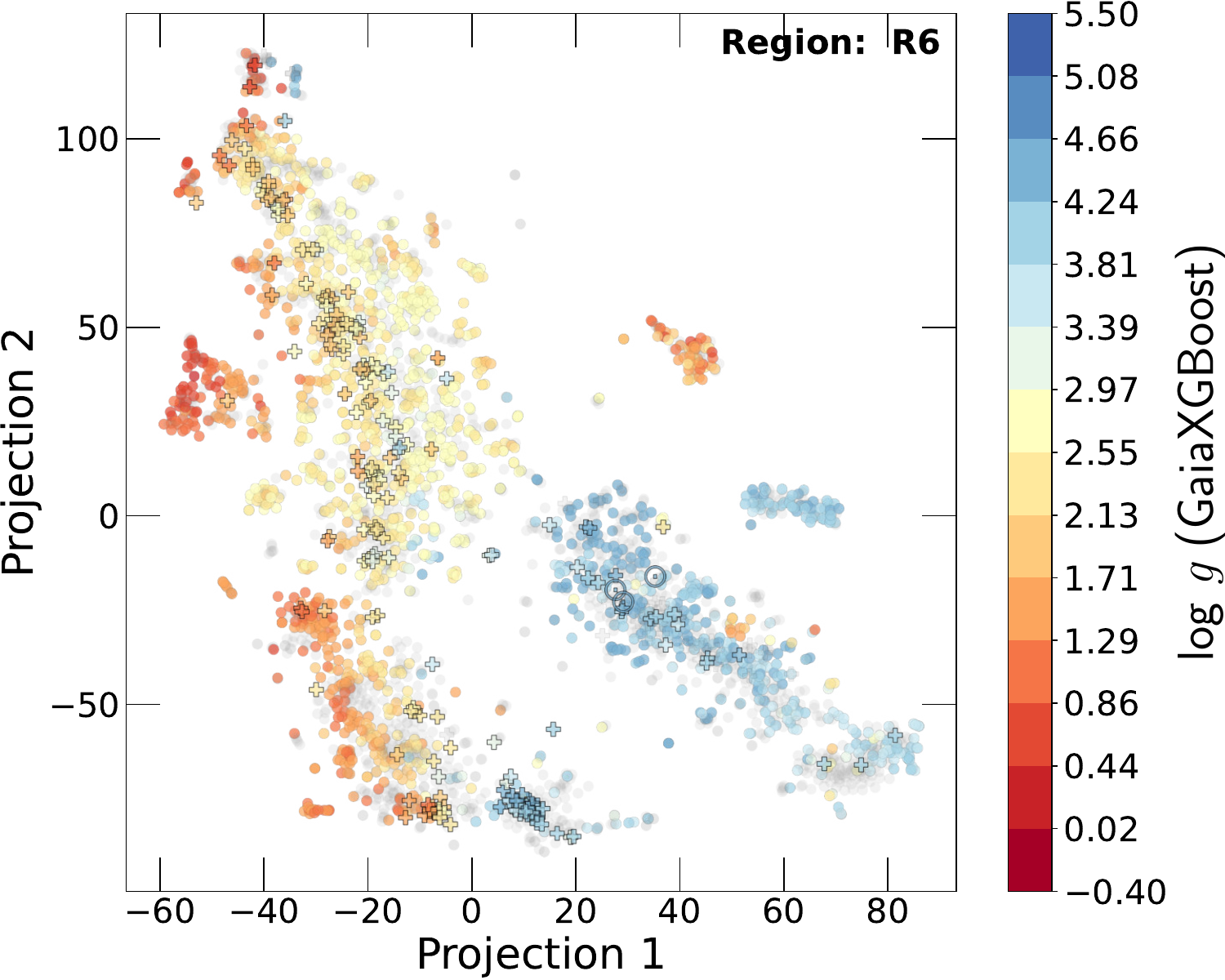} & \includegraphics[width=0.3\textwidth]{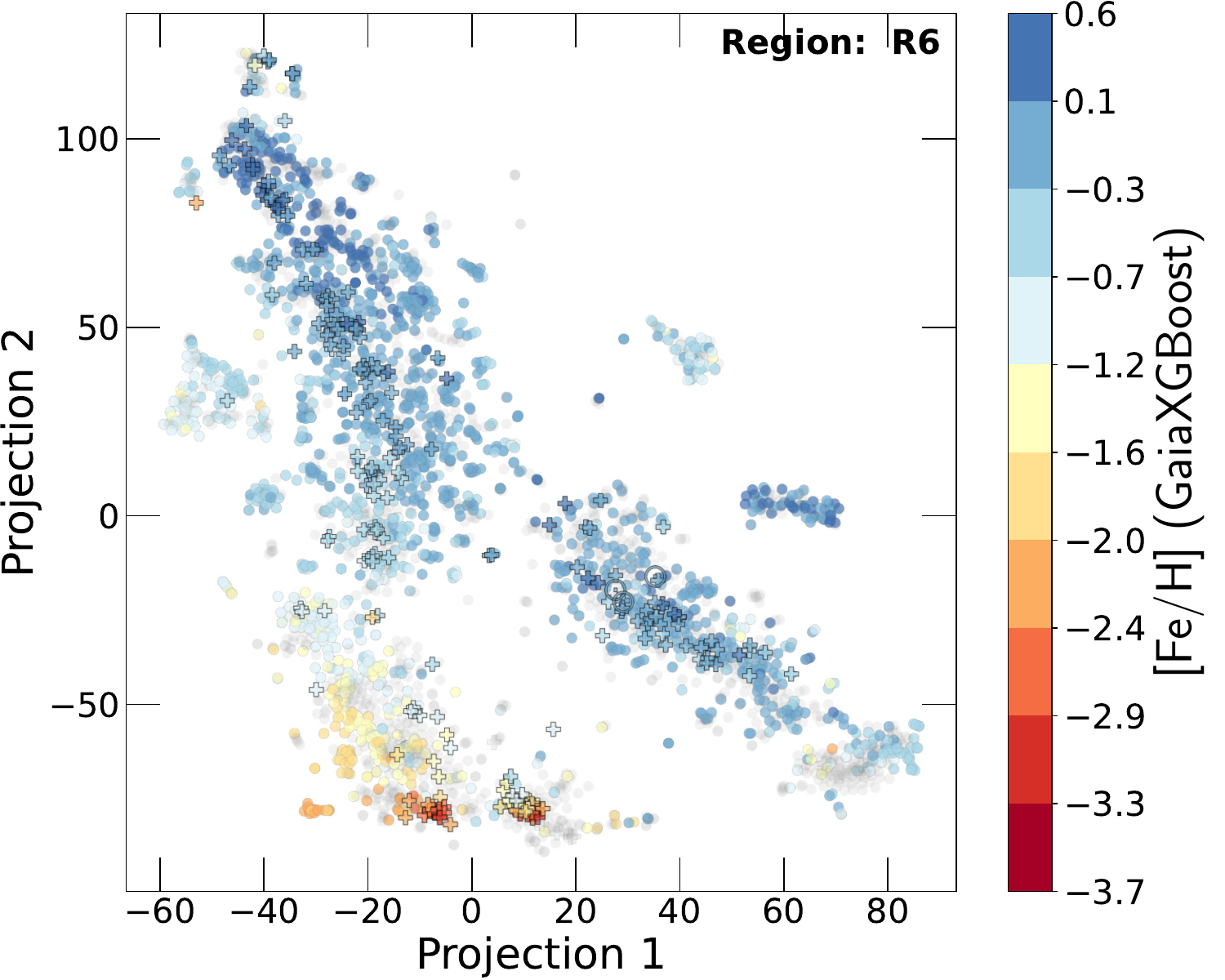} \\
        \includegraphics[width=0.3\textwidth]{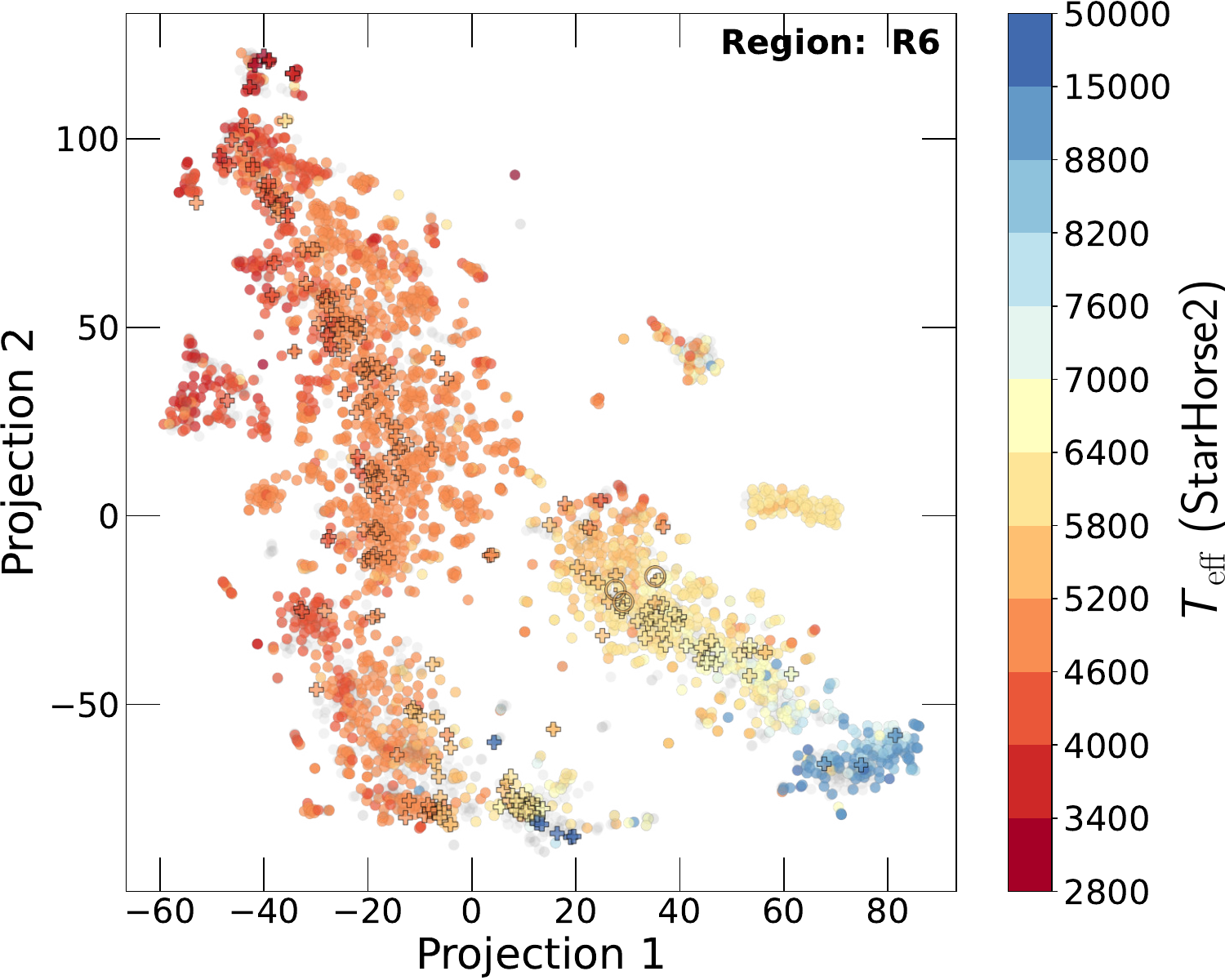} &
        \includegraphics[width=0.3\textwidth]{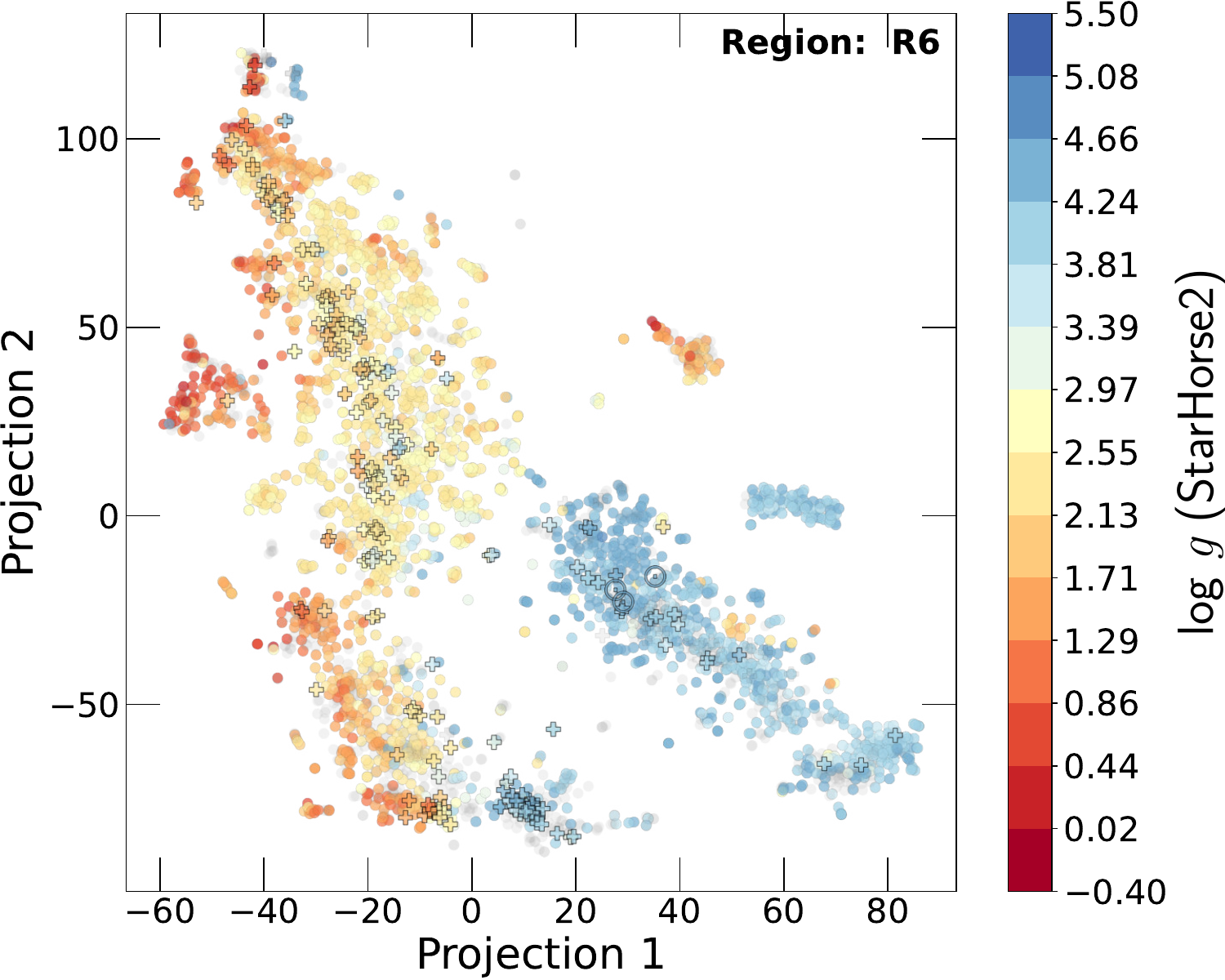} &
        \includegraphics[width=0.3\textwidth]{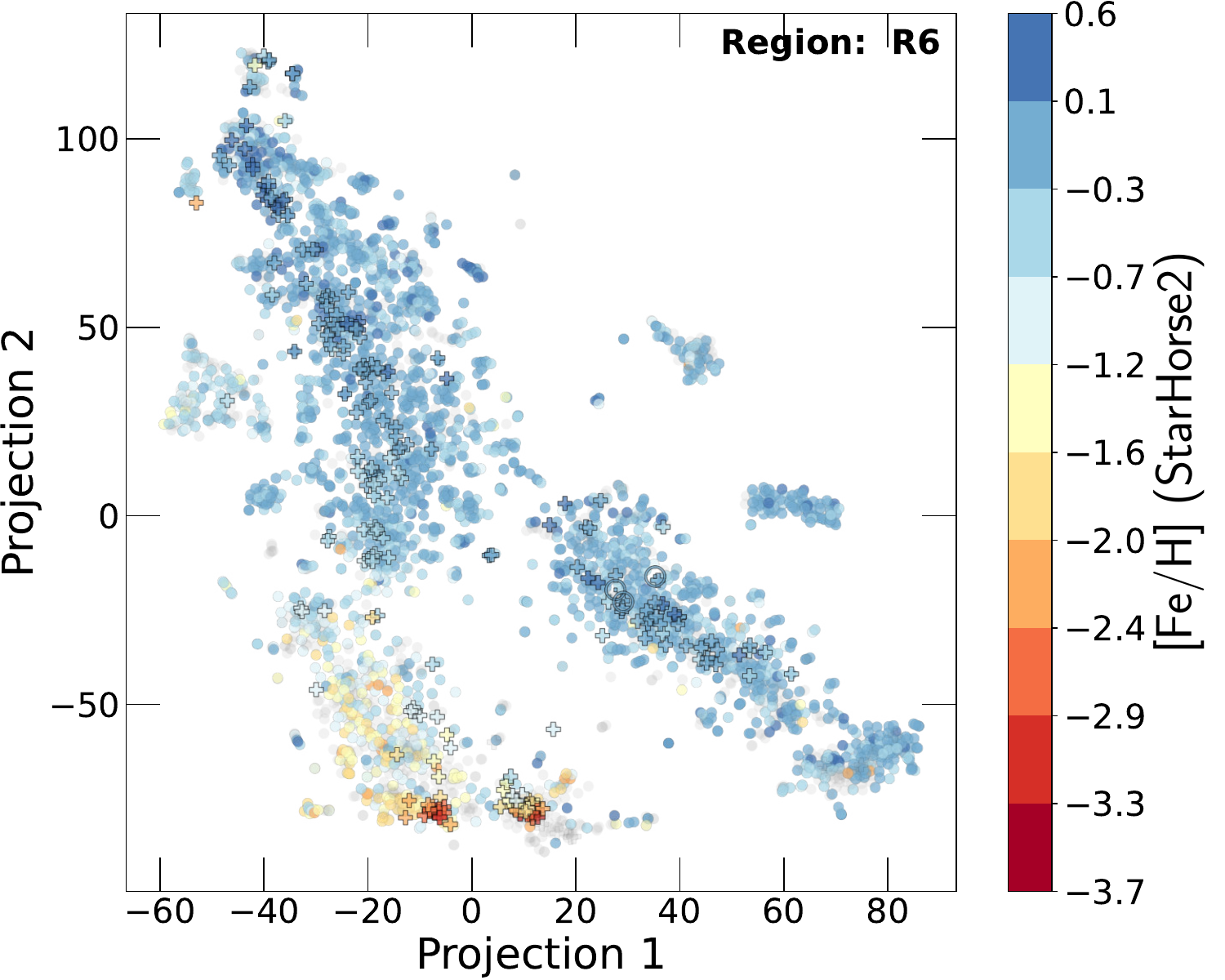} \\
        \includegraphics[width=0.3\textwidth]{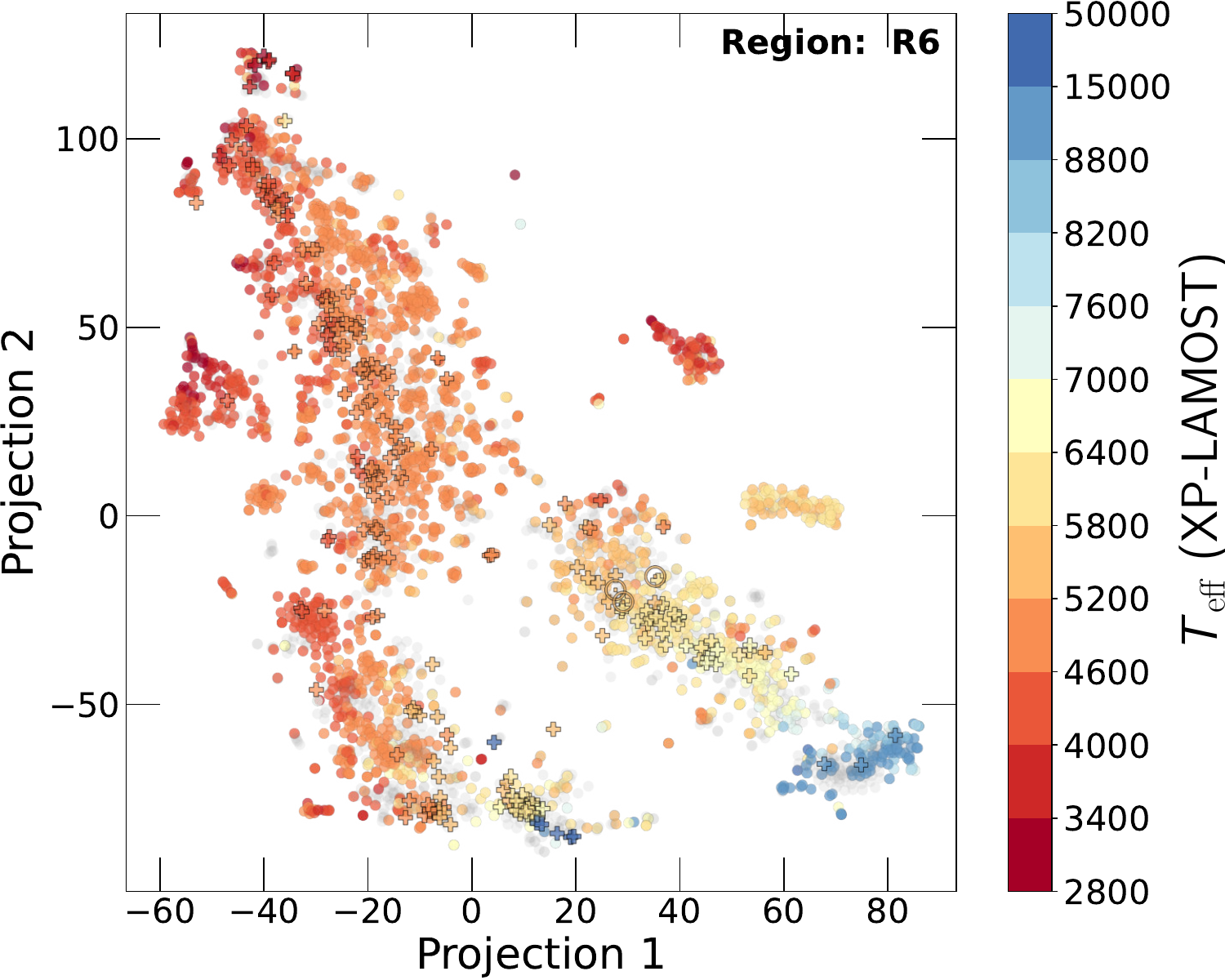} &
        \includegraphics[width=0.3\textwidth]{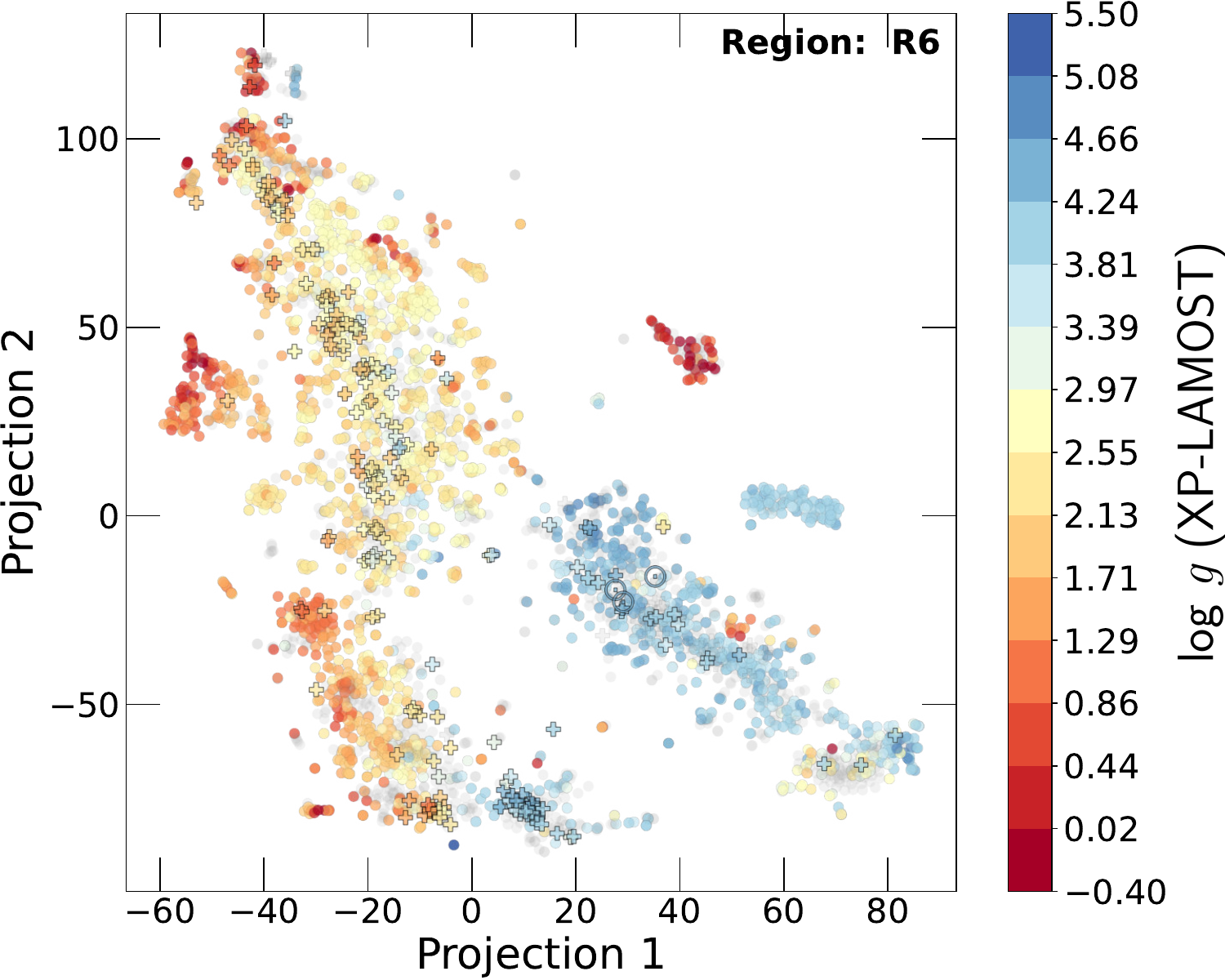} &
        \includegraphics[width=0.3\textwidth]{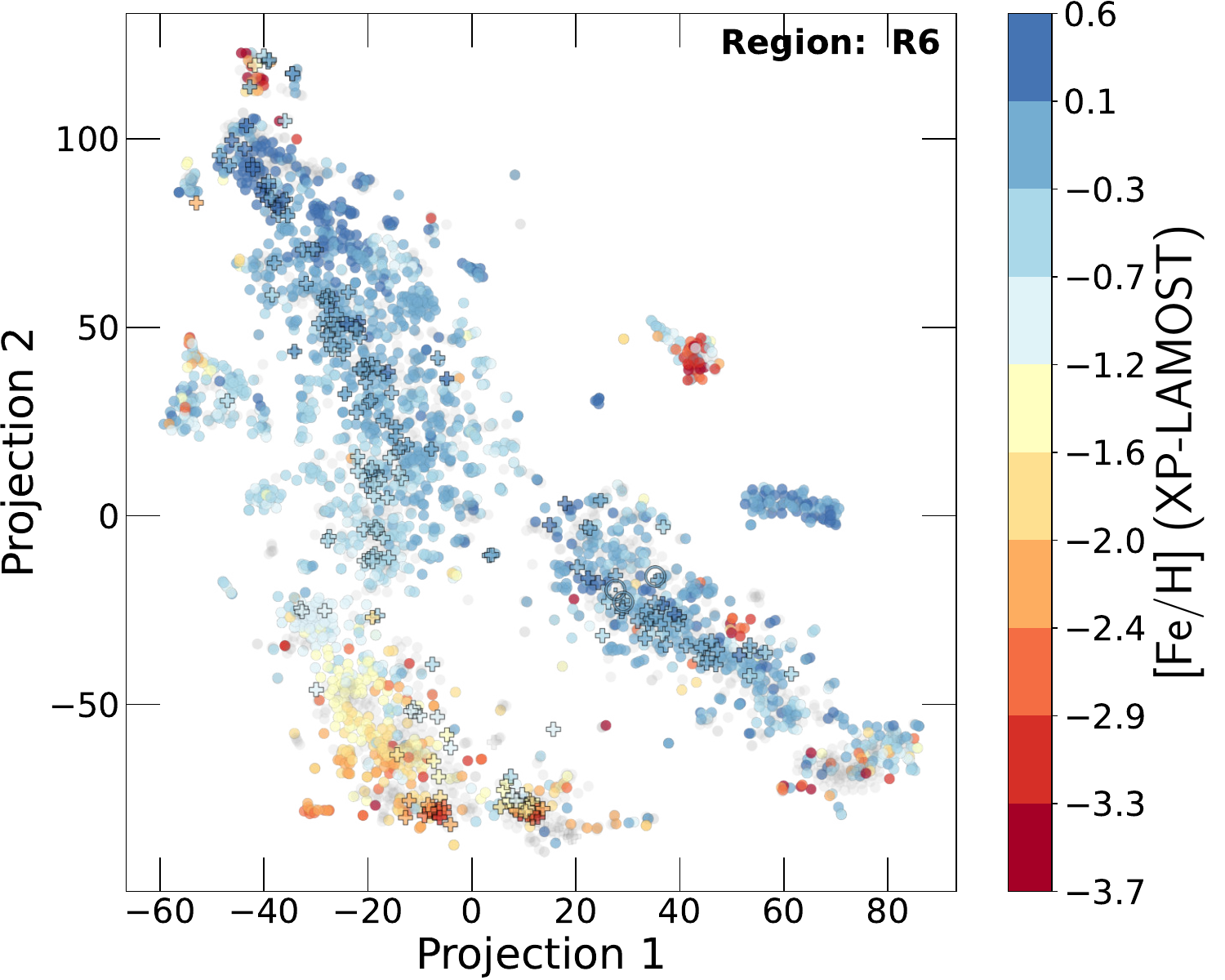} \\
    \end{tabular}
    \caption{Projection maps for the region R6 coloured with the atmospheric parameters of four different catalogues.}
\end{figure*}

\onecolumn
\section{Histograms of $S_{thresh}$ for all regions and all catalogues}
\label{app:B}

\begin{figure*}[htbp]
    \centering
    \begin{tabular}{ccc}
        \includegraphics[width=0.28\textwidth]{Plots/simil_thresh_hists/simil_thresh_histograms_R2_ges.pdf} & \includegraphics[width=0.28\textwidth]{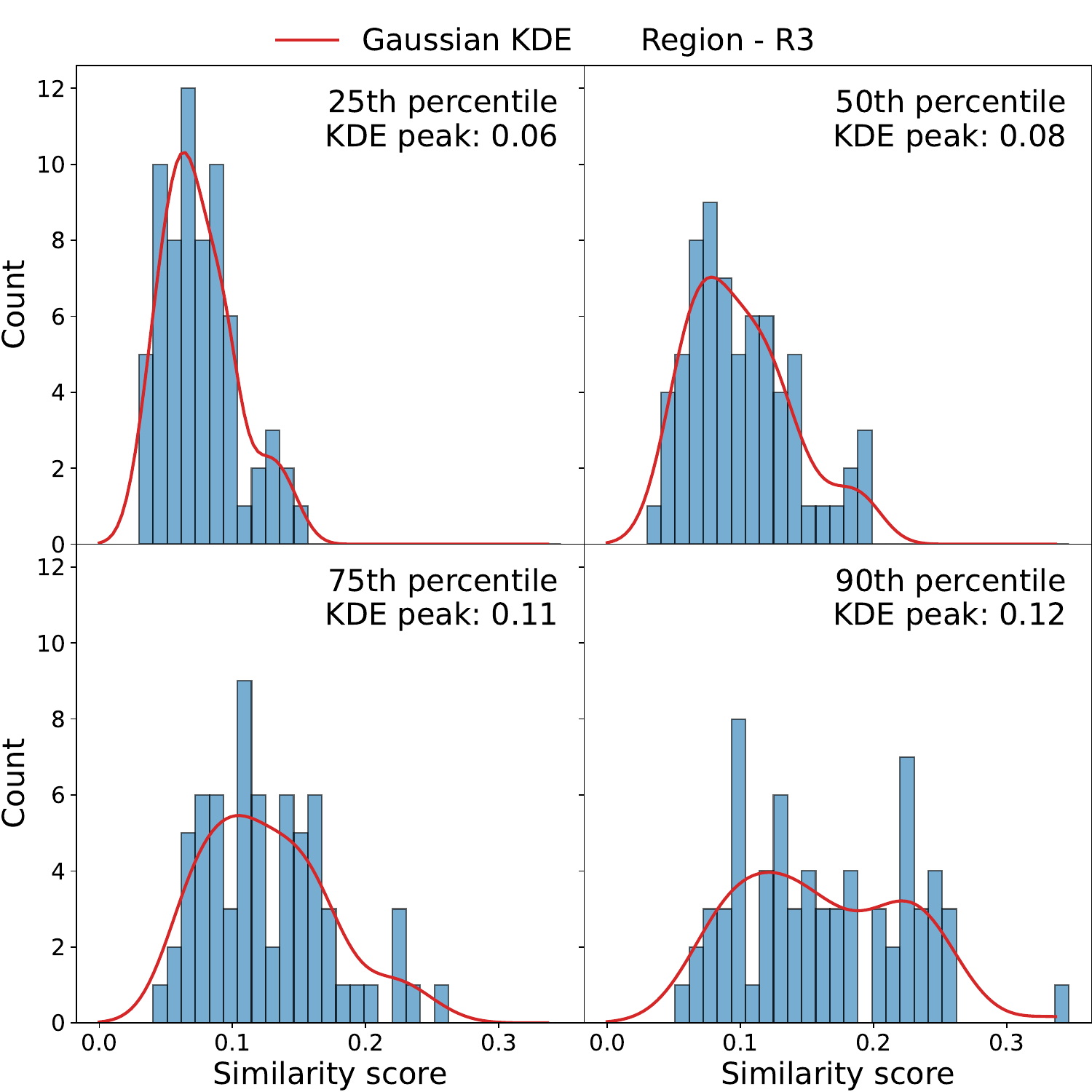} & \includegraphics[width=0.28\textwidth]{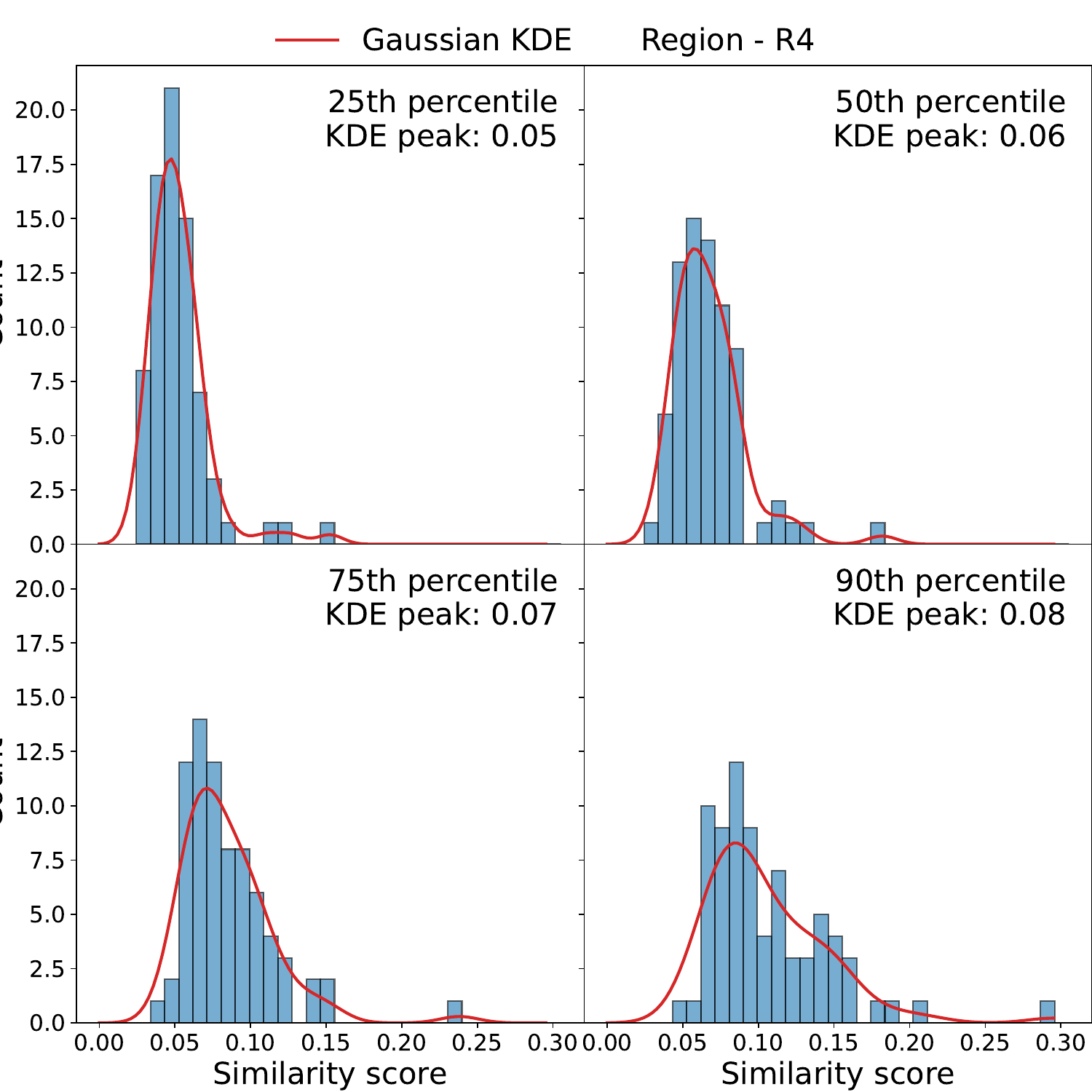}  \\ \includegraphics[width=0.28\textwidth]{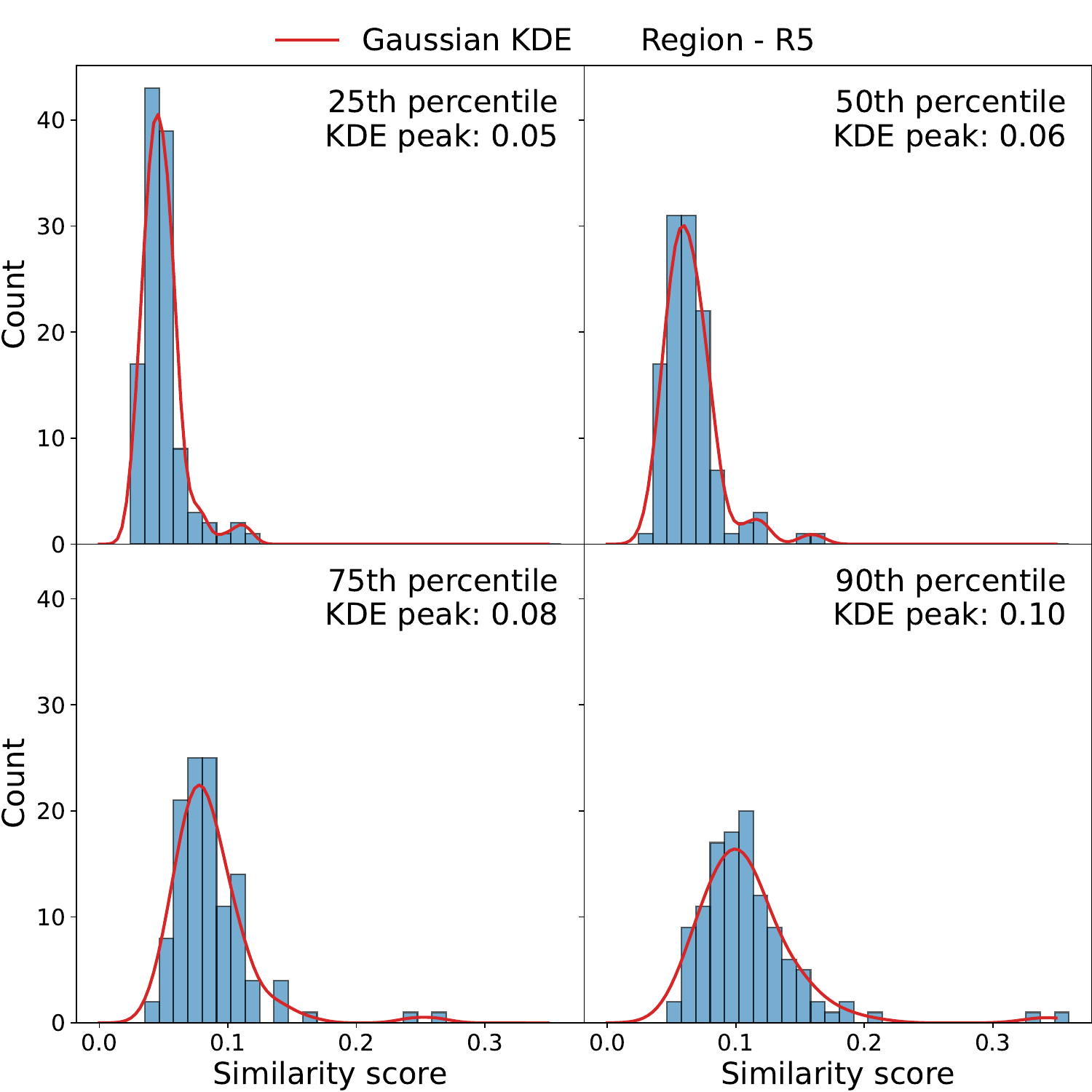} &        \includegraphics[width=0.28\textwidth]{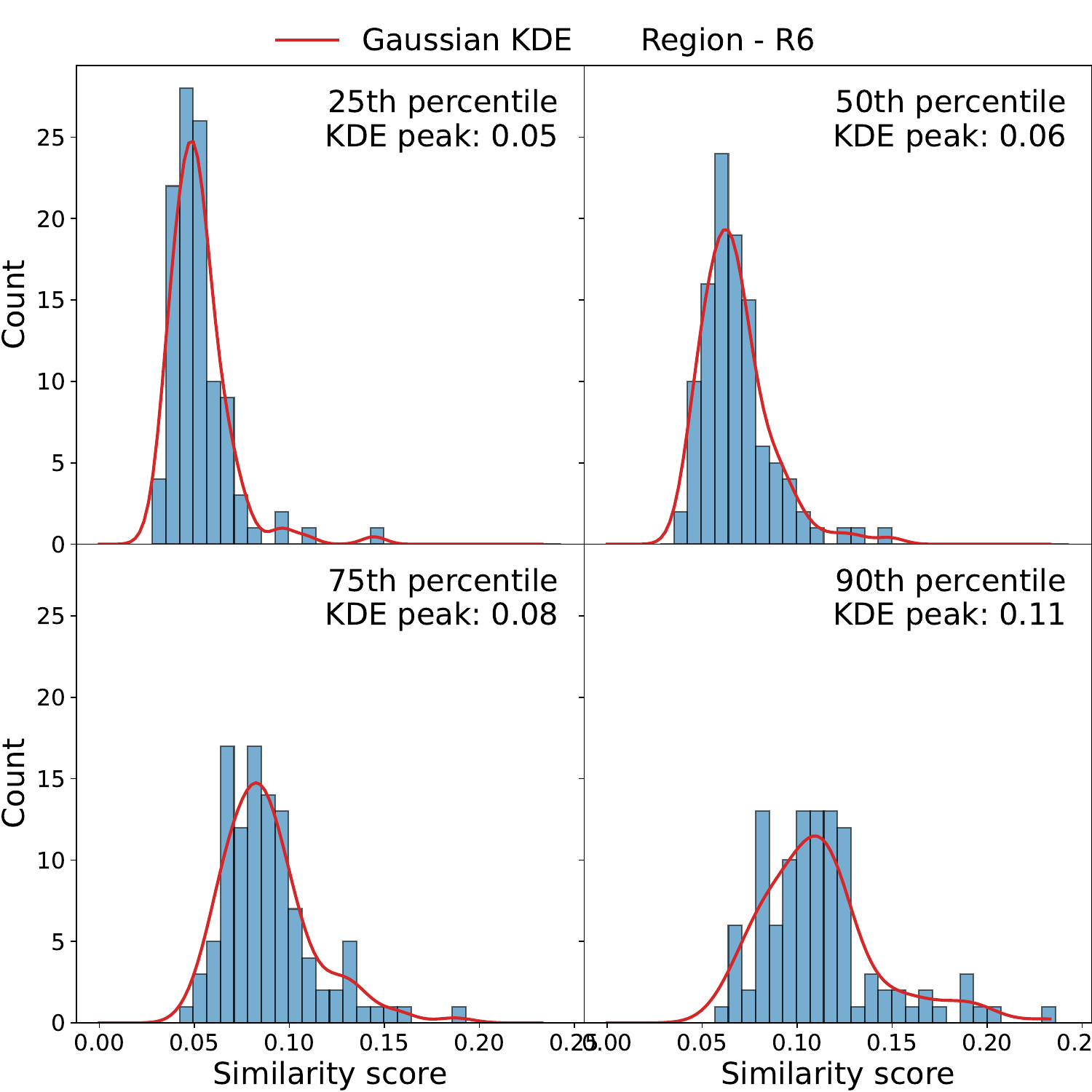} & 
        \\
    \end{tabular}
    \caption{Histogram of the percentiles (25th, 50th, 75th, 90th) of similarity scores for all spectral regions using the GES catalogue.}
\end{figure*}

\begin{figure*}[htbp]
    \centering
    \begin{tabular}{ccc}
        \includegraphics[width=0.28\textwidth]{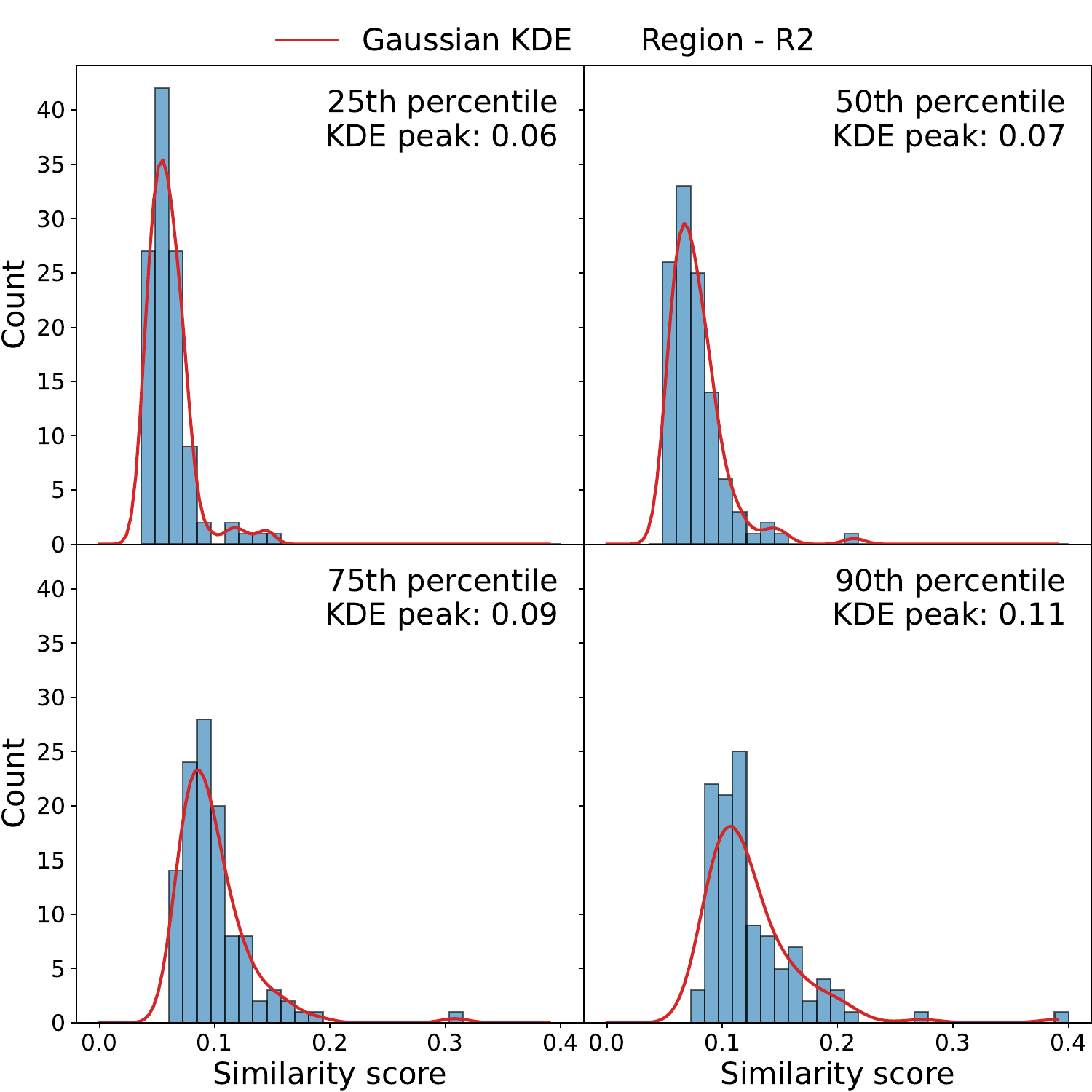} & \includegraphics[width=0.28\textwidth]{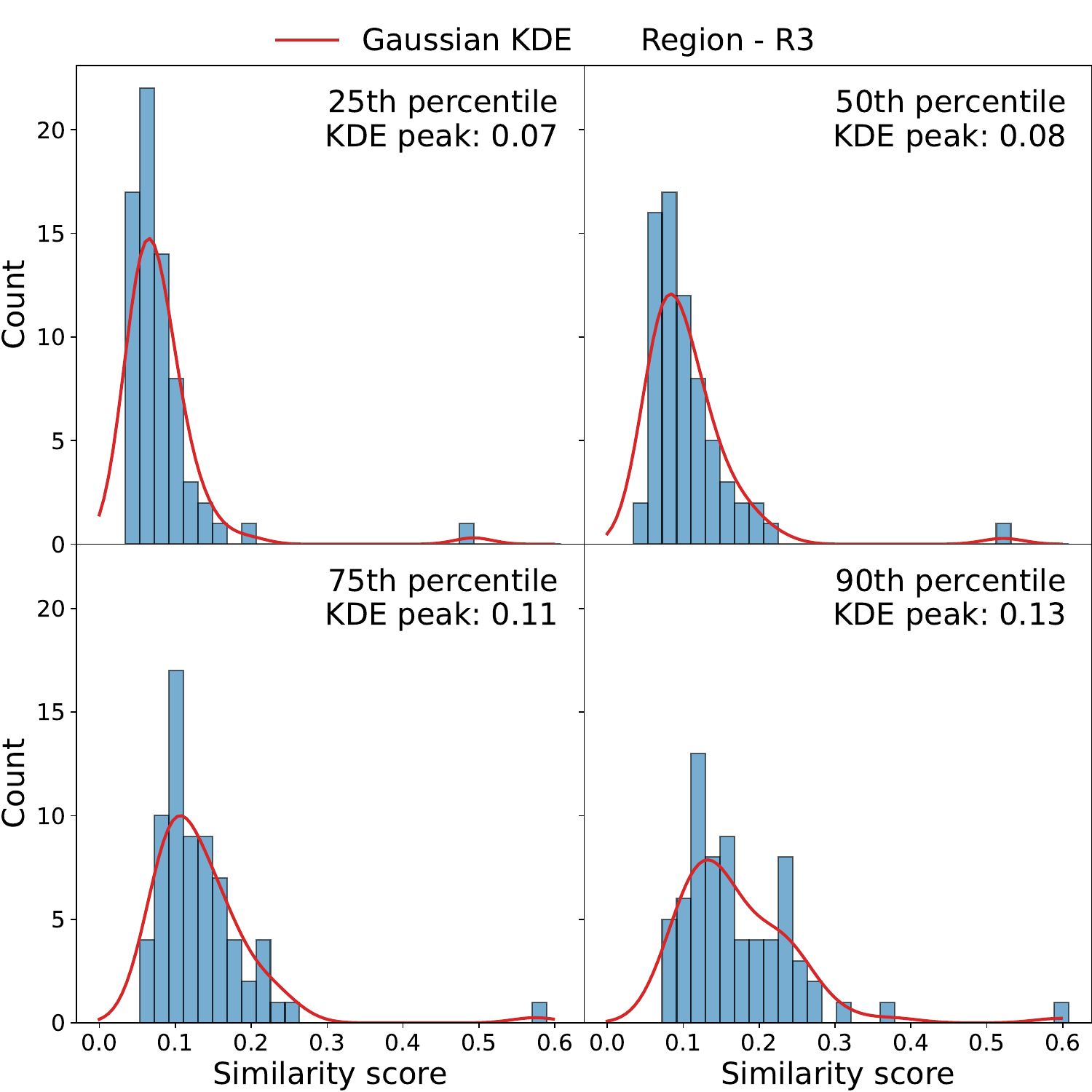} & \includegraphics[width=0.28\textwidth]{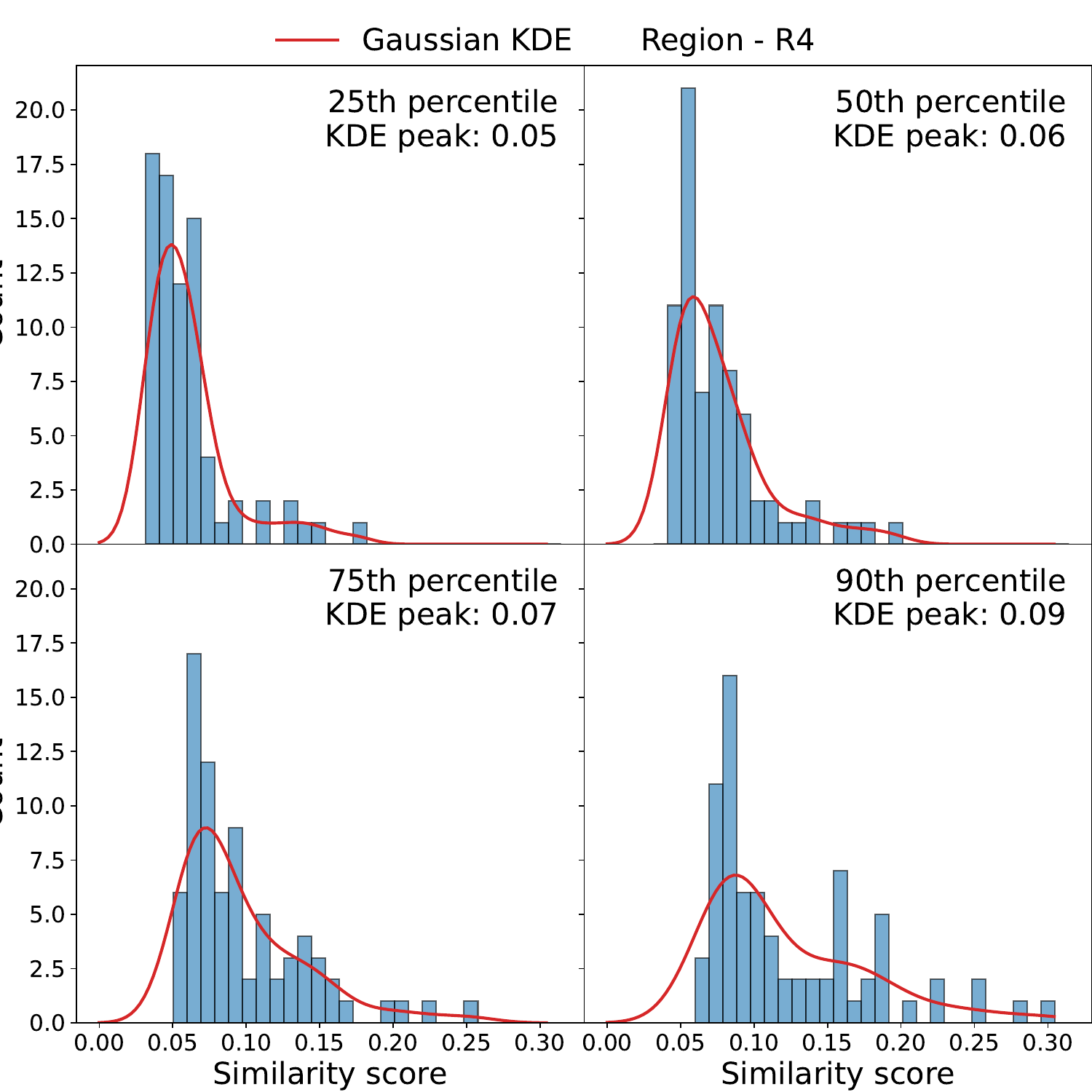}\\ \includegraphics[width=0.28\textwidth]{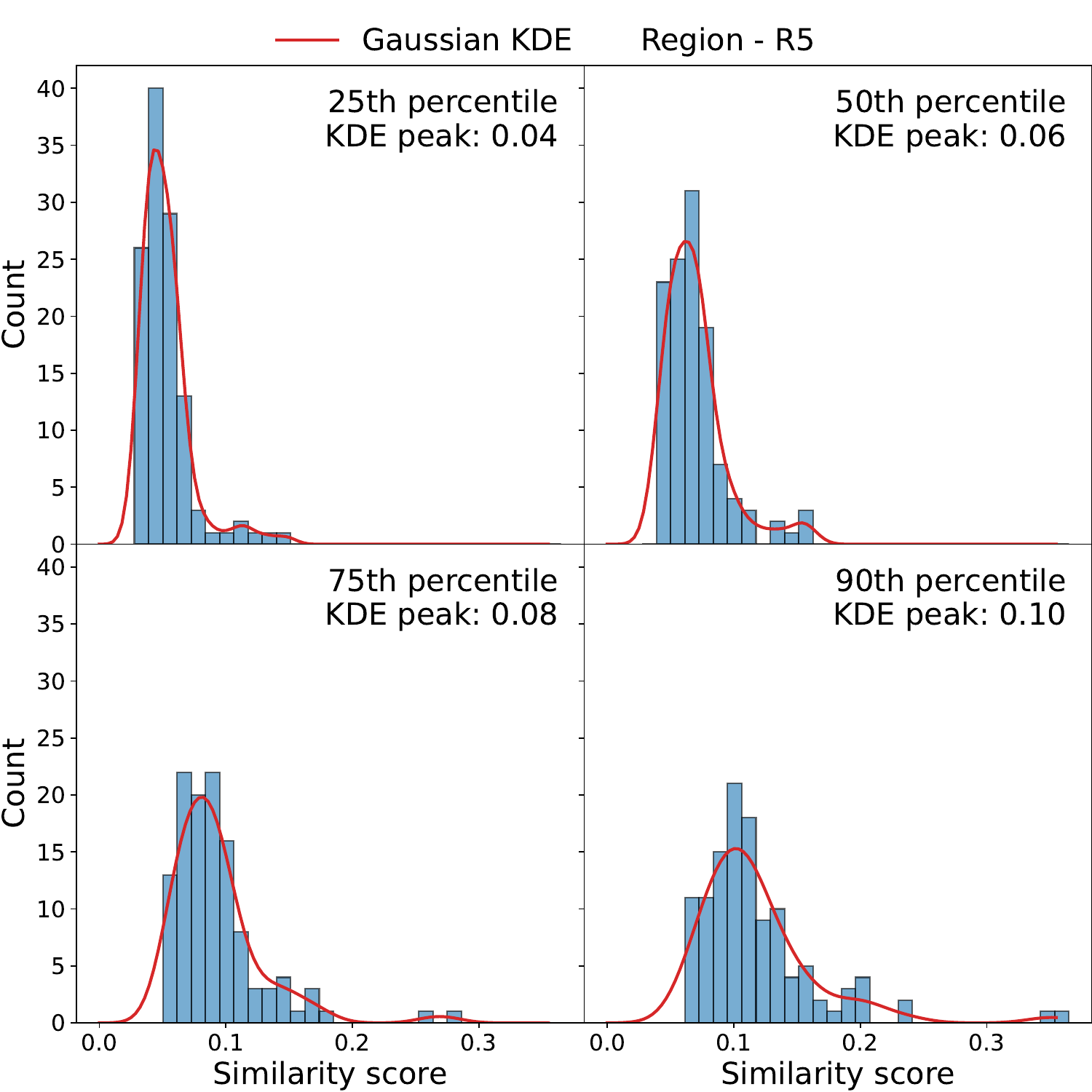} &  \includegraphics[width=0.28\textwidth]{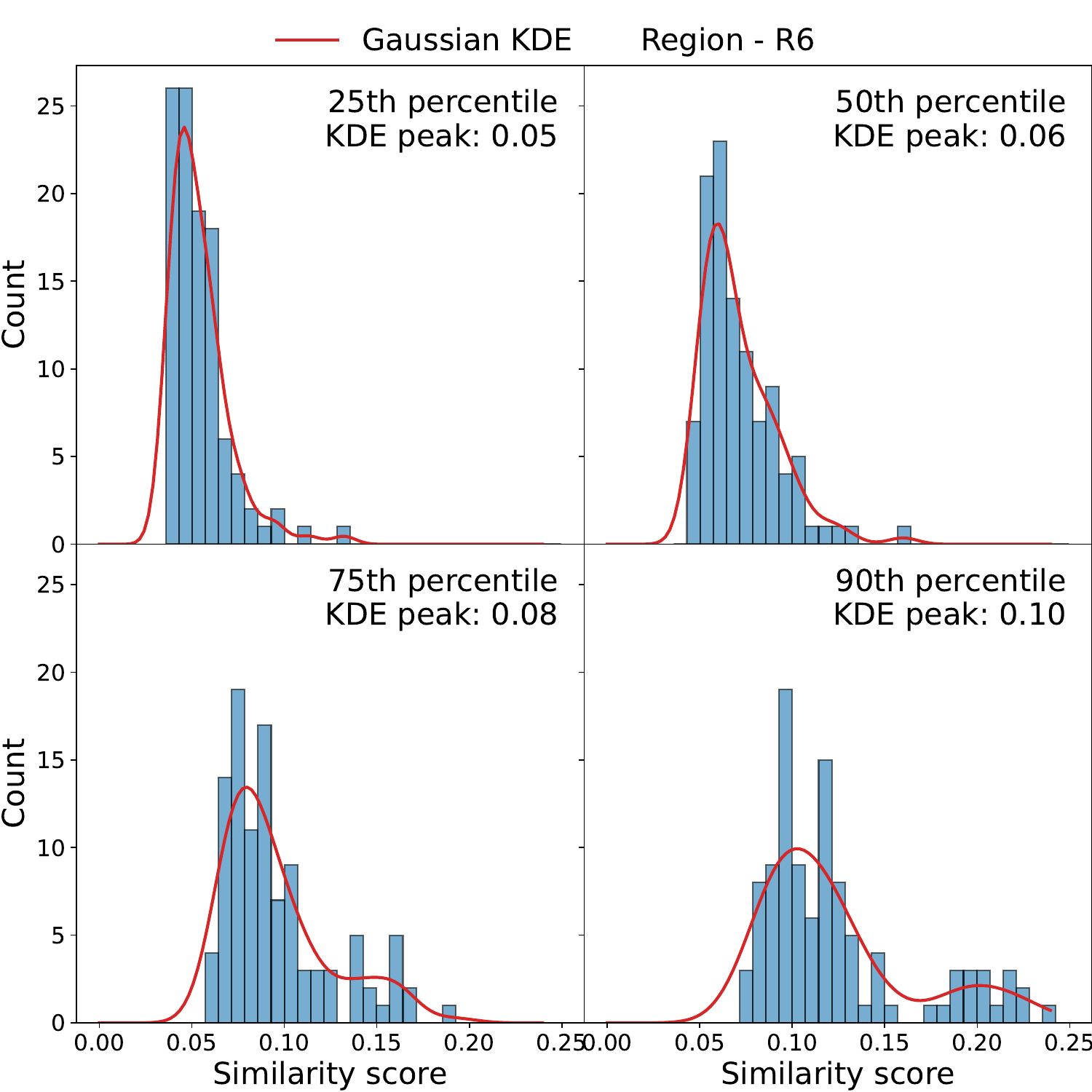} & 
        \\
    \end{tabular}
    \caption{Histogram of the percentiles (25th, 50th, 75th, 90th) of similarity scores for all spectral regions using the XGBoost catalogue.}
\end{figure*}

\begin{figure*}[htbp]
    \centering
    \begin{tabular}{ccc}
        \includegraphics[width=0.29\textwidth]{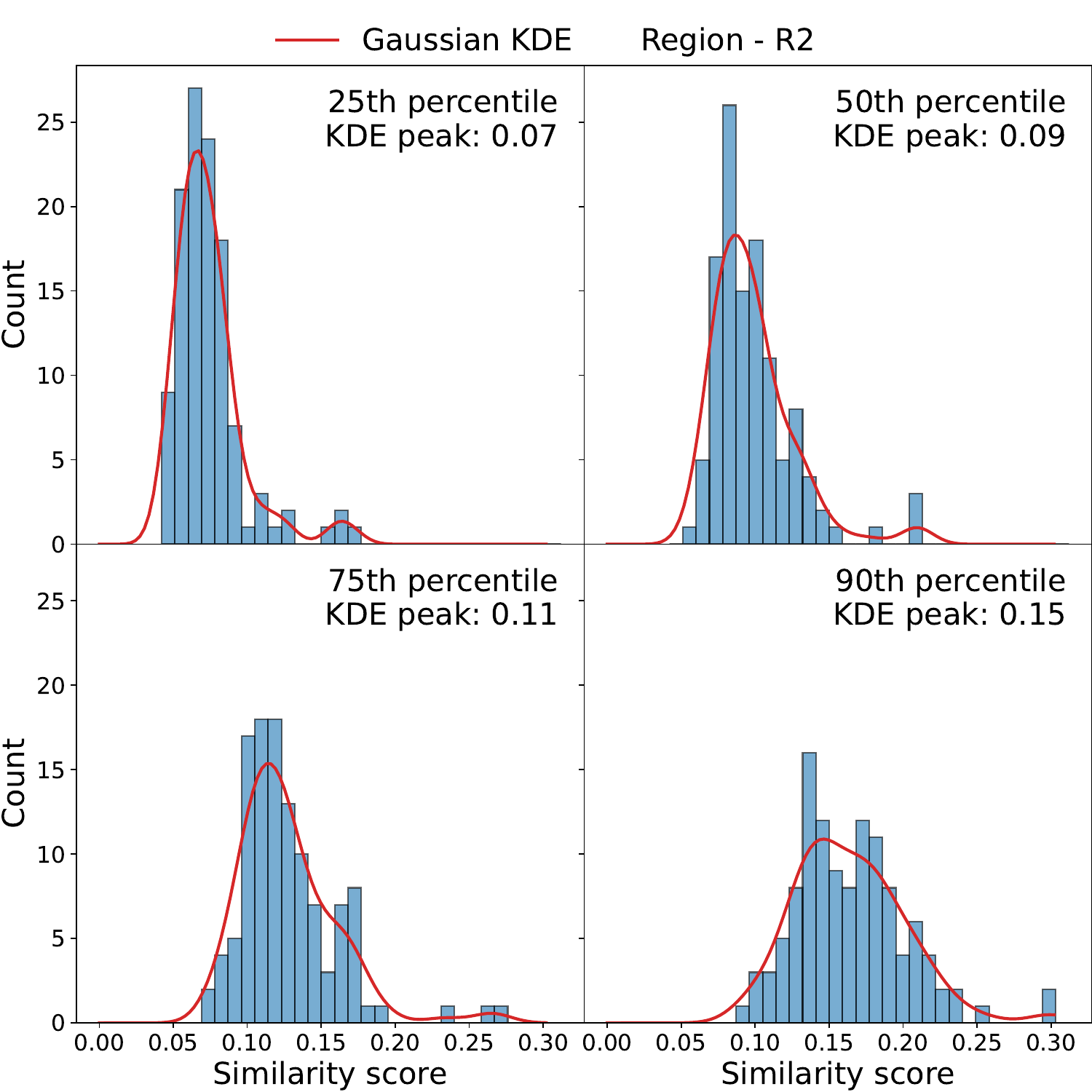}& \includegraphics[width=0.29\textwidth]{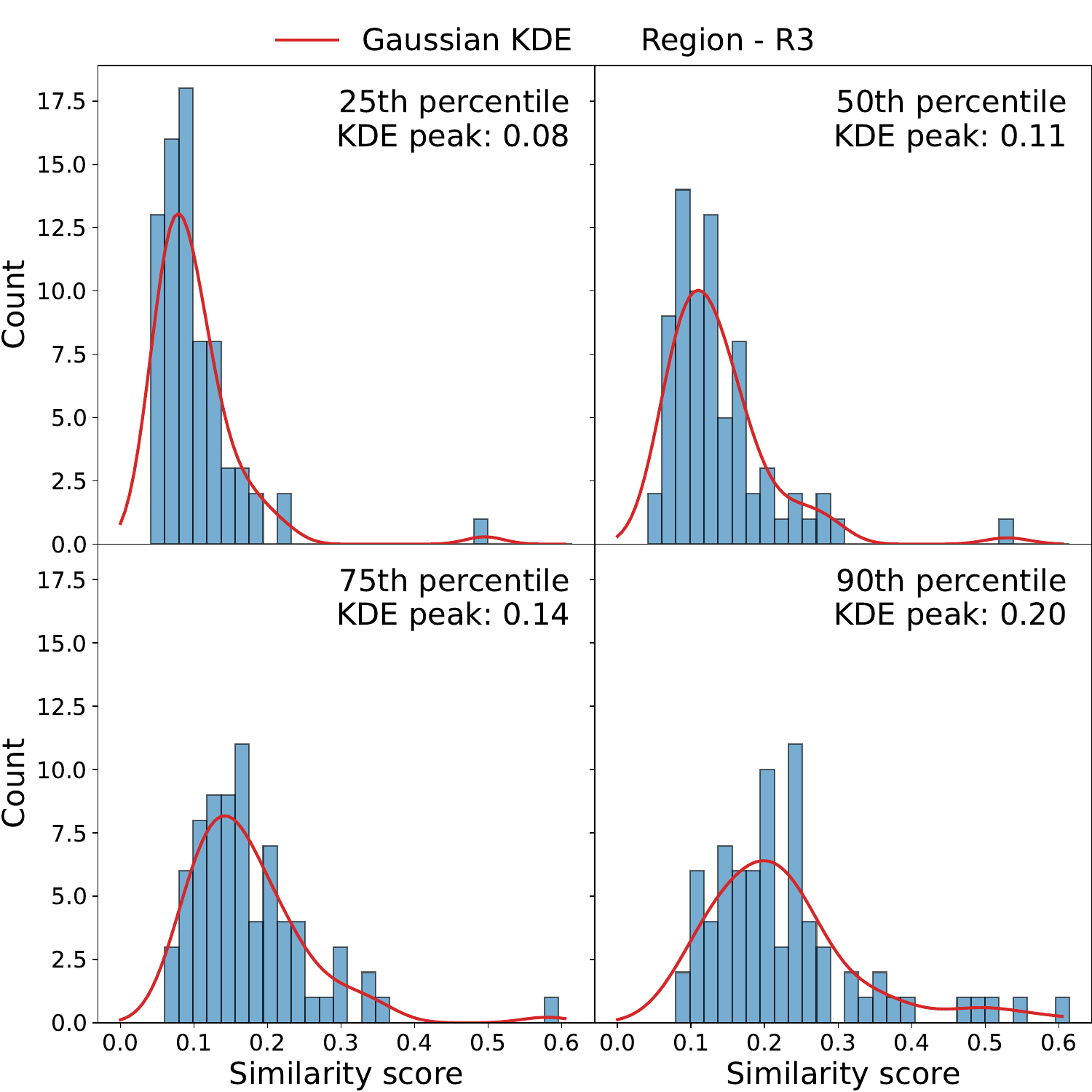}& \includegraphics[width=0.29\textwidth]{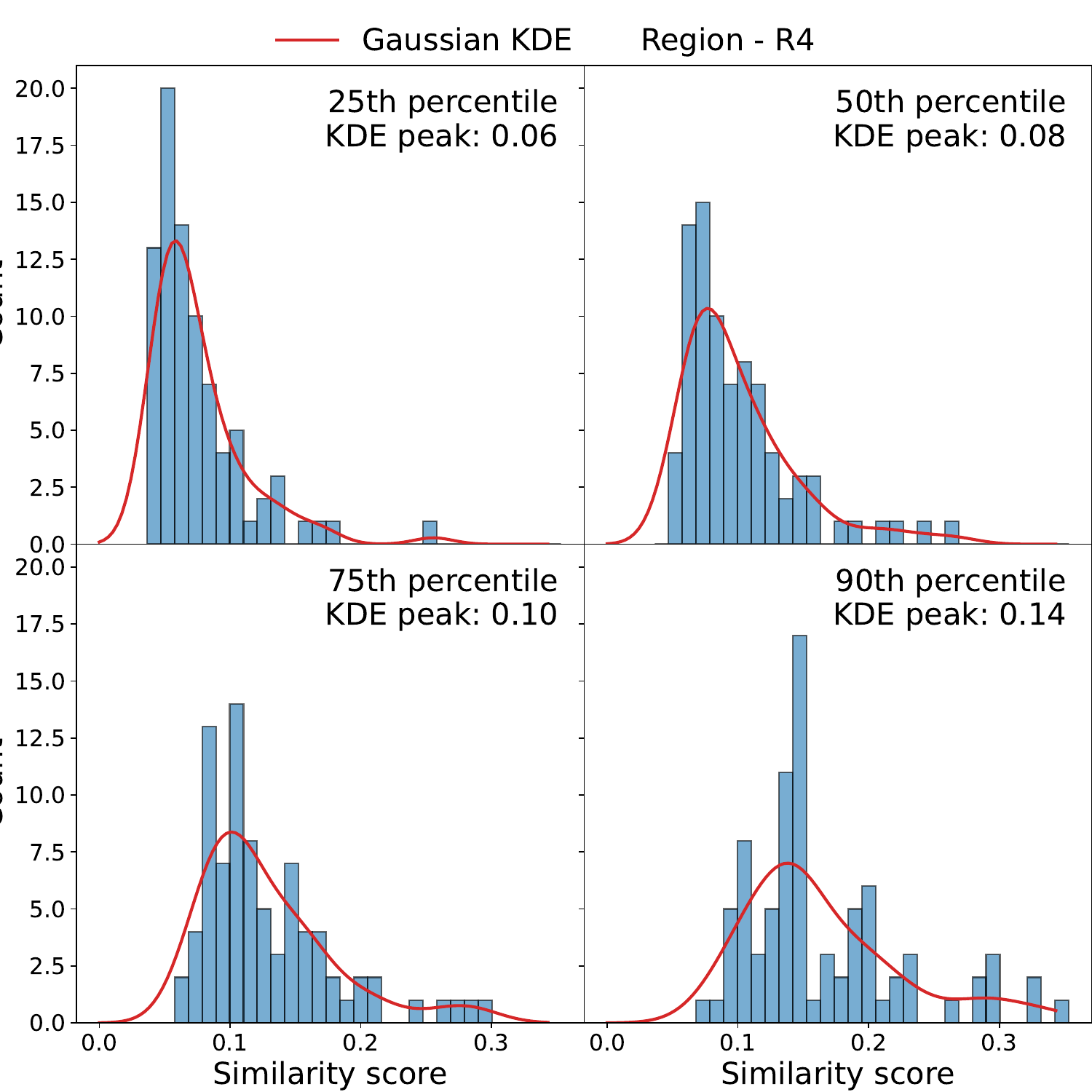}\\ \includegraphics[width=0.29\textwidth]{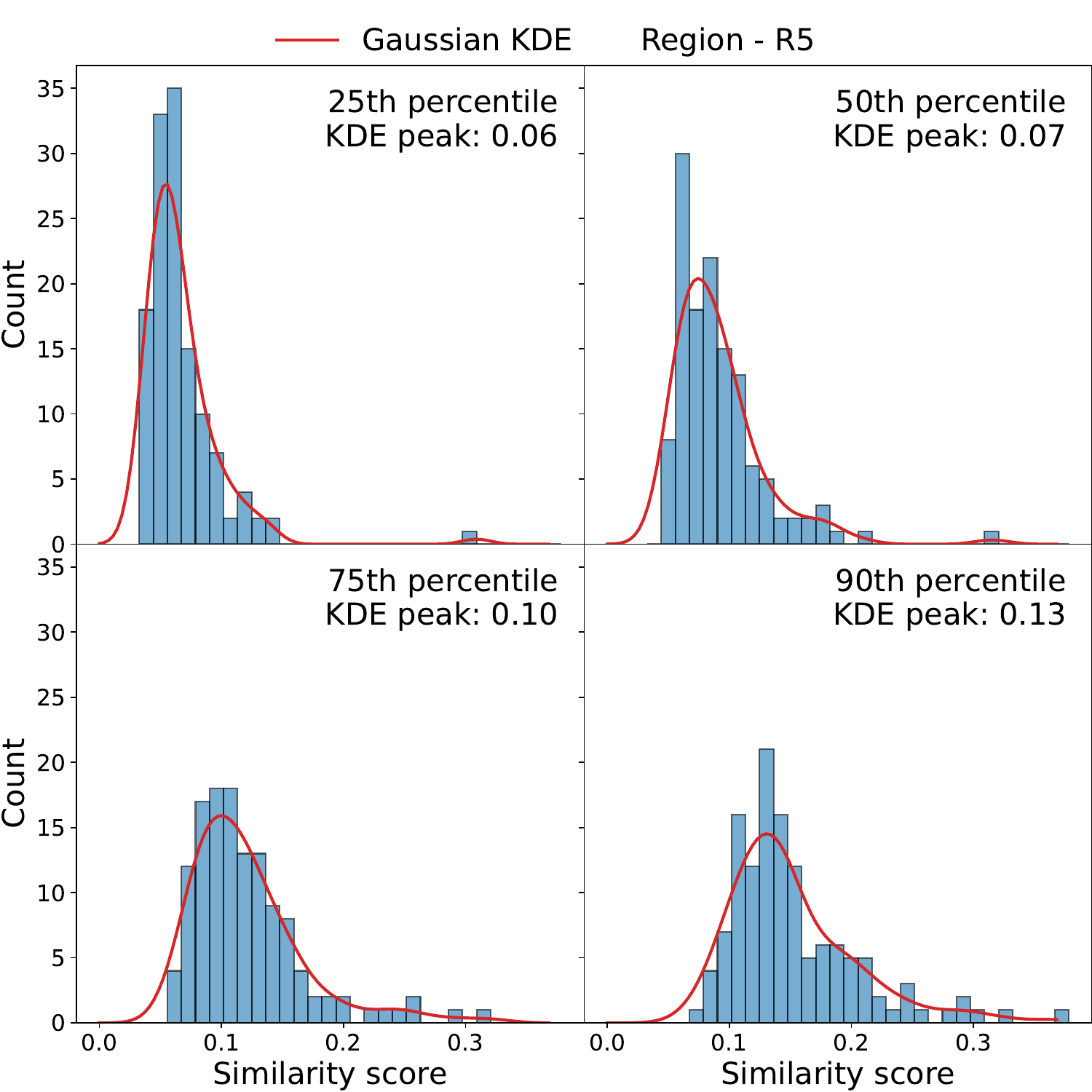}& \includegraphics[width=0.29\textwidth]{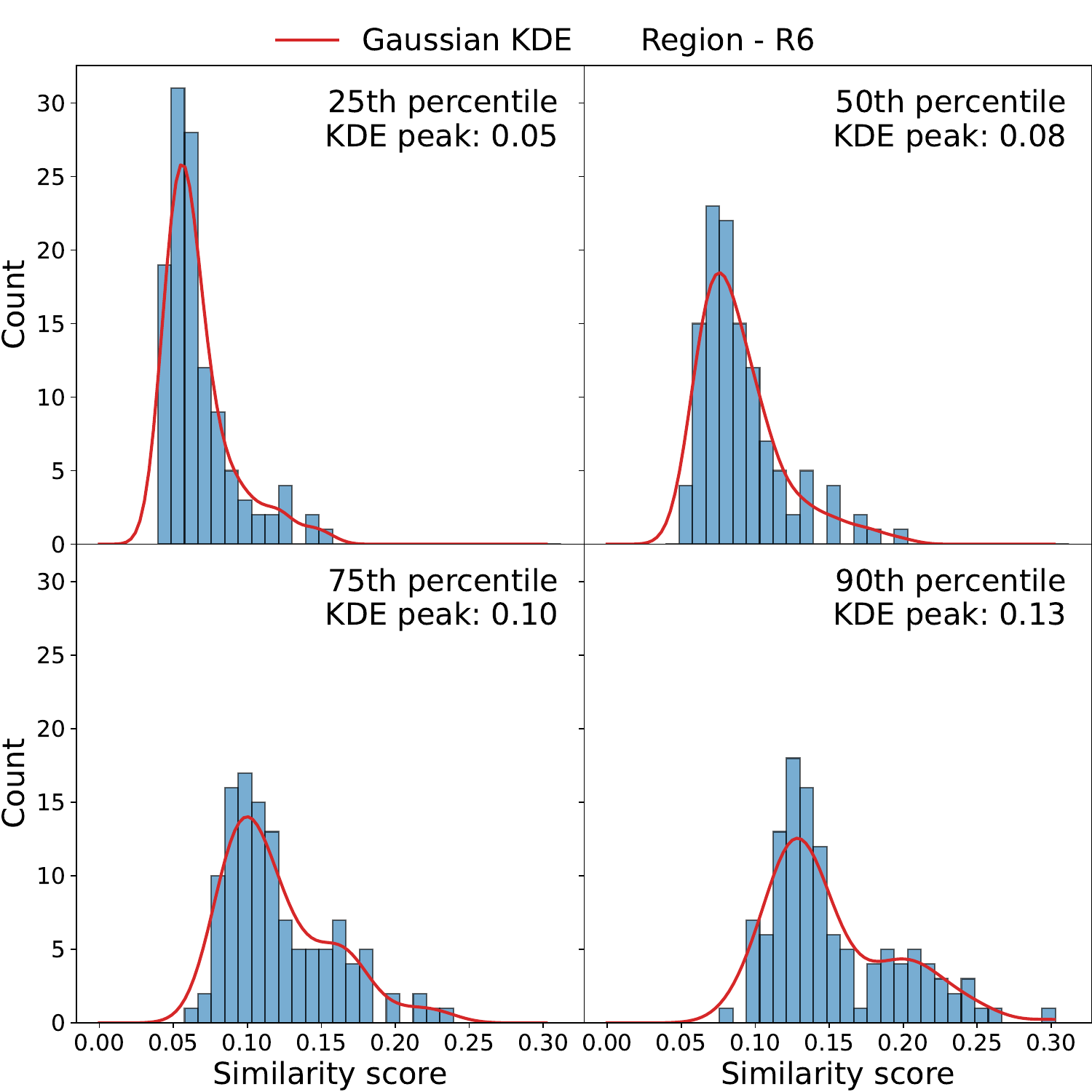} & 
        \\
    \end{tabular}
    \caption{Histogram of the percentiles (25th, 50th, 75th, 90th) of similarity scores for all spectral regions using the StarHorse2 catalogue.}
\end{figure*}

\begin{figure*}[htbp]
    \centering
    \begin{tabular}{ccc}
        \includegraphics[width=0.29\textwidth]{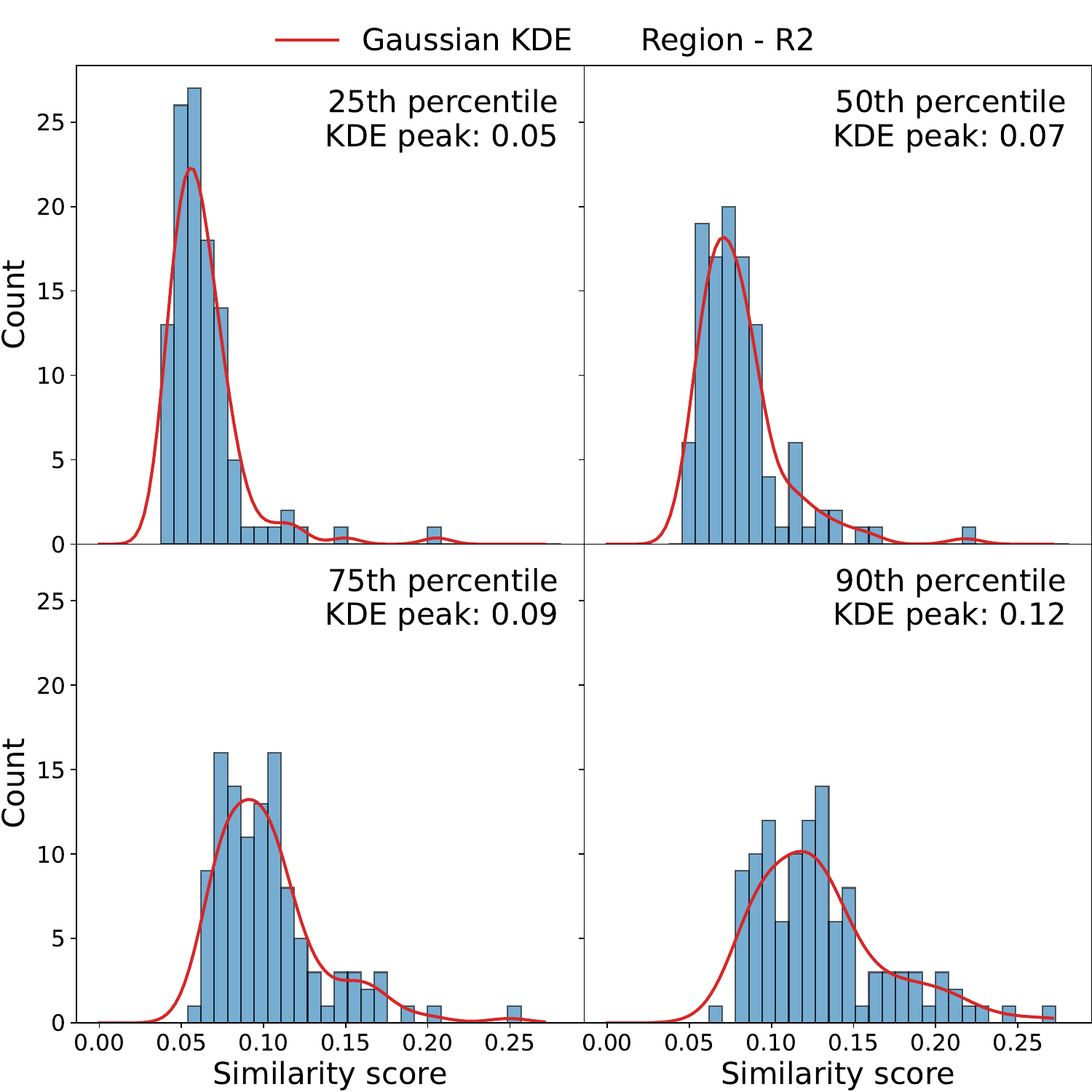} & \includegraphics[width=0.29\textwidth]{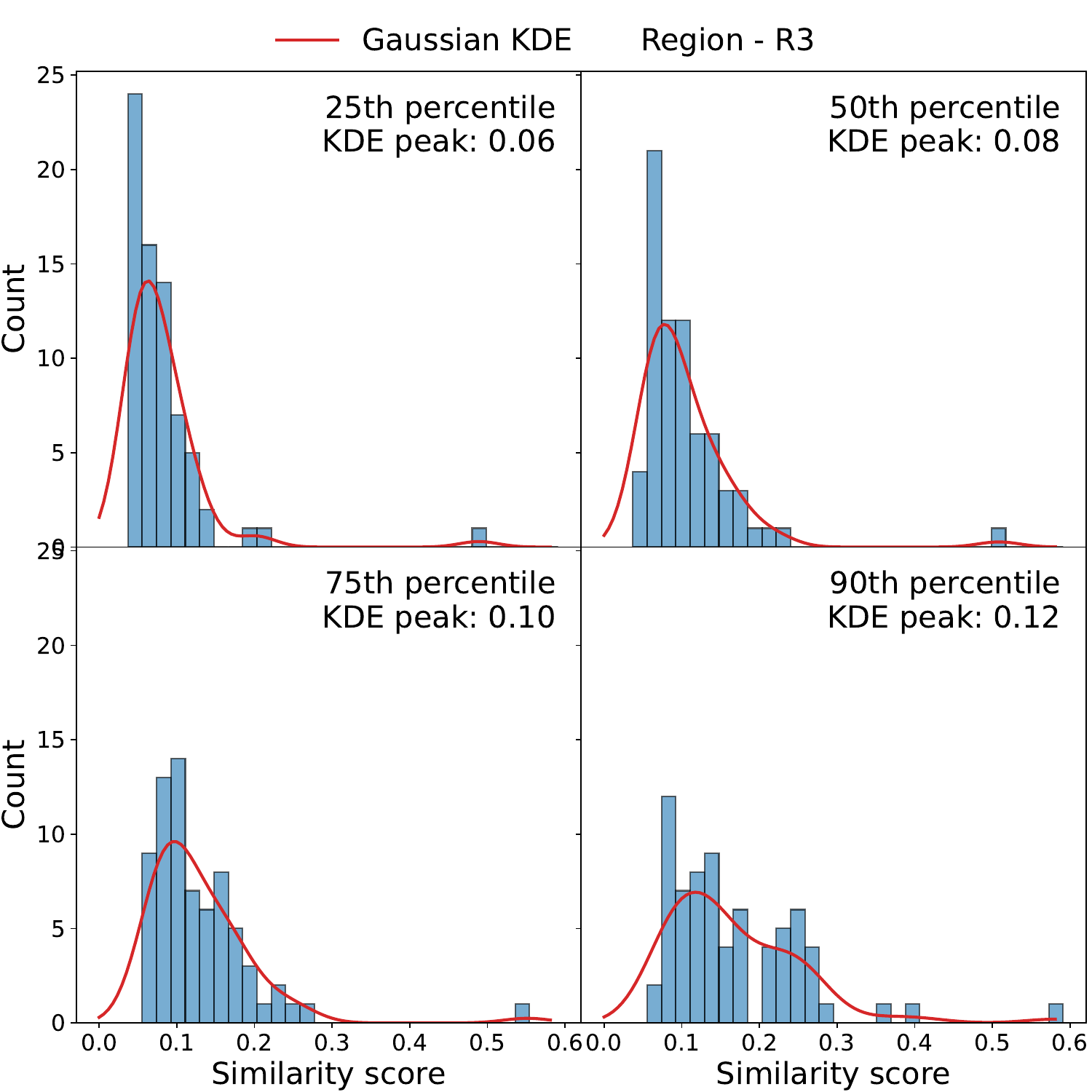} & \includegraphics[width=0.29\textwidth]{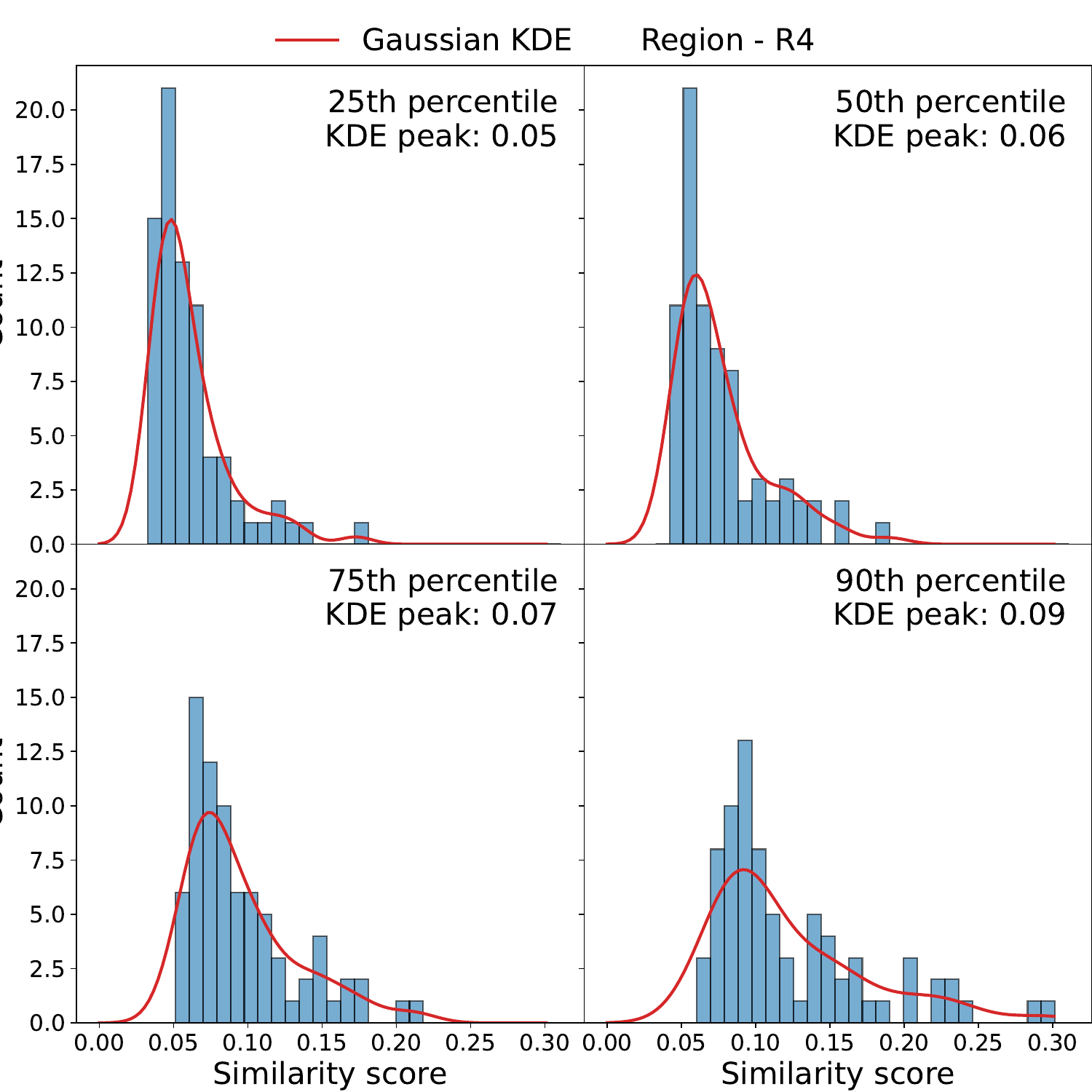}  \\ \includegraphics[width=0.29\textwidth]{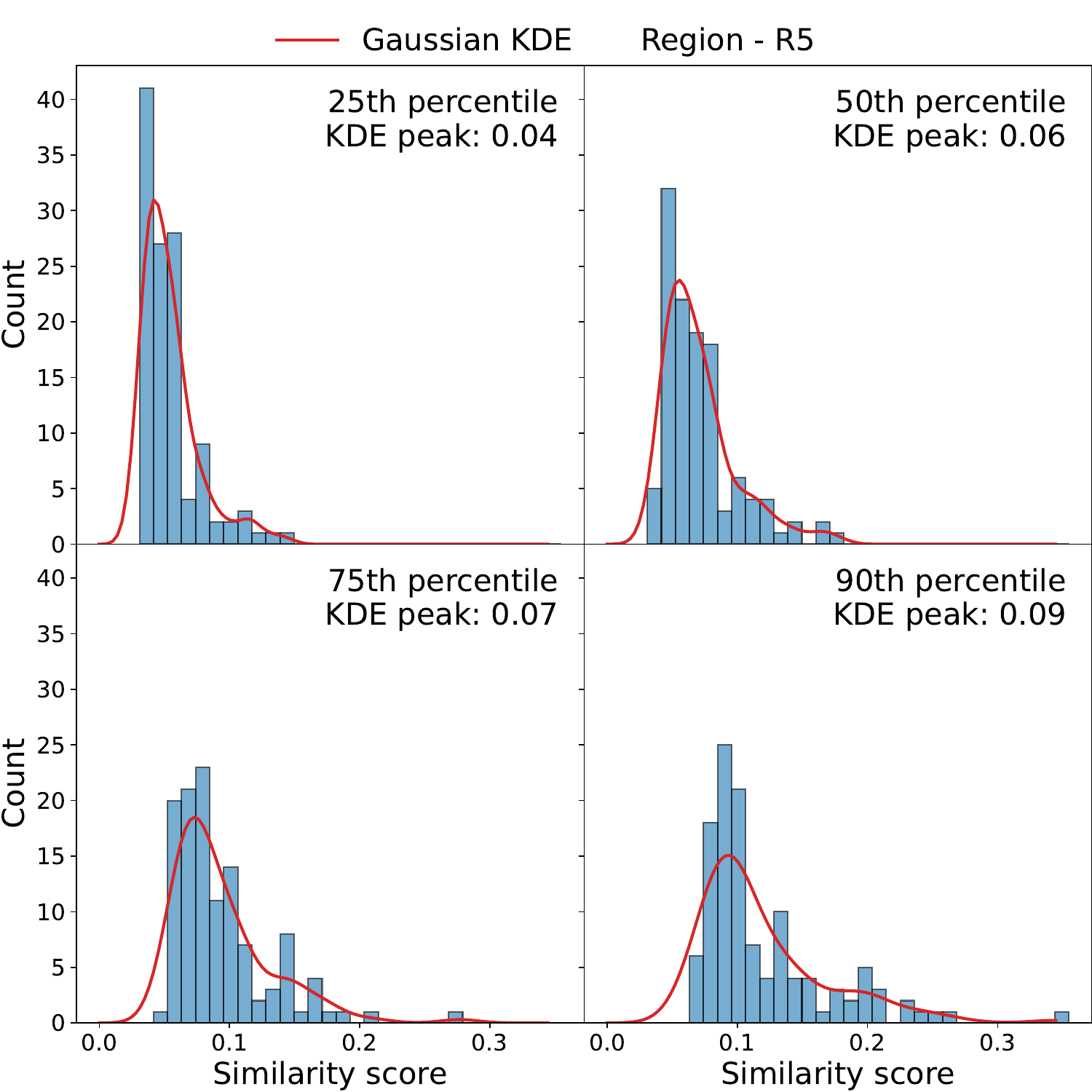} &     \includegraphics[width=0.29\textwidth]{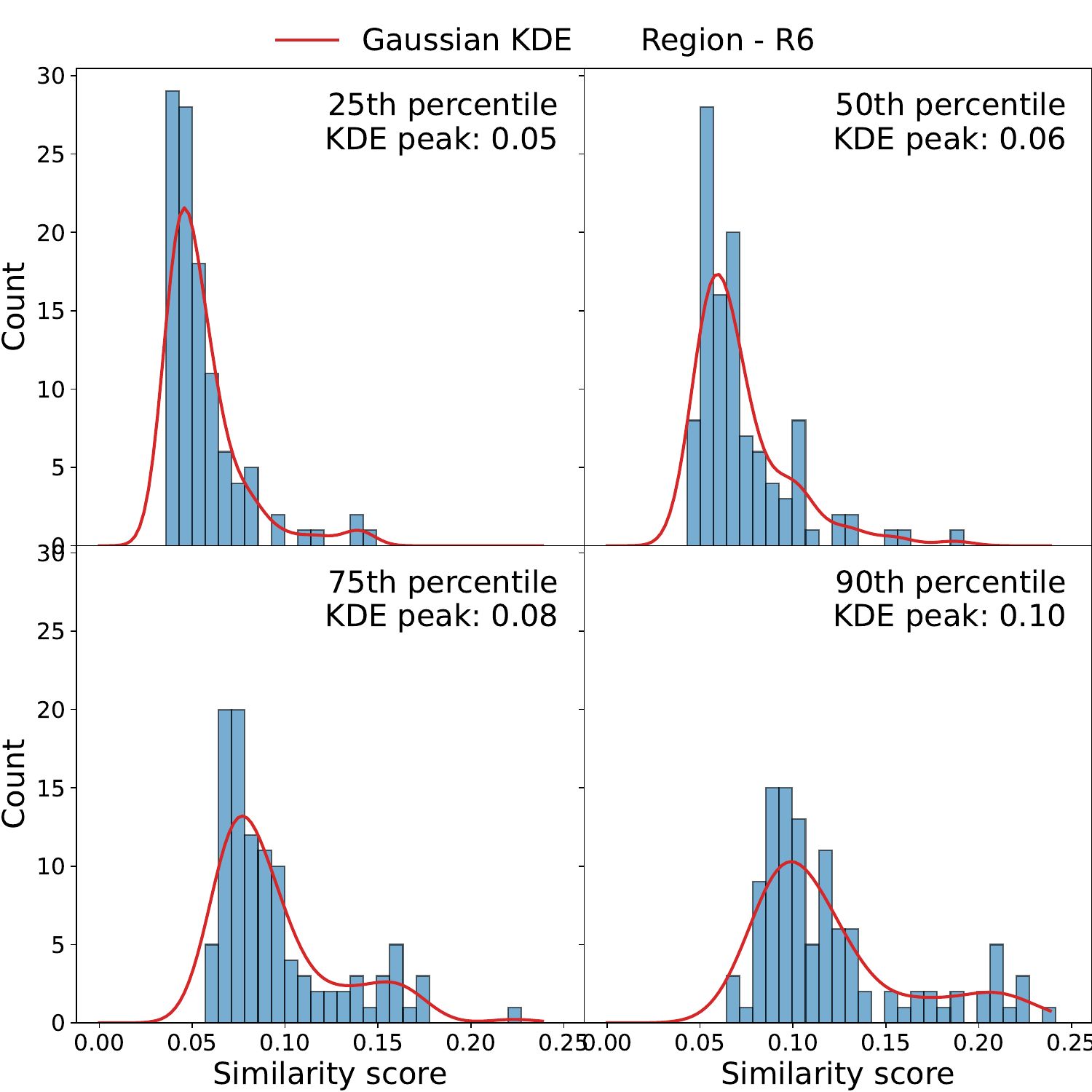}
        \\
    \end{tabular}
    \caption{Histogram of the percentiles (25th, 50th, 75th, 90th) of similarity scores for all spectral regions using the XP-LAMOST catalogue.}
\end{figure*}

\end{appendix}

\end{document}